\documentclass[superscriptaddress,
reprint,
 amsmath,amssymb,
 aps,
floatfix,
]{revtex4-2}

\usepackage{svg}
\usepackage{physics}
\usepackage{braket}
\usepackage{tikz}
\usepackage{graphicx}
\usepackage{dcolumn}
\usepackage{bm}
\usepackage{xcolor}
\usepackage{soul}
\usepackage[normalem]{ulem}
\usepackage{tabularx}
\usepackage{comment}

\newcommand{\mbf}{\mathbf}

\usepackage{soul}

\newcounter{stage}
\newenvironment{stage}[1]
  {\noindent\hrulefill\\
    \refstepcounter{stage}\textbf{Stage \Roman{stage}:} #1}
  {\hrulefill}

\newcounter{build}
\newenvironment{build}[1]
  {\noindent\hrulefill\\
    \refstepcounter{build}\textbf{Build} #1}
  {\hrulefill}

\newcommand{\MG}[1]{{\color{teal} #1}}

\begin{document}


\title{Heralded photonic graph states with inefficient quantum emitters}

\author{Maxwell Gold}
\thanks{These two authors contributed equally}
 \affiliation{Department of Physics, University of Illinois Urbana-Champaign, Urbana, IL 61801}
 
\author{Jianlong Lin}%
\thanks{These two authors contributed equally}

 \affiliation{Department of Electrical and Computer Engineering, University of Illinois Urbana-Champaign, Urbana, IL 61801}
 

\author{Eric Chitambar}
 \affiliation{Department of Electrical and Computer Engineering, University of Illinois Urbana-Champaign, Urbana, IL 61801}
 
 \author{Elizabeth A. Goldschmidt}%
 \email{goldschm@illinois.edu}
 \affiliation{Department of Physics, University of Illinois Urbana-Champaign, Urbana, IL 61801}

\date{\today}

\begin{abstract}
Quantum emitter-based schemes for the generation of photonic graph states offer a promising, resource efficient methodology for realizing distributed quantum computation and communication protocols on near-term hardware. We present a heralded scheme for making photonic graph states that is compatible with the typically poor photon collection from state-of-the-art coherent quantum emitters. We demonstrate that the construction time for large graph states can be polynomial in the photon collection efficiency, as compared to the exponential scaling of current emitter-based schemes, which assume deterministic photon collection. The additional overhead here consists of an extra spin qubit plus one additional spin-spin entangling gate per photon added to the graph. While the proposed scheme requires both non-demolition measurement and efficient storage of photons in order to generate graph states for arbitrary applications, we show that many useful tasks, including measurement-based quantum computation, can be implemented without these requirements. As a use-case of our scheme, we construct a protocol for secure two-party computation that can be implemented efficiently on current hardware. Estimates of the fidelity to produce graph states used in the computation are given assuming current and near-term fidelities for highly coherent quantum emitters.

\end{abstract}

\maketitle


\clearpage

\section{\label{sec:intro} Introduction}
The vast promise of quantum information technologies for faster and more secure computational and information systems relies on entanglement as a primary resource. Traditional gate-based quantum computing requires the ability to perform sequences of joint operations across multiple qubits. An alternative paradigm, known as measurement-based quantum computation (MBQC), realizes universal computation through sequences of single qubit measurements made on an initially prepared entangled resource state \cite{Raussendorf-2001a}. This is an appealing approach for photonic quantum systems as single photon rotations and measurements are straightforward using commercial optical elements, and fast efficient routing solutions are readily available \cite{Browne-2005a}. Furthermore, the sequential nature of photon emission allows entangled states to be built from photons emitted at different times by the same emitter \cite{Lindner-2009a}. By using entangled emitters, it then becomes possible to simulate entangling gates between photons and overcome the difficulty of realizing direct photon-photon interactions \cite{Economou-2010a}. Indeed, it is known that computationally useful graph states of photons can be generated using either a small number of coherent quantum emitters \cite{Buterakos-2017a, Russo-2019a}, a combination of a single emitter and fusion gates \cite{Hilaire-2022a, Meng-2023a, Thomas-2024a}, or a single quantum emitter in a feedback scheme \cite{Zhan-2020a}. 

All of these aforementioned schemes assume a photon is successfully added to the graph every time an emitter is excited, and we thus term this class of schemes as being ``deterministic''. For realistic systems with less than perfect emission and collection efficiency, the time to make a graph thus scales exponentially in the size of the graph because any failure to detect a photon triggers a restart of the whole protocol. In general, highly efficient collection from individual quantum emitters is a challenge that remains largely out of reach today, making such deterministic schemes impractical for generating even moderately sized graph states ($10-100$ photons) on near-term hardware \cite{Schwartz-2016a, Schupp-2021a, Thomas-2022a,  Cogan-2023a}. 

We introduce in Sec. \ref{sec:EtA} a method for photonic graph generation based on coherent emitters that uses an ``emit-then-add" approach, where each photon is attached to the graph only following confirmation of its emission and collection. In this way, a lost photon does not truncate the rest of the graph and the time to produce a photonic graph state scales polynomially in the size of the graph. Compared to the deterministic protocols, one additional spin is required in order to allow the emitter to be disconnected from the graph until photon collection is confirmed. Furthermore, for each photon added to the graph, our scheme requires one additional spin-spin entangling operation plus one mid-circuit measurement and reset (MCMR) of the emitting spin. (Note that these only occur upon heralding of successful photon emission, they are not required on every attempt.)

Typically, confirming the presence of a photon in a particular optical mode requires destructive measurement of the photon. For arbitrary applications of photonic graph states where the photons are measured in any order and in any basis, one would thus require  quantum non-demolition (QND) measurement \cite{Munro-2005a, Xiao-2008a, Yanagimoto-2023a}, in order to determine successful collection of the photon before adding it to the larger graph, plus storage of the photon until it can be measured in the desired basis. Such a scheme is largely out of reach today, though we discuss in Appendix \ref{app:APP} a method for implementing this on current hardware based on entanglement swapping with a separately produced entangled photon pair. 

Importantly, we show here that a large class of graph state protocols (including MBQC) admit projective measurements of each photon before adding it to the larger graph. In this way we show how to build a heralded virtual graph state of photons that never exist at the same time. Large virtual graph states can be built quickly with high fidelity in this way. We introduce an example use case in Sec. \ref{sec:MPC}, a particular protocol that uses multiple copies of a small virtual graph state to implement secure two-party computation on arbitrarily large inputs and requires just a single quantum emitter and two auxiliary spin qubits. We demonstrate how the coherence and entanglement fidelity requirements for this protocol are likely met by current state-of-the-art to near-term trapped ion and trapped neutral atom systems.

\section{\label{sec:EtA} Emit-then-add}
Given the generally poor collection efficiency shown by highly coherent quantum emitters that can be entangled with additional spin qubits we propose a general methodology, dubbed ``emit-then-add,'' for constructing photonic graph states to circumvent this issue. A single emitter is initialized into an unentangled state and excited to produce a single photon that is entangled with its long-lived internal spin state and collected with some overall efficiency $\eta_e$. This can be implemented with a wide variety of quantum emitters including laser-cooled atoms or ions, quantum dots, and defects or dopants in wide-bandgap semiconductors \cite{Lindner-2009a, Bock-2018a, Economou-2006a, Cogan-2023a, Gimeno-Segovia-2019a, Rielander-2016a}. Photons can be encoded in various degrees of freedom including polarization, time-bin, and frequency \cite{Covey-2023a, Ward-2022a, Krutyanskiy-2023a, Jayakumar-2014a, Senellart-2017a}. In addition to the single emitter, we require a set of auxiliary spin qubits that can be controllably entangled in a pairwise way (with the emitter and with each other) via local and deterministic two-qubit spin-spin entangling gates. We note that these auxiliary spins are never used for emission of photons for the graph and can be different physical qubits from the emitter (i.e., a different atomic species in an atom or ion system or nearby nuclear spins coupled to a defect or dopant emitter in solid-state) \cite{Anand-2024a, Bruzewicz-2019a, Bradley-2019a}. We see that our protocol works best for an emitter optimized for collection efficiency and fast and high-fidelity preparation and readout, while the auxiliary spins are optimized for coherence.

Deterministic schemes for constructing photonic graph states, such as those in \cite{Lindner-2009a, Russo-2019a}, center their building operations on the emitting spin itself, while the central feature of our method is to add a photon to the graph only following a logical herald of its successful emission and collection. Successfully adding each photon to the graph state requires entangling the emitter with an auxiliary spin, and then measuring it out of the graph and reinitializing it to prepare for the next attempted photon emission. Following a correction to the photon and auxiliary spin based on the outcome of this measurement, the state of the system is as if the auxiliary spin had directly emitted the photon itself. In this sense, we construct a graph state of only heralded photons, where we have split the role of the emitter in deterministic schemes into an emitting spin that generates the photons and an auxiliary spin that stores the quantum information. Thus, we see that emit-then-add admits the construction of arbitrary photonic graph states using the same methods described in \cite{Russo-2019a} with minimal additional overhead. Namely, the graph state itself is built using a sequence of two-qubit and local Clifford gates on the spins and only local Clifford gates on the photons. The local gates on each photon can all be combined and applied as a single rotation prior to using the graph state in subsequent applications.

The key improvement of our scheme is that uncollected photons do not disturb the state of the graph that is being built. When successive photon collection is required for building a graph on a single emitter, any failed detection traces the photon out of the system and truncates the graph, which necessitates restarting construction from scratch. In typical deterministic quantum emitter-based schemes, any inefficiency in collection therefore leads to exponentially poor scaling with graph size. Given current hardware, this severely limits the size and rate of generation for photonic graph states, particularly those that require multiple spin qubits. We note that there are loss tolerance and percolation thresholds that improve this scaling, but they require much better collection efficiency than is feasible with near-term systems \cite{Varnava-2006a, Morimae-2012a, Morley-Short-2017a, Pant-2019a, Bartolucci-2023a, Lobl-2024a}. In our scheme, any failed detection simply results in the reinitialization of the emitting spin without any disturbance to the overall graph under construction. 

For general purpose applications, we require a method to detect the presence of a photon that is non-destructive. This can be met with any form of QND measurement that heralds spin-photon entanglement. In Appendix \ref{app:APP} we provide an example for implementing such a scheme on current hardware, utilizing entanglement swapping and a separately produced entangled photon pair. Here, a pair of entangled photons is probabilistically produced, which can be implemented via standard nonlinear optical processes, such as spontaneous parametric down conversion (SPDC) \cite{Schneeloch-2019a, Zhang-2021a}. One member of the photon pair, the signal photon, is wavelength and bandwidth matched to the emitter photon, and the entangled pair is encoded in the same degree of freedom as the emitter photon. The emitter photon and signal photon are sent to a joint measurement apparatus which, upon a successful measurement outcome, projects the emitter spin and the unmeasured photon, the idler photon, onto an entangled Bell state \cite{Pan-1998a}. Since only certain measurement outcomes correspond to entanglement between the emitter spin and idler photon, the procedure is repeated until a successful herald is flagged. While this procedure works in principle, it has some major drawbacks including a significant reduction in rate due to the probabilistic nature of generating entangled photon pairs as well as the introduction of additional infidelities due to factors including  multi-pair production and imperfect heralding efficiency.  

\begin{figure}
    \centering
    \includegraphics[width=0.48\textwidth]{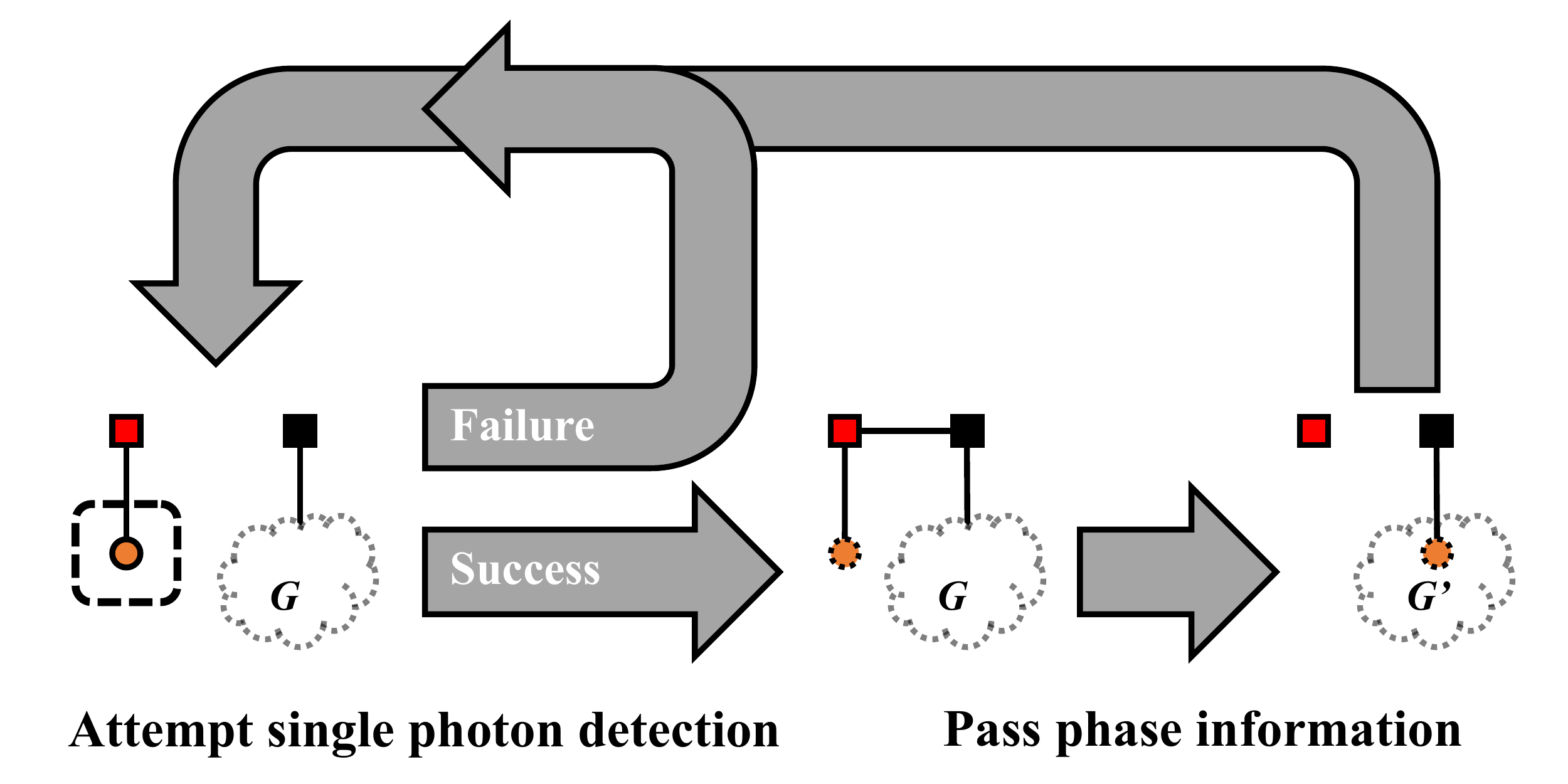}
    \caption{\label{fig:VGS-scheme} A construction scheme for heralding virtual graph states, with emit-then-add. A quantum emitter (red) is repeatedly excited until a single photon (orange) is successfully detected. Upon detection, logic is executed which passes the conditional phase information written on to the emitter to an auxiliary memory spin (black), which encodes a virtual graph state (represent by dashed boundaries).
    }
\end{figure}

Instead, we show that for many applications, including MBQC, non-destructive measurement is not required. For these applications the nodes in the graph are measured sequentially with the choice of measurement basis on one node depending on outcomes of previous ones. It is, thus, not necessary to build the full graph state before beginning these measurements, as an emitter photon can be destructively measured as soon as the correct basis for measurement is determined \cite{Houshmand-2018a}.  This measurement thus serves doubly as a prescribed step in the MBQC protocol and as a herald for the emitter photon. Upon failure to detect the photon, we reinitialize the emitter and attempt generation again, as we would with some QND measurement. Upon successful detection, we perform all required logic on the emitter and auxiliary spins, measure the emitter out of the graph, and reinitialize the emitter to attempt generation of the next photon. This simpler emit-then-add scheme is shown in Fig. \ref{fig:VGS-scheme}, and it has the advantage that it requires no storage of the emitter photon prior to measurement. The graph state being constructed is therefore ``virtual'' in the sense that not all photons within the graph need exist at the same time.

There are limitations, however, for when this scheme can be employed. Namely, a photon can be destructively measured immediately upon generation provided that two conditions are satisfied: (1) the correct measurement basis for that photon, as set by the protocol, is determined prior to its emission, and (2) this measurement is either Pauli $Z$ or of the form $\cos\phi X+\sin\phi Y$.  Condition (1) can be met for MBQC, as the emission order can be chosen to match the measurement order of the computation, albeit at the cost of using extra auxiliary spins and two-qubit gates in some cases \cite{Li-2022a}.  The second condition can also be met in principle, since the specified gate sets are sufficient for universal MBQC \cite{Raussendorf-2001a, Raussendorf-2003a}.

To understand condition (2) in more detail, note that if the full graph state were built such that the emitter is measured before its emitted photon, then each photon would have a local Clifford error of the form $UZ^m$, where $U$ is a fixed Clifford and $m\in\{0,1\}$ is determined by the decoupling measurement on the emitter spin.  If $M$ is the measurement to be subsequently performed on the photon in the MBQC protocol, then when correcting for the Clifford error, the effective measurement would be $M'=U Z^m MZ^m U^\dagger$.  When $M=Z$, then $M'=UMU^\dagger$; or when $M=\cos\phi X+\sin\phi Y$, then $M'=(-1)^m UMU^\dagger$.  In the first case, the dependence on $m$ is completely removed, and the photon can be equivalently measured with $UMU^\dagger$ immediately after it is emitted in this simpler scheme. In the second case, it can also be immediately measured with $UMU^\dagger$, but now one must perform a bit flip on the classical measurement outcome if $m=1$; this is because an overall $-1$ factor on a spin observable simply flips the spin-up/spin-down outcomes. In total, the correct computation can still be attained, because protocols admitting only measurements of this form write their outcomes as a conditional phase on their neighbors in the graph. In our scheme, this conditional phase information is imparted on the emitter with each successful single photon detection, and subsequently passed into memory on one of the auxiliary spins.  Computation is then achieved by passing conditional phases into memory in a specific pattern, performing logic between auxiliary spins, and classical post-processing for observables of the specified form.

While satisfying conditions (1) and (2) above is sufficient for universal MBQC using virtual graph states, this is often not the most efficient method in terms of the overall number of photons used to drive the computation.  For example, building out the graph to a certain depth and measuring in a different sequence than the emission order can lead to more compact gate implementations \cite{Raussendorf-2003a}. Also, using MBQC measurements outside of the x-y plane allows for more general forms of information flow \cite{Browne-2007a}. Nevertheless, as we show below, this simpler scheme can enable dramatically faster construction and higher fidelity with no photon storage required. For emit-then-add schemes incorporating some form of QND measurement instead, photon storage is required for at least the time to perform the spin-spin entangling gates and the decoupling measurement on the emitter. For the remainder of this section, we discuss in more detail the advantages of emit-then-add, specifically in the context of this simplified construction scheme for heralding virtual graph states that contain $n_p$ photons. 

\begin{figure*}
    \centering
    \includegraphics[width=0.8\textwidth]{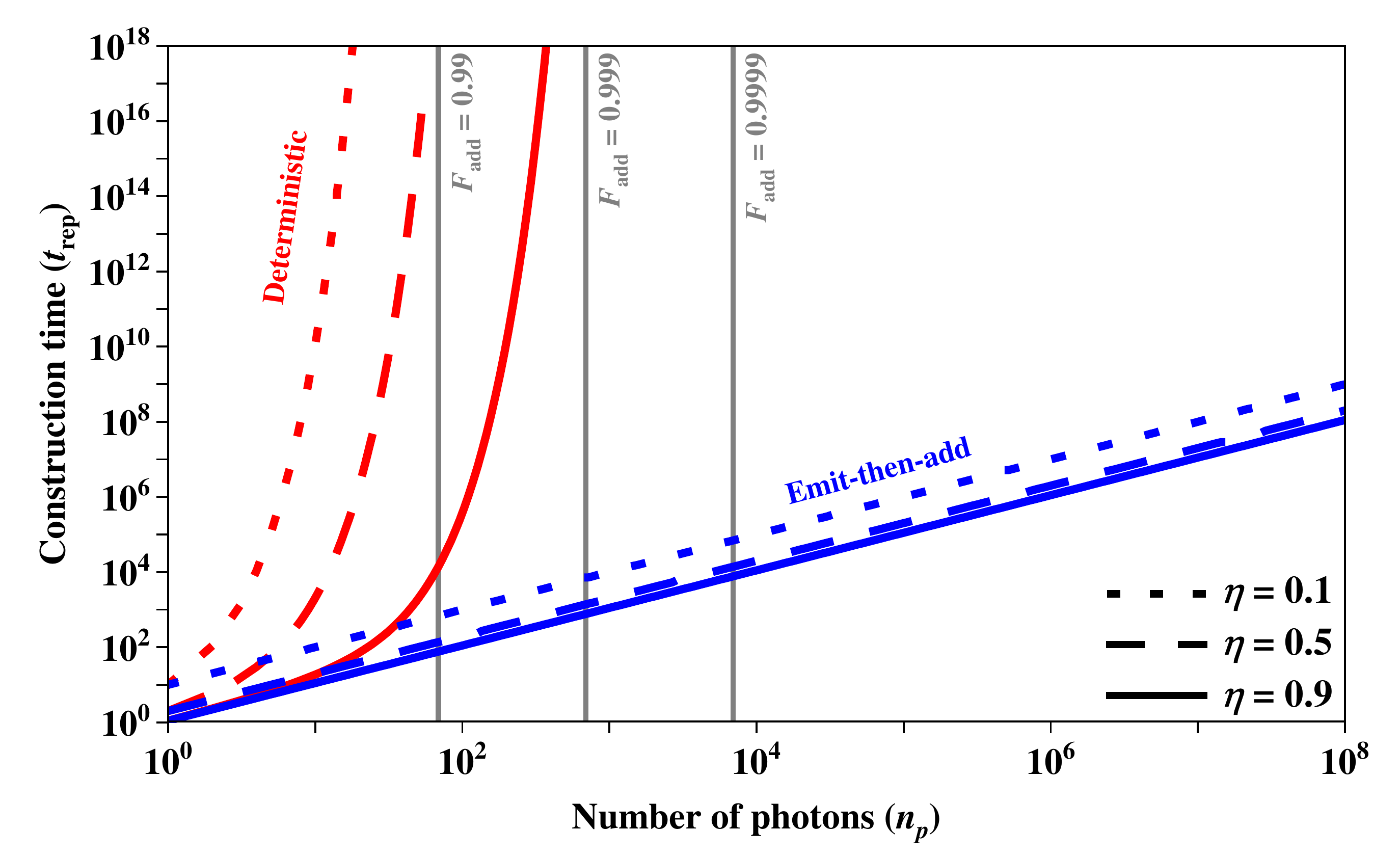}
    \caption{\label{fig:scaling} Time to make photonic graph states of size $n_p$ in units of the repetition period. Principal deterministic schemes (red) and our emit-then-add (blue) scheme are compared. The three curves of each color represent emitter photon collection efficiencies $\eta_{e}\in\{0.1, 0.5, 0.9\}$. The polynomial scaling in our schemes allows for the construction of larger photonic graphs. Cutoffs (grey, vertical) are depicted where the additional infidelity associated with emit-then-add precludes the construction of larger graph states with fidelities better than $1/2$, for $F_{\text{add}}\in\{0.99, 0.999, 0.9999\}$. Despite these cutoffs, graphs states of $10-100$ photons are achievable on realistic timescales, with only moderate requirements on $F_{\text{add}}$.}
\end{figure*}

\textbf{Overhead.} It is known that the sequential nature of photon emission imposes nontrivial resource requirements when constructing graph states, specifically in terms of the number of required emitters and two-qubit spin-spin entangling gates needed to build out the graph \cite{Li-2022a}. As discussed in the introduction, constructing graph states with emit-then-add further adds to this overhead. The total number of spins we require is only one more than the deterministic case, as each auxiliary spin in our scheme replaces an emitter in deterministic schemes and only a single emitting spin is required. In practice, more emitting spins can be employed in parallel to speed up the construction. As the emitting spin in our scheme must remain disjoint from the graph for each excitation, an additional two-qubit spin-spin entangling gate is required per successfully heralded virtual photon, in order to entangle the emitter with the larger graph. In Appendix \ref{app:CON}, we demonstrate efficient construction subroutines for building certain graph states that directly employ the methods of \cite{Russo-2019a}.

\textbf{Fidelity.} In addition to overhead, there are infidelities that affect the final graph state with emit-then-add that are not present in the deterministic scheme, namely any infidelity in the additional two-qubit spin-spin entangling gate and mid-circuit measurement and reset of the emitter. We capture all of this, plus the initial fidelity of the emitter-photon entanglement, in a parameter, $F_{\text{add}}$, for the total fidelity of the process to add a photon to the graph. Note that photon detection inefficiency does not affect the fidelity because failing to detect a photon simply triggers a reset. Any infidelity in measuring the state of the photon does affect $F_{\text{add}}$, but this can be minimal for photons (and has the same effect as for the deterministic schemes for photonic graph state production).

Importantly for our scheme, when building a graph all auxiliary spins must remain coherent for the entire time they are a part of the graph, which is much longer than the $n_p$ repetition cycles over which the graph is generated in the deterministic scheme. Thus, moving from the deterministic scheme to one proposed here effectively means moving from a scheme limited by the photon collection efficiency to a scheme limited by spin coherence (denoted $\tau$). In modeling the effects of this dephasing, we note that the emitter is reinitialized with each attempt to add a new photon to the larger graph, and hence it is only required to remain coherent for time to add it to the larger graph, denoted $t_{\text{add}}$. For the auxiliary spins, on the other hand, the average time for the successful addition of a new photon to the larger graph goes as $t_{\text{rep}}/\eta_e$ where $t_{\text{rep}}^{-1}$ is the experimental repetition rate. Hence, the total contribution to the final state fidelity from the emitter and any auxiliary spins is
\begin{subequations}
    \begin{eqnarray}
        F_D^{(e)}(n_p) &&= \left(\frac{1}{2}\left(1+e^{-(t_{\text{rep}}+t_{\text{add}})/\tau}\right)\right)^{n_p}, \label{eq:emt-dephasing-main}\\
        \langle F_D^{(s)}(n_p)\rangle&&= \frac{1}{2}\left(1+e^{-n_p (t_{\text{rep}} /\eta_e+t_{\text{add}})/\tau}\right),\label{eq:aux-dephasing-main}
    \end{eqnarray}
\end{subequations}
where we include a superscript to denote the emitter $(e)$ and auxiliary spins $(s)$, respectively. For simplicity in the above, we have used the same coherence time $\tau$ for the emitter and auxiliary spins in the system, though this may not be the case for hybrid systems \cite{Bruzewicz-2019a, Anand-2024a}. 

Given typical photon collection efficiencies, experimental repetition rates, and entangling gate and MCMR speeds, we expect to be in the regime of $t_{\text{rep}}\ll t_{\text{add}}\lesssim t_{\text{rep}}/\eta_e\ll\tau$ where the emitting spin must have a coherence time longer than the time to add a single photon to the graph, while the auxiliary spins must have much longer coherence times that account for all the failed attempts as well as repeated entangling gates and MCMR operations. Thus, a realistic implementation of this scheme requires auxiliary spins that have exceedingly large coherence times, with the requirement increasing for increasing graph size. Trapped ion and neutral atom systems have been shown to host second-scale coherence times \cite{Bluvstein-2022a, Wang-2021a}, which should allow generation of moderately sized graph states (10-100 photons), even with the relatively slow ($\lesssim\text{ms}$) entangling gate and MCMR times that are typical of these systems.

\textbf{Scaling.} Finally, the most significant advantage of emit-then-add comes in scaling up the size of graph states using state-of-the-art to near-term hardware. The average time to successfully generate a $n_p$-photon graph via the deterministic scheme, by directly collecting photons from an emitter in $n_p$ subsequent excitation events, is $\mathcal{O}(\eta_{e}^{-n_p})$, where $\eta_e$ is the emitter collection efficiency, because any failed detection event truncates the graph. For our emit-then-add scheme, a failed detection of the emitter photon simply triggers reinitialization of the emitter and another attempt at photon emission, while the graph under construction remains unaffected. Therefore, a $n_p$-photon graph state will be created over a time that is $\mathcal{O}(n_p\eta_{e}^{-1})$.

As a comparison, the time to make a $n_p$-photon graph state (in units of the repetition period for exciting the emitter) is shown in Fig. \ref{fig:scaling} for $\eta_{e}\in\{0.1,0.5,0.9\}$ for deterministic (red), and emit-then-add (blue) schemes. We do not include the time to perform gates and MCMR operations of the spins for these estimates, and we assume that the emitter coherence does not limit the graph size for either scheme (i.e. $\tau/n_p\gg t_{\text{rep}}/\eta_e+t_{\text{add}}$). We include cutoffs where the additional infidelity, captured by $F_{\text{add}}$ defined above, precludes the construction of larger graph states. Explicitly, given the current state-of-the-art for trapped ion and neutral systems \cite{Evered-2023a, Srinivas-2021a, Clark-2021a, Gaebler-2021a, Norcia-2023a}, the vertical gray lines in Fig. \ref{fig:scaling} represent where the fidelity of the final graph state generated with emit-then-add falls below $1/2$.

\section{\label{sec:MPC} Two-party computation with graph states}

\begin{figure*}
    \centering
    \includegraphics[width=0.6\textwidth]{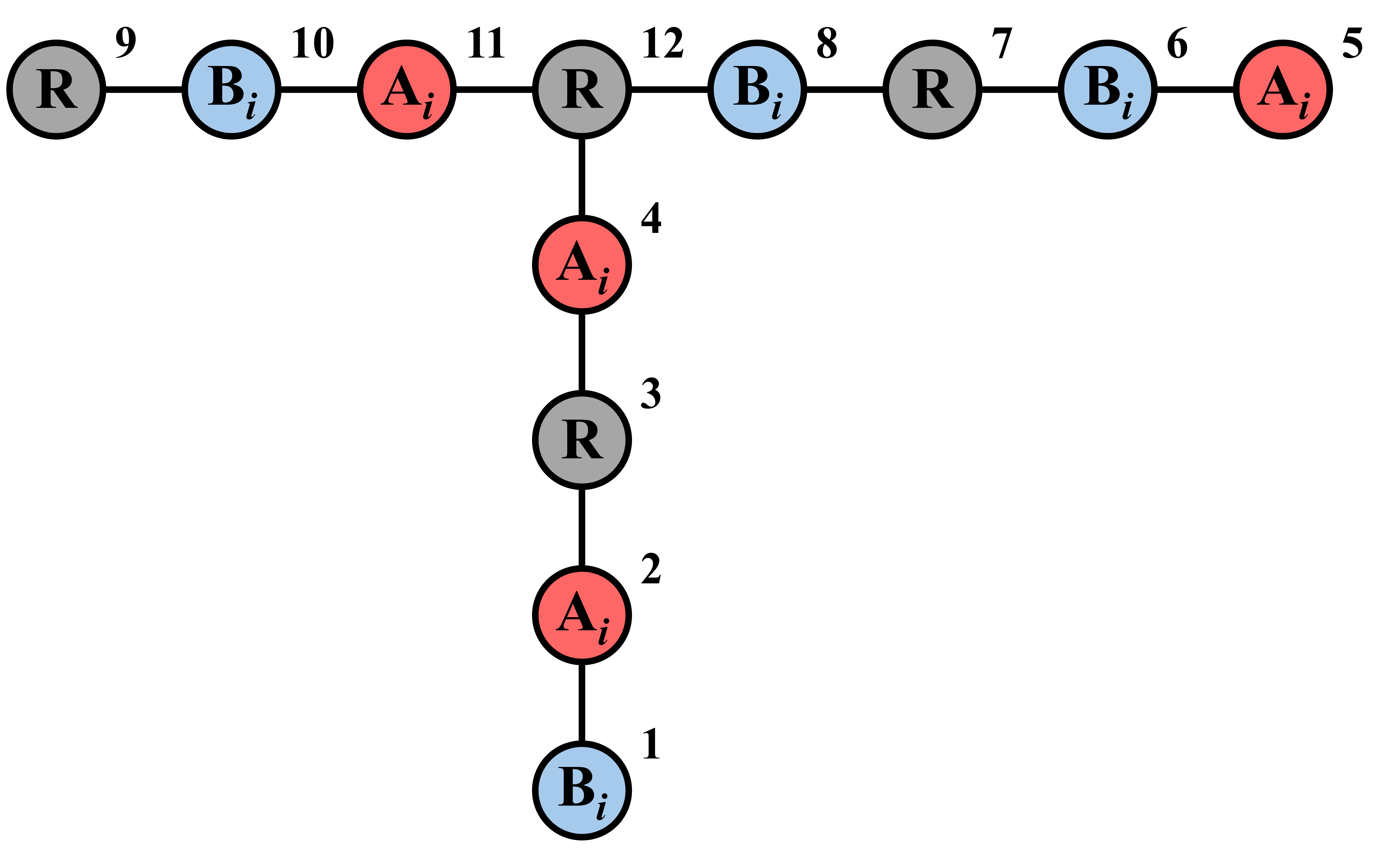}
    \caption{\label{fig:G} For each requisite bit conjunction in Stage I, the 12-qubit graph state $\ket{G}$ is distributed to $\mathbf{R}$ (the Referee) and pair of parties labeled $\mathbf{A}_i$ (Alice), $\mathbf{B}_i$ (Bob). When the measurement sequence detailed by the protocol is implemented honestly a correlation is generated between the parties, such that $a_ib_i=m_{i,12}+\alpha_i+\beta_i$, where $\alpha_i$ and $\beta_i$ are locally computed by Alice and Bob, respectively, from openings made in Stage II. Hence, we utilize each copy of the state to achieve a homomorphic encryption of a bit conjunction $a_ib_i$, where Alice, Bob, and the Referee each posses an additive share. The numeric script above each qubit reflects an example photon emission order for the generation of the graph state.}
\end{figure*}

As one application of our emit-then-add scheme, we describe a method for securely computing an arbitrary Boolean function $f:\{0,1\}^{\times n}\to\{0,1\}$ of either two parties or a restricted class of multi-party functions. As a special type of MBQC, our protocol, $\mathcal{P}$, performs the calculation through a sequence of measurements on distributed graph states, followed by classical broadcasting and local processing.  Only Pauli measurements are needed, so $\mathcal{P}$ can be implemented without the need for photonic memory, where the generation of the requisite resource state happens in parallel with the computation. Furthermore, the size of each graph state required for $\mathcal{P}$ is fixed at $12$ photons, regardless of the number of parties or the size of their inputs, which bodes well for state-of-the-art to near-term hardware.

Secure multi-party computation (MPC) is a task in which two or more parties compute some function on their individually held variables without revealing the values of the variables to each other \cite{Yao-1982a, Goldreich-1987a}.  For example, in Yao’s famous millionaire problem, two parties want to determine whose bank account has the most money without actually revealing how much money is in each account.  MPC is a deeply-studied topic in both classical and quantum cryptography, and a variety of MPC protocols have been proposed achieving different levels of security and relying on different operational assumptions \cite{Zhao-2019a}.  

We propose a method for restricted MPC that includes all two-party computations and requires only two rounds of public communication in the form broadcasting, or \textit{openings}, regardless of the size of the computational input. Our protocol follows a well-known approach of decomposing an MPC into \textit{offline} and \textit{online} phases \cite{Beaver-1992a, Nielsen-2012a, Damgard-2012a}. In the offline phase, some universal computational resource is distributed to all the parties. Crucially, this resource does not depend on the particular function being computed other than its input size.  Then in the online phase this resource is used to compute some chosen function of the parties’ inputs. For example, in the classical setting one well-known computational resource is a special form of shared randomness known as “Beaver triples,” which can be used to efficiently compute logical AND gates in the online phase \cite{Choudhury-2017a}. The problem of MPC then reduces in part to secure and efficient offline methods for distributing shared randomness, such as Beaver triples, which directly enable secure online computation. In a similar spirit, our protocol involves distributing certain quantum graph states in the offline phase which then enable the computation of a logical AND in the online phase, through quantum measurement and classical post-processing. 

Beyond its relatively low communication costs, a significant advantage of our protocol is that the parties can, in principle, use self-testing methods to unconditionally verify that some untrusted Source is faithfully distributing the correct graph state \cite{Takeuchi-2019a, Unnikrishnan-2022a}, an ability that does not exist for classical sources of shared randomness. Furthermore, we prove that our online phase offers unconditional \textit{privacy} against an arbitrary malicious adversary, with access to a general quantum instrument \cite{Davies-1970a}, in the sense that honest participation reveals no information about a party's input, other than what can be inferred from the evaluation of the final computed function. 

Note that in our discussion of the protocol here we do not offer any means of ensuring fairness, such \textit{security with abort} or \textit{guaranteed output delivery}. We do not claim that the full MPC here is unconditionally secure in this sense, and a classical means of ensuring fairness for the broadcasts in our online phase can be employed with computational security guarantees \cite{Gordon-2015a, Damgard-2020a}. Instead, the novelty of this work is in the ability to disseminate a specific online correlation with unconditional security. In a similar vein as quantum key distribution, we demonstrate how quantum states offer the ability to distribute information-theoretic secure shared randomness, in our case, in a form that directly enables MPC.

\textbf{Overview.} Suppose that $N$ parties $\mathbf{P}_1,\cdots,\mathbf{P}_N$ wish to compute some Boolean function $f(\mathbf{x}_1,\cdots,\mathbf{x}_N)$, where $\mathbf{x}_k$ is a string of bits representing the input for party $\mathbf{P}_k$. In addition to correctness, the evaluation of $f$ should be done securely such that the parties learn no more information about the individual $\mathbf{x}_1,\cdots,\mathbf{x}_N$ beyond their own input and what is revealed in the function value $f(\mathbf{x}_1,\cdots,\mathbf{x}_N)$.  To achieve this task, we propose a method of delegated computation in which a non-collaborating Referee, $\mathbf{R}$, is introduced to assist in the computation of $f(\mathbf{x}_1,\cdots,\mathbf{x}_N)$. To maintain privacy, $\mathbf{R}$ also should not learn any more information about the $\mathbf{x}_k$ beyond what is implied by the computed value $f(\mathbf{x}_1,\cdots,\mathbf{x}_N)$, nor does $\mathbf{R}$ reveal any more information to the other parties, other than what is necessary to evaluate $f$ securely.

We utilize the fact that every Boolean function $f$ can be expressed in an algebraic normal form (ANF), which presents $f$ as a sum (mod $2$) of different variable conjunctions.  That is, we can write $f=\sum_{i=1}^\mathfrak{R} \mathfrak{c}_i$, where each $\mathfrak{c}_i$ is the logical AND of a certain group of input variables. By combining variables belonging to the same party, every $\mathfrak{c}_i$ becomes the conjunction of at most $N$ variables, each one belonging to a different party. In this work, we restrict attention to functions $f$ that admit an ANF whose conjunctions involve no more than two variables. This covers the entire class of two-party functions, but also includes certain multi-party functions, such as the three-party majority function $\varphi_{3}(x,y,z)=xy+xz+yz\mod 2$, which outputs the majority value among inputs $x,y,z\in\{0,1\}$. In general, the functions we consider have the form $f=\sum_{i=1}^{\mathfrak{R}}a_ib_i+\sum_{k=1}^N z_k$, where $a_i$ and $b_i$ are the input bits of each quadratic conjunction, belonging to different parties, and $z_k$ is the input bit of the linear part of $f$, belonging to party $\mathbf{P}_k$. We let $\Tilde{\mathbf{x}}_k\subset\{a_i,b_i,z_k\}_{i=1,k=1}^{\mathfrak{R},N}$ denote all the inputs of $f$ belonging to $\mathbf{P}_k$.  Furthermore, if each $\Tilde{\mathbf{x}}_k$ is no more than $M$ bits, then $\mathfrak{R}\leq \binom{N}{2}(M-1)^2$.

The offline phase of $\mathcal{P}$ calls for the distribution of $\mathfrak{R}$ copies of the graph state $\ket{G}$ depicted in Fig. \ref{fig:G}, one copy for each conjunction $a_ib_i$ in $f$.  The generation of each copy of $\ket{G}$ requires only two auxiliary spins in the experimental schemes we propose \cite{Li-2022a}. We denote the $i^{\text{th}}$ copy as $\ket{G_i}$, and their distribution can be conducted in parallel. The specific qubits in each $\ket{G_i}$ are given to the parties $\{\mbf{A}_i,\mbf{B}_i\}\subset\{\mbf{P}_1,\cdots,\mbf{P}_N\}$ who, respectively, have inputs $\{a_i,b_i\}$ to the conjunction $a_ib_i$.  We assume without loss of generality that each party receives qubits belonging to at least one $\ket{G_i}$, an assumption that can be trivially ensured by adding conjunctions to $f$ of inputs that are identically zero.  This is a technical requirement of our protocol since the $\ket{G_i}$ will ultimately be used to generate one-time pad bits, and each party needs at least one. In practice, the $\ket{G_i}$ can be generated by an untrusted quantum Source, and their correctness can be certified using self-testing methods (see Appendix \ref{app:SECc}). Specific steps for building $\ket{G}$ in the experimental schemes given in Sec. \ref{sec:EtA} are presented in Appendix \ref{app:CONb}.

The online phase of $\mathcal{P}$ thereafter is split into two stages. Stage I involves measuring Pauli observables, $X,Y,Z$, on $\mathfrak{R}$ copies of the graph state $\ket{G}$, depicted in Fig. \ref{fig:G}, following the sequence below. 

\begin{stage}\label{stage:I}
    \textbf{Obtaining correlations from quantum states}.
    
    \noindent\textit{Input}: For $i\in\{1,\cdots,\mathfrak{R}\}$, the following measurement sequence is performed by $\{\mathbf{A}_i,\mathbf{B}_i\}\subset\{\mbf{P}_1,\cdots,\mbf{P}_N\}$,  along with $\mathbf{R}$, on copy $\ket{G_i}$. $\mathbf{A}$ inputs $a_i$.
    \begin{enumerate}
        \item[(I.1).] $\mathbf{B}_i$ and $\mathbf{A}_i$ measure $Z$ on qubits $1$ and $5$, respectively, obtaining measurement outcomes $m_{i,1}$ and $m_{i,5}$.
        \item[(I.2).] $\mathbf{A}_i$, $\mathbf{B}_i$, and $\mathbf{R}$ measure $X$ on qubits $2$, $6$, and $9$, respectively, obtaining measurement outcomes $m_{i,2}$, $m_{i,6}$, and $m_{i,9}$
        \item[(I.3).] $\mathbf{R}$ measures $Z$ on qubits $3$ and $7$, and $\mathbf{B}_i$ likewise measures $Z$ on qubit $10$. They obtain measurement outcomes $m_{i,3}$, $m_{i,7}$, and $m_{i,10}$.
        \item[(I.4).] $\mathbf{A}_i$ applies $Z^{m_{i,2}}$ to qubit $4$ and measures $W^{a_i}Z(W^\dagger)^{a_i}$ on qubits $4$ and $11$ thereafter, where $W\equiv(iX)^{1/2}$. She obtains measurement outcomes $m_{i,4}$ and $m_{i,11}$. Note that $WZW^\dagger=Y$
        \item[(I.5).] $\mathbf{B}_i$ applies $Z^{m_{i,6}}$ to qubit $8$ and measures $W^{m_{i,10}}Z(W^\dagger)^{m_{i,10}}$ thereafter, obtaining measurement outcome $m_{i,8}$.
        \item[(I.6).] $\mathbf{R}$ measures $V^{m_{i,9}}X(V^\dagger)^{m_{i,9}}$ on qubit $12$, where $V\equiv(-iZ)^{1/2}$, obtaining measurement outcome $m_{i,12}$. Note that $VXV^\dagger=Y$.
    \end{enumerate}
\end{stage}

When all the measurements through step (I.3) are made as specified, the parties obtain correlated measurement outcomes satisfying the relationships
\begin{subequations}
    \begin{eqnarray}
        m_{i,1}+m_{i,2}+m_{i,3}&&=0, \\
        m_{i,5}+m_{i,6}+m_{i,7}&&=0, \\
        m_{i,9} + m_{i,10}&&=0, \quad \left(s_i\equiv m_{i,9}=m_{i,10}\right).
    \end{eqnarray}
\end{subequations}
When the subsequent measurements through step (I.6) are made as specified, the Referee's final outcome satisfies
\begin{equation}
    a_ib_i = m_{i,12} + \alpha_i + \beta_i,
\end{equation}
where
\begin{subequations}
    \begin{eqnarray}
        \alpha_i &&= a_i(m_{i,1}+b_i+1)+ (m_{i,4}+m_{i,11})\\
        \beta_i &&= (m_{i,5} + a_i)s_i + m_{i,8}.
    \end{eqnarray}
\end{subequations}
The end result of each iteration $i$ of the measurement sequence leaves Alice, Bob, and the Referee with additive homomorphic shares of the bit conjunction $a_ib_i$. The only temporal restrictions on the sequence, are that measurements in steps (I.1)-(I.3) need be performed prior to those in steps (I.4)-(I.6). Furthermore, we can understand both $\alpha_i$ and $\beta_i$ as two layers of data that are nested together by one-time pads. For $\alpha_i$, the bit $(m_{i,4}+m_{i,11})$ serves as a pad (known only to Alice) for the two multipliers $a_i$ and $(m_{i,1}+ b_i + 1)$; while within the latter multiplier, the bit $m_{i,1}$ serves as a pad (known only to Bob) for $b_i + 1$. A similar interpretation holds for $\beta_i$.

The padded structure of the bit values obtained in Stage I enables the calculation of each $\alpha_i$ and $\beta_i$ using public communication (i.e. ``openings'') and local processing in Stage II of $\mathcal{P}$, without revealing any information about the corresponding $a_i$ or $b_i$. A second round of communication thereafter opens only the collective sum $\sum_{i=1}^{\mathfrak{R}}a_ib_i$ of the bit conjunctions encrypted in Stage I, while also adding the linear term $\sum_{k=1}^N z_k$, thereby allowing the secure evaluation of $f$.

\begin{stage}\label{stage:II}
    \textbf{Public communication and classical post-processing.}

    \noindent\textit{Input}: Let $\mathcal{A}_k\subset\{1,\cdots,\mathfrak{R}\}$ denote the set of conjunctions in which $\mathbf{P}_k$ played the role of $\mathbf{A}_i$, and similarly, let $\mathcal{B}_k\subset\{1,\cdots,\mathfrak{R}\}$ denote the set of conjunctions in which $\mathbf{P}_k$ played the role of $\mathbf{B}_i$. $\mathbf{P}_k$ inputs their bits from $\{\alpha_i\}_{i\in\mathcal{A}_k}$ and $\{\beta_i\}_{i\in\mathcal{B}_k}$, along with their bit $z_k$. $\mathbf{R}$ inputs their bits from $\{m_{i,12}\}_{i=1}^{\mathfrak{R}}$.
    \begin{enumerate}
        \item[(II.1).] For each $i\in\{1,\cdots,\mathfrak{R\}}$,
        \begin{enumerate}
            \item[(a).] $\mathbf{A}_i$ opens $c_{i,A}\equiv m_{i,5} + a_i$ and $\mathbf{B}_i$ opens $c_{i,B}\equiv m_{i,1}+b_i+1$. 
            \item[(b).] $\mathbf{A}_i$ locally computes $\alpha_i=c_{i,B}a_i + (m_{i,4}+m_{i,11})$ and $\mathbf{B}_i$ locally computes $\beta_i=c_{i,A}m_{i,10}+m_{i,8}$.
        \end{enumerate}
        \item[(II.2).] For each $k\in\{1,\cdots,N\}$, $\mathbf{P}_k$ opens $\Gamma_k\equiv z_k +\sum_{i\in\mathcal{A}_k}\alpha_i +\sum_{i\in\mathcal{B}_k}\beta_i$. Concurrently, $\mathbf{R}$ opens $\Gamma_\mathbf{R}\equiv\sum_{i=1}^{\mathfrak{R}}m_{i,12}$.
        \item[(II.3.)] All the parties locally compute $f=\Gamma_{\mathbf{R}}+\sum_{k=1}^N\Gamma_k$.
    \end{enumerate}
\end{stage}

It should be noted that by parallelization, Stage II can be performed using just two rounds of simultaneous communication between the parties. Indeed, each $\mathbf{P_k}$ needs to broadcast at most two public messages, the first being no more than $\log \mathfrak{R}_1$ bits, and the second being just one bit.  When run in parallel, all the step (II.1) messages can be broadcast concurrently, and likewise for the step (II.2) messages.  Moreover, this classical communication can be done after all quantum measurements are performed since the choice of local measurement at each step in $\mathcal{P}$ does not depend on the classical message of any other party.  Experimentally this is very desirable since it means that $\mathcal{P}$ can be implemented in three parts: (i) entanglement distribution (offline phase), (ii) local measurement of qubits (online phase Stage I), and (iii) classical post-processing (online phase Stage II).  Without the use of quantum memory it is not possible to cleanly separate parts (i) and (ii), and in practice each state $\ket{G_i}$ can be measured in a streaming manner.  This type of implementation does not affect the performance or security of the protocol.

\begin{figure}
    \centering
    \includegraphics[width=0.45\textwidth]{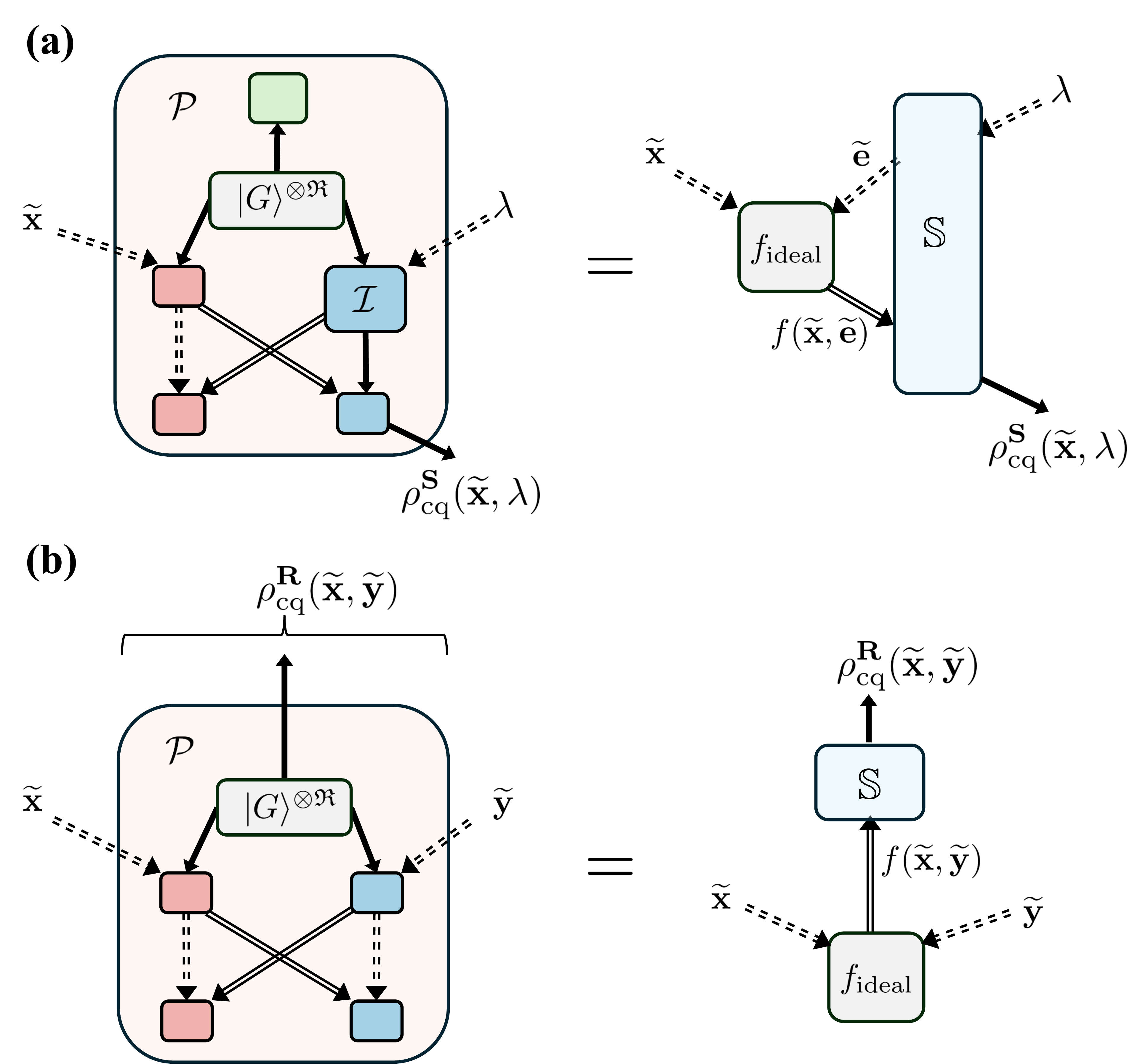}
    \caption{\label{fig:MaliciousAttack} Malicious attacks on $\mathcal{P}$ by $\mathbf{S}$ and $\mathbf{R}$. Real (left) vs. ideal (right) instantiations of the attacks are depicted, where boxes (red, blue, green) indicate local processing by a given party ($\mathbf{P},\mathbf{S},\mathbf{R}$). (a) General attack by $\mathbf{S}$. The attack is described by a quantum instrument $\mathcal{I}$ that acts on the qubits of $\ket{G}^{\otimes\mathfrak{R}}$ distributed to $\mathbf{S}$, with possible dependence on some side-information $\lambda.$ For every $\Tilde{\mathbf{x}}$ of $\mathbf{P}$, the attack will generate for $\mathbf{S}$ a classical-quantum state $\rho_{\text{cq}}^{\mathbf{S}}(\Tilde{\mathbf{x}},\lambda)$ that depends on all public communication in the protocol. We show there exists a simulator $\mathbb{S}$ in place of party $\mathbf{S}$ in the ideal world that submits some input $\Tilde{\mathbf{e}}$ to $f_{\text{ideal}}$ and then uses the output $f(\Tilde{\mathbf{x}},\Tilde{\mathbf{e}})$ to generate the same $\rho_{\text{cq}}^{\mathbf{S}}(\Tilde{\mathbf{x}},\lambda)$ achieved in the real world. (b) General attack by $\mathbf{R}$. We again show there exists a simulator $\mathbb{S}$ that takes the places of $\mathbf{R}$ in the ideal world, receiving just the function output $f(\Tilde{\mathbf{x}},\Tilde{\mathbf{y}})$ and outputting the classical-quantum state $\rho_{\text{cq}}^{\textbf{R}}(\Tilde{\mathbf{x}},\Tilde{\mathbf{y}})$ achieved in the real world, as shown to the right.}
\end{figure}

\textbf{Security.} We now turn to analyze the security of $\mathcal{P}$. At a high level, our security claim is that if $\mathbf{R}$ plays honestly, then $\mathcal{P}$ reveals no more information about $\Tilde{\mathbf{x}}_k$ of an honest party $\mathbf{P}_k$, beyond what can be inferred from the function values $f(\Tilde{\mathbf{x}}_1,\cdots,\Tilde{\mathbf{x}}_N)$. Here, an honest party is anyone who executes the steps of $\mathcal{P}$ faithfully. Our protocol also offers some security against a cheating $\mathbf{R}$. Namely, if all $N$ parties follow $\mathcal{P}$ honestly, then $\mathbf{R}$ is also unable to infer any information about the inputs beyond what can be computed from $f(\Tilde{\mathbf{x}}_1,\cdots,\Tilde{\mathbf{x}}_N)$.

Note that for $N$ parties $\mathbf{P}_1,\cdots,\mathbf{P}_N$ these claims allow for the possibility that any subset of $N-1$ parties are colluding against a single party. The colluding parties are allowed to share an unlimited amount of classical and quantum communication over side channels, and they can thus collectively be viewed as a single party $\mathbf{S}$. In this case, the task essentially reduces to a problem between the two parties $\mathbf{P}$ and $\mathbf{S}$, and $\mathbf{R}$. We let $\Tilde{\mathbf{x}}$ denote the input of $\mathbf{P}$ and $\Tilde{\mathbf{y}}$ as the total input of all the colluding parties constituting $\mathbf{S}$. The goal, then, is to securely compute $f(\Tilde{\mathbf{x}},\Tilde{\mathbf{y}})$ with an \textit{honest majority} in $\{\mathbf{P},\mathbf{S},\mathbf{R}\}$.

In this work we adopt a simulation based notion of security. Consider first an ideal world in which there exists some device $f_{\text{ideal}}$ that privately receives inputs $\Tilde{\mathbf{x}}$ and $\Tilde{\mathbf{y}}$ from $\mathbf{P}$ and $\mathbf{S}$, respectively, and then outputs $f(\Tilde{\mathbf{x}},\Tilde{\mathbf{y}})$ to $\mathbf{P}$, $\mathbf{S}$, and $\mathbf{R}$. These two worlds are described more thoroughly in Appendix \ref{app:SECa}. Intuitively, we want the ideal world to reveal at least as much information about the inputs $(\Tilde{\mathbf{x}},\Tilde{\mathbf{y}})$ than what is revealed to any dishonest party acting maliciously in the real world. 

This intuition is made precise through the notion of a simulator. A simulator, $\mathbb{S}$, in this setting is some device that can take the place of a party $\mathbf{X}\in\{\mathbf{P},\mathbf{S},\mathbf{R}\}$ and interact with the ideal functionality $f_{\text{ideal}}$ from the viewpoint of party $\mathbf{X}$. Consider now any action taken by a potentially dishonest party in $\mathcal{P}$. Our security definition is that there must exists a simulator $\mathbb{S}$ interacting with $f_{\text{ideal}}$ that exactly reproduces what the dishonest party obtains from $\mathcal{P}$. This operationally captures the idea that the adversary learns no more information about an honest party's input, other than what can be inferred directly from an ideal evaluation of $f$. Fig. \ref{fig:MaliciousAttack} depicts
the two forms of malicious attacks for which we prove that $\mathcal{P}$ is secure. A formal security definition and full proof are provided in Appendix \ref{app:SECa}.

\textbf{Performance.} To handle experimental errors in the above protocol, we can employ a simple repetition code to suppress the effects of any infidelity in our ability to make each copy of $\ket{G}$ in the offline phase of $\mathcal{P}$. By determining the total bit error probability associated with each share of $a_ib_i$, an arbitrarily small total bit error probability $\epsilon_f$ on the output $f$ can be chosen to set the number of repetitions required for Stage I. We use this information to estimate the lower bound rate of computation at which $N$ parties can compute $f$ on their $M$-bit inputs. Details are provided in Appendix \ref{app:SECb}. 

\begin{figure}
    \centering
    \includegraphics[width=0.48\textwidth]{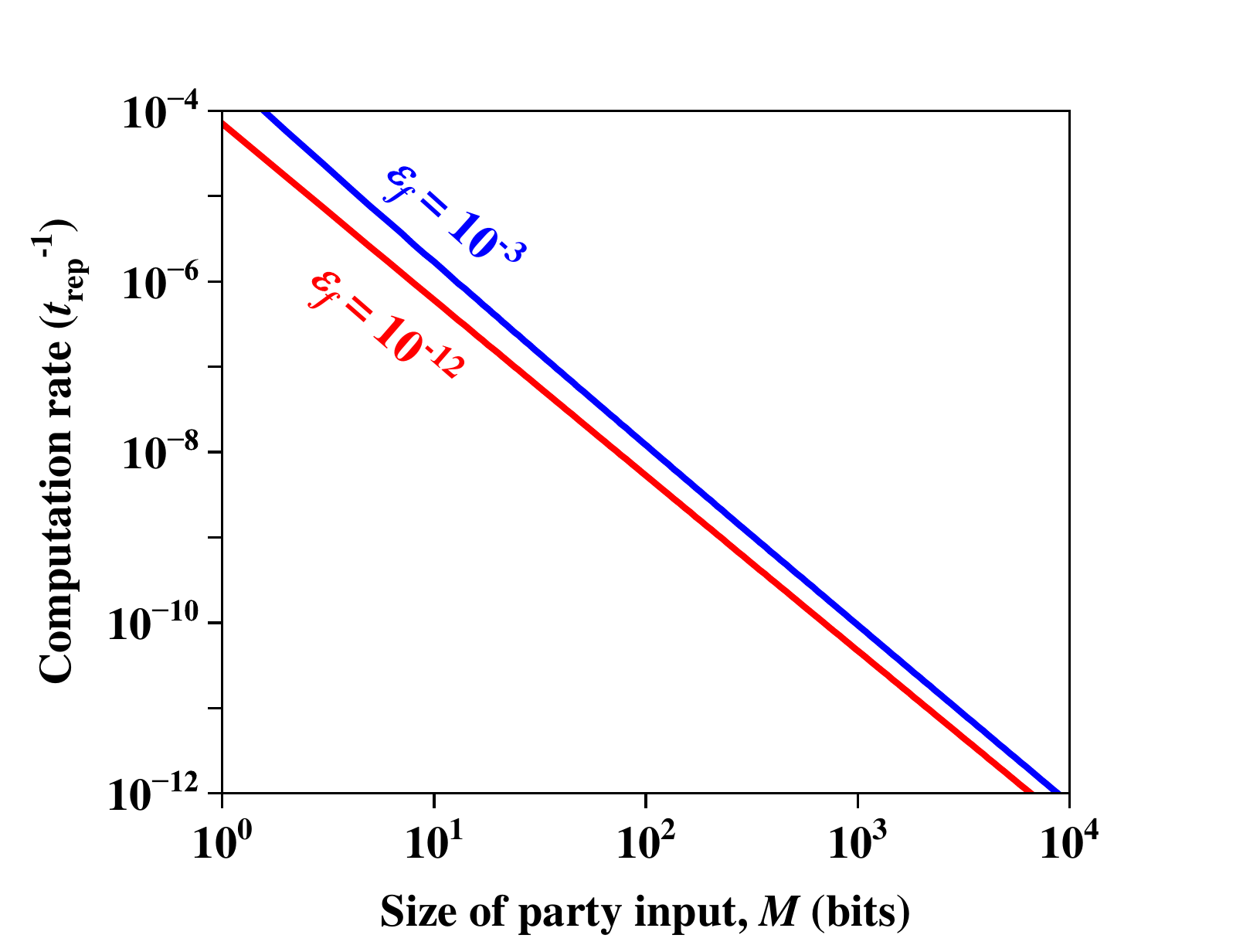}
    \caption{\label{fig:RateVsET} Error corrected two-party computation rate versus the number of required conjunctions in $f$, shown for two acceptable error probabilities on the computation ($\epsilon_f$). We assume the use the our scheme with destructive measurements and no photonic memory. The rate is expressed in units of the repetition rate of excitation of the chosen quantum emitter, $t_{\text{rep}}^{-1}$, which can be in the range of $10^6-10^9~\rm{s}^{-1}$ for highly coherent emitters.  The individual bit error probability for each copy of $\ket{G}$ is a pessimistic 0.157, assuming $\eta_e=0.1$, and $F_{\text{add}}=0.99$ (see. Appendix \ref{app:APP} and \ref{app:SECb} for details). We see that even these parameters allow virtually unlimited reduction in the total error probability with minimal change in the overall rate.}
\end{figure}
Functionally, our protocol allows for the secure implementation of any two-party Boolean function, $f$, by repeatedly generating and distributing copies of the 12-qubit state $\ket{G}$. The number of required copies depends (polynomially) on the size of the input, $M$, and the desired final error probability, $\epsilon_f$. A pair of auxiliary spins can be employed to generate each copy constructed in parallel, and additional emitters can be employed to speed up the construction. Furthermore, these spins need only remain coherent for at most the time to make each copy of $\ket{G}$, a requirement easily met with current trapped ion or atom array systems. Using the fidelity estimates established in Sec. \ref{sec:EtA} and detailed in Appendix \ref{app:APP}, we can conservatively estimate the maximum possible bit error probability associated with each copy from the compliment of the probability that no bit error occurs, that is, the compliment of the fidelity to produce $\ket{G}$ in our proposed emit-then-add scheme. In Fig. \ref{fig:RateVsET}, we plot the lower bound rate of computation at which our protocol can operate with error correction, in units of $R_{\text{rep}}$, against the size of each parties input, $M$, for total acceptable error probabilities $\epsilon_f\in\{10^{-3}, 10^{-12}\}$. We assume that $M$ here is the number of bits each party needs to perform a conjunction on, neglecting the single linear bit they each input as well.

\section{\label{sec:conc} Discussion and conclusion}
We have introduced a new paradigm for generation of photonic graph states using coherent quantum emitters using an emit-then-add approach. This enables generation of such states without requiring near unity generation, collection, and detection efficiency of the photons. Many state-of-the-art quantum emitter platforms display excellent performance in the other required metrics including coherence time, gate fidelity, and MCMR fidelity, but exhibit poor ($\lesssim10\%$) overall photon emission and detection probability due to fundamental challenges. Our scheme is thus much better suited to current and near term hardware than other deterministic schemes in the literature that rely on detection of all emitted photons. It is a toolbox for making large entangled photonic states for MBQC and other applications that are limited by spin decoherence rather than photon collection efficiency. We demonstrated this advantage in scaling, and introduced an application of our scheme for secure two-party computation. 

Furthermore, the scheme introduced here naturally lends itself to various extensions and modifications to increase functionality. First, adding multiplexing, such as between many arrays of atoms or ions, would increase the generation rate with a linear factor in the degree of multiplexing. Other hybrid approaches that combine the virtual graph states discussed here with more traditional graph states can add additional functionality. Further theoretical work remains to be done to determine more applications that are suited to this scheme. 

\begin{acknowledgments}
We acknowledge helpful discussions with Kejie Fang and Vito Scarola. E.C. thanks Christian Schaffner and Ian George for some helpful explanations on multi-party computation and security. This work was supported by the NSF Quantum Leap Challenge Institute on Hybrid Quantum Architectures and Networks (NSF Award No. 2016136).
\end{acknowledgments}

\appendix

\begin{widetext}

\section{\label{app:GSP} Preliminaries on graph states}

For an arbitrary graph $G=(V,E)$ with vertices $V=\{a_1,\cdots,a_n\}$ and edge set $E\subset V\times V$, consider the $n$-qubit operator obtained by performing a controlled$-Z$ gate, $CZ_{a,b}$, between every $(a,b)\in E$.  We denote this global operator by
\begin{equation}
    U_G=\prod_{(a,b)\in E}CZ_{a,b}.
\end{equation}
The graph state associated with the graph $G$ is the $n$-qubit state 
\begin{equation}
    \ket{G}=U_G\ket{+}^{\otimes V}.
\end{equation}
Note that $\ket{+}^{\otimes n}$ is stabilized by $n$ commuting operators $\{X_a\}_{a\in V}$.  Hence the stabilizer of $\ket{G}$ can be understood by examining the how the $X_a$ transform under $U_G$. Since $CZ_{a,b} (X_a) CZ_{a,b}=X_{a}Z_b$, it follows that the stabilizer of $\ket{G}$ is generated by the operators $\{K_a\}_{a\in V}$, where
\begin{eqnarray}
    K_a&&=U_GX_a U_G \nonumber\\
    &&=X_a\prod_{b\in N_a}Z_b\nonumber\\
    &&=X_a\prod_{b\in V}Z_b^{\Gamma_{a,b}}\quad\quad\forall a\in V, \label{eq:correlators}
\end{eqnarray}
and $\Gamma$ is the adjacency matrix of $G$.

For any $n$-qubit graph state $\ket{G}$, we can generate an orthonormal basis for $\mathbb{C}_2^N$, called the associated graph basis.  The basis vectors have the form
\begin{equation}
    Z^{\mathbf{r}}\ket{G}\quad\text{where}\quad Z^{\mathbf{r}}:=Z^{r_1}_1\otimes Z_2^{r_2}\otimes\cdots\otimes Z_N^{r_N},
\end{equation}
and we will call $\mathbf{r}=(r_1,r_2,\cdots,r_n)\in\mathbb{Z}_2^n$ a conditional phase vector and each bit $r_k$ \textit{phase information} for qubit $k$.  To see that these states are orthogonal, let $\mathbf{r}$ and $\mathbf{r}'$ be two distinct conditional phase vectors, and suppose their bit values differ in position $a$.  Then $K_a$ will anti-commute with $Z^{\mathbf{r}}Z^{\mathbf{r}'}$, and so 
\begin{eqnarray}
    \bra{G}Z^{\mathbf{r}}Z^{\mathbf{r}'}\ket{G}&&=\bra{G}Z^{\mathbf{r}}Z^{\mathbf{r}'}K_a\ket{G}=-\bra{G}K_aZ^{\mathbf{r}}Z^{\mathbf{r}'}\ket{G}=-\bra{G}Z^{\mathbf{r}}Z^{\mathbf{r}'}\ket{G}.
\end{eqnarray}
We will be particularly interested in how the graph basis states transform under the local Pauli measurements of $Y$ and $Z$.  For a binary vector $\mathbf{r}\in \mathbb{Z}_2^n$ and subset of nodes $S\subset V$, let $\mathbf{r}-S$ denote the vector of length $n-|S|$ obtained from $\mathbf{r}$ by removing the coordinates in $S$.  Suppose that for an initial graph basis state $Z^{\mathbf{r}}\ket{G}$, either $Y$ or $Z$ is measured on qubit $a$ and outcome $m_a\in\{0,1\}$ is received.  The initial state transforms as follows:
\begin{subequations}
        \begin{eqnarray}
            Z_a:&& \quad Z^{\mathbf{r}}\ket{G}\mapsto \left(\prod_{b\in N_a}  Z_b^{m_a}\right)Z^{\mathbf{r}-a}\ket{G-a}, \label{Eq:Z-measurement-post-correction}\\
            Y_a:&& \quad Z^{\mathbf{r}}\ket{G}\mapsto \left(\prod_{b\in N_a} \left(-iZ_b\right)^{1/2}Z_b^{r_a+m_a}\right)Z^{\mathbf{r}-a}\ket{\tau_a(G)-a}, \label{Eq:Y-measurement-post-correction}
        \end{eqnarray}
    \end{subequations}
where $\tau_a(G)$ is the local complementation of $G$ at vertex $a$, i.e. $\tau_a(G)$ is the graph $(V, E\Delta E(N_a,N_a))$, and $\tau_a(G)-a$ is the graph obtained by removing $a$ from $\tau_a(G)$ \cite{Hein-2004a}. Explicitly,
\begin{equation}
    \ket{\tau_a(G)}=\left(-iX_a\right)^{1/2}\left(\prod_{b\in N_a}\left(iZ_b\right)^{1/2}\right)\ket{G}, \label{eq:local-comp}
\end{equation}
describes a set of single-qubit rotations comprising a local complementation. Note that the $Y_a$ post-measurement state can always be transformed back to the associated graph basis by performing $(iZ_b)^{1/2}$ on each $b\in N_a$. For the special case of measurement at a leaf in the graph, a vertex with only a single neighbor such that $\tau_a(G)=G$, the $Y_a$ post-measurement state after rotation back to the graph basis is effectively a $Z_a$ post-measurement state up to the conditional phase flip $Z_b^{r_a}$.

\subsection{\label{app:GSPa} Phase transmission along a chain}

Consider the effect of locally measuring along some linear chain in graph $G$.  Let $a_1,a_2,a_3,\cdots, a_n, a_{n+1}$ denote the constituent qubits, with $a_1$ being the first node in the chain and $a_{n+1}$ being the final node, which is connected to the remainder of the graph and left unmeasured in this sub-routine.  The specific measurement sequence consists of measuring $Y$ on $a_1$ and $(-iZ)^{1/2}Y(iZ)^{1/2}=-X$ on $a_k$ for all $k=2,\cdots,n-1$.  If the total state prior to measurement is a graph basis state $Z^{\mathbf{r}}\ket{G}$ and $(m_{a_1},\cdots,m_{a_{n-1}})$ is the binary sequence of outcomes from these measurements, then by Eq. \eqref{Eq:Y-measurement-post-correction} the overall state evolution is
\begin{eqnarray}
    Y_{a_1},(-X_{a_{2}}),\cdots,(-X_{a_{n-1}}):\quad Z^{\mathbf{r}}\ket{G}&&\mapsto [-iZ_{a_n}]^{1/2}Z_{a_n}^{\Omega}Z^{\mathbf{r}-\{a_1,\cdots,a_{n-1}\}}\ket{G-\{a_1,\cdots,a_{n-1}\}}, \label{Eq:MeasurementAlongChain}
\end{eqnarray}
where $\Omega=\sum_{i=1}^{n-1}(r_{a_i}+m_{a_i})\mod 2$.  Crucially, each $m_{a_k}$ is a random bit uncorrelated from $\mathbf{r}$ and any other measurement data.  Hence, we can think of each $m_{a_k}$ as a one-time pad that is added to the conditional phase information $r_{a_k}$ when passing from node $a_k$ to $a_{k+1}$.

\subsection{\label{app:GSPb} Phase transmission at a fork}
Consider the effect of measuring two qubits, $a$ and $b$, that are connected to a single common node $c$.  Suppose the total state prior to measurement is a graph basis state $Z^{\mathbf{r}}\ket{G}$. If $m_a$ and $m_b$ denote the binary outcomes when either $Y$ or $Z$ is measured on both qubits, the respective state transformations are given by
\begin{subequations}
    \begin{eqnarray}
            Y_a,Y_b:\quad Z^{\mathbf{r}}\ket{G}&&\mapsto Z_c^{r_a+r_b+1}Z_c^{m_a+m_b}Z^{\mathbf{r}-\{a,b\}}\ket{G-\{a,b\}},\label{Eq:joint-Y}\\
            Z_a,Z_b:\quad Z^{\mathbf{r}}\ket{G}&&\mapsto Z_c^{m_a+m_b}Z^{\mathbf{r}-\{a,b\}}\ket{G-\{a,b\}}\label{Eq:joint-Z},
    \end{eqnarray}
\end{subequations}
up to an overall phase. Hence, the key difference between the two measurements is that $Y_a,Y_b$ transfers the conditional phase flip $Z_c^{r_a+r_b+1}$ onto the state $\ket{G-\{a,b\}}$ while $Z_a,Z_b$ does not.

\section{\label{app:CON} Constructing graph states with emit-then-add}

Building graph states with our emit-then-add scheme necessitates additional experimental overhead from typical deterministic quantum emitter-based schemes, in both the number of qubits required and entangling operations between them. We demonstrate through an inductive argument that these additional resource costs scale at worst linearly. Subsequently, we present a resource-efficient set of subroutines used to produce the states for our MPC protocol, described in Sec. \ref{sec:MPC}. As the whole of our protocol is Clifford, we can employ emit-then-add, without the need for a photonic memory, to produce and measure these graph states efficiently. We briefly discuss how phase corrections, resulting from these construction subroutines, are handled classically in our scheme. The subroutines we introduce are employed in Appendix \ref{app:APP} to estimate the fidelity to make the states used in our protocol. In what follows, a superscript $(p)$, $(s)$, or $(e)$ denotes the kind of physical qubit associated with the relevant subspace on which an operator acts, as a photon, auxiliary spin, or emitter, respectively.

\subsection{\label{app:CONa} Additional overhead}

Let $\ket{G}$ define an existing graph state in which there is at least a single edge between an auxiliary spin and the set of photons previously added to the graph. We label each of these photons by an emission order $1,\,\cdots, m-1$. Let $V_m$ define the additional vector space describing the emitter and the next photon, $m$, to be added to the graph, both of which start in $\ket{0}$. The set of generators which stabilizes the collective vector space  $V_G+V_m$, consisting of the graph and subsequent emitter-photon pair, has the form
\begin{equation} \label{eq:stab1}
    S_{V_G+V_m}=\big\langle \cdots, \cdots Z_k^{(p)}\cdots X^{(s)}, Z^{(e)}, Z_m^{(p)} \big\rangle,
\end{equation}
where $\langle\cdots\rangle$ denotes a set of generators and the notation $\cdots Z_k^{(p)}\cdots$ is used to keep track of an arbitrary edge between the auxiliary spin and a photon previously added to the graph at some emission step $k<m$. The subcircuit depicted in Fig. \ref{fig:HPSEpump} demonstrates an example of how to transfer entanglement (or conditional phase information) from the emitter-photon sub-system to $\ket{G}$, with a single two-qubit spin-spin entangling gate and local complementation. In implementing the example, we transform the stabilizer of the combined vector space in Eq. \eqref{eq:stab1} as
\begin{eqnarray}
    S_{V_G+V_m}'&&=\big\langle \cdots, (-1)^{c_m}\cdots Z_k^{(p)}\cdots X_m^{(p)}X^{(s)}, (-1)^{c_m}Z_m^{(p)}Z^{(s)}, (-1)^{c_m}Z^{(e)} \big\rangle \label{eq:stab2}
\end{eqnarray}
where $c_m\in\{0, 1\}$ is a classical bit value conditioned on the measurement of the emitter.

The result of these operations produces the same stabilizer we would have arrived at had we instead pumped the auxiliary spin itself. With these additional operations, any graph state accessible in the deterministic scheme can be constructed with emit-then-add. As we restrict those auxiliary spins already entangled with any previously added photons from being pumped, it follows that one additional spin, the emitter, and one two-qubit spin-spin entangling gate per photon in $G$ are the minimum additional overhead in our proposed schemes for making arbitrary graph states. Furthermore, we can define a new vector space $V_G'$, containing the existing graph state and a newly entangled photon, with a dimension that increases by one with each new photon. The new space $V_G'$ is stabilized by a unique set of generators that can be rotated back to a graph state basis of the same general form as Eq. \eqref{eq:stab1}, now defined by a new graph state $\ket{G'}$.
\begin{figure}
    \centering
    \includegraphics[width=0.535\textwidth]
    {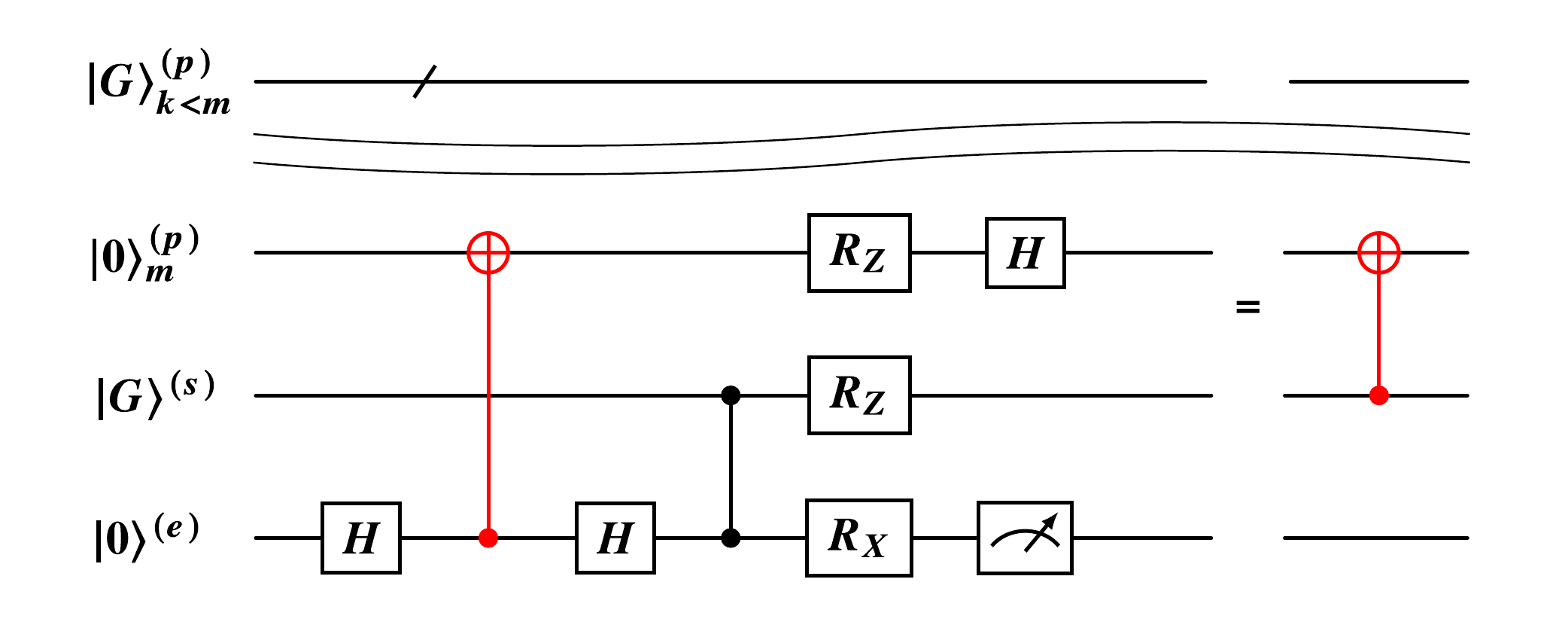}
    \caption{\label{fig:HPSEpump} A subcircuit equivalent to a $CX_{s,p}$ gate, up to a conditional phase correction, for transferring entanglement in our proposed scheme. This example ``emit-then-add'' step replaces every pumping gate in typical deterministic schemes for generating arbitrary photonic graph states.  Entanglement between a photon (p) and a coherently pumped emitter (e) (represented by a red $CX_{e,p}$ gate) is exchanged to an auxiliary spin (s) via a two-qubit entangling $CZ_{e,s}$ gate and local complementation. The emitter is measured out thereafter and reinitialized for the next iteration of the procedure. Rotations about $X$ and $Z$ are by $\pi/2$ and $-\pi/2$, respectively, as noted in Eq. \eqref{eq:local-comp}. The measurement of the emitter is with respect to the $Z$ basis. All previously added photons at iterations $k<m$ are unaffected.}
\end{figure}

\subsection{\label{app:CONb} Construction subroutines}

\begin{figure*}
    \centering
    \includegraphics[width=1.0\textwidth]{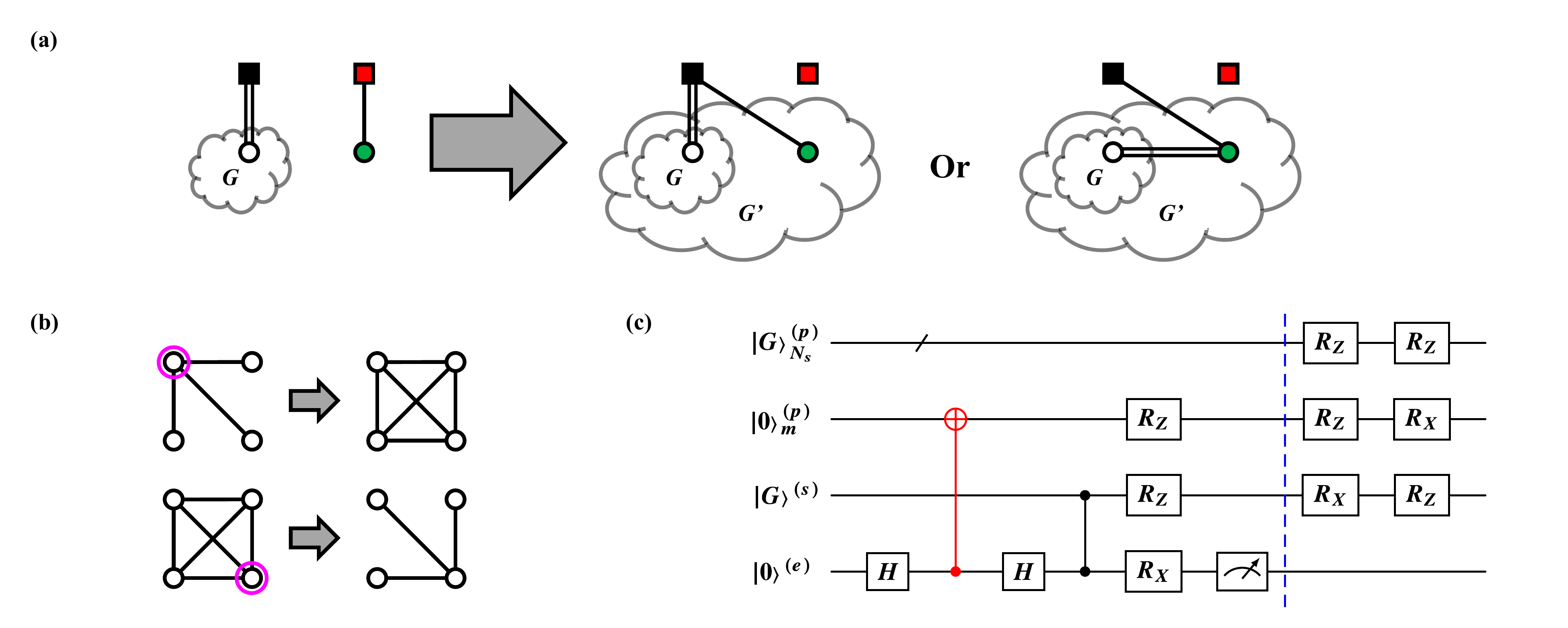}
    \caption{\label{fig:Pass} Passing-subroutine for adding new photons to an existing graph, $G$. (a) A graph transformation of passing a new photon (green) from the emitter (red) to an auxiliary spin (black), which is connected to one or more previously added photons (white, denoted with a double edge for the multiplicity). This subroutine consists of two variations: (left) leaving all previously added photons invariant, (right) transplanting those edges to the newly added photon. (b) An example of how local complementations on a target qubit (magenta circle) can be used to add or remove edges. (c) A quantum subcircuit which implements the graph transformation. The right variation of the graph transformation above is achieved with the additional two local complementations (blue dashed line, depicting the deviation). Rotations and measurements follow the same notation as in Fig. \ref{fig:HPSEpump}.} 
\end{figure*}

We also offer a pair of subroutines that simplifies the construction and overhead for the graph state $\ket{G}$, employed in our MPC. Representations of the graph transformations associated with the two subroutines, along with example circuits, are depicted in Fig. \ref{fig:Pass} and \ref{fig:Patch}. These transformations can be performed successively with no additional operations, transforming the previous graph $G$ built on the auxiliary spin to a new graph $G'$ with any new photons sharing an edge to the auxiliary spin. Despite their intended application in our MPC, we make no assumptions about the measurement of the photons in these subroutines, such that they can be applied generally across experimental implementations.

One ``passing''-subroutine, shown in Fig. \ref{fig:Pass} consists of two variations: ``join'' and ``extend''. The join-subroutine adds a new photon to an existing graph and leaves all previously existing edges invariant. The extend-subroutine transfers all edges from the auxiliary spin to the new photon. The two are achieved without or with the additional two local complementations depicted at the end of the example circuit, respectively. Both variations only act on previously added photons in the neighborhood of the auxiliary spin, $N_s$. Repeatedly applying the join-subroutine or extend-subroutine on a single auxiliary spin produces a star graph or linear cluster state, respectively for each variation. Additionally, either variation of this subroutine can be appropriately implemented between auxiliary spins to connect subgraphs.

We note briefly that in practice with each implementation of the passing-subroutine, the newly added photon to the graph and the auxiliary spin each carry a conditional phase that is a byproduct of the decoupling measurement made on the emitter, as shown in Eq. \eqref{eq:stab2}. This byproduct phase determines the precise basis state of the graph and may require correction for general MBQCs. In the implementation of our scheme without photonic memory, photons are measured before they are decoupled from the emitter and hence any requisite phase corrections need be commuted after each measurement. The Clifford nature of the measurements employed in our MPC simplifies all of these corrections to bit flips that can be handled classically. Furthermore, correction of this phase is not always necessary as certain measurements made by the parties destroy this phase information, while other measurements allow the parties to absorb this phase information into their own pad. Conversely, conditional measurements, such as the ones made by Alice in step (I.4) of Stage I of $\mathcal{P}$, couple these byproduct phases to the phase information input into the computation. Therefore, classical communication is required here between Alice and the Source. A simple solution is for the Source to make public the outcomes of each of these decoupling measurements. 

The other ``patching''-subroutine, shown in Fig. \ref{fig:Patch}, serves to attach two subgraphs $G_1$ and $G_2$ by a common edge between photons. This process requires additional two-qubit spin-spin entangling gates from the passing-subroutine. For further simplicity we only consider the case where $G_1$ and $G_2$ each have a single edge to all previously added photons in their respective subgraphs, though in general this subroutine fully connects all of the photons at the first layer in each subgraph. Operations in this subroutine are restricted locally to only the emitter, spin, and the photons we ultimately require to share an edge, labeled in the figure by arbitrary emission steps $j,k$ in $1,\cdots,n_p$. This subroutine mirrors the one employed in the production of large 2D cluster states in \cite{Russo-2019a}, and can be applied in either scheme we propose for the same purpose.

\begin{figure*}
    \centering
    \includegraphics[width=1.0\textwidth]{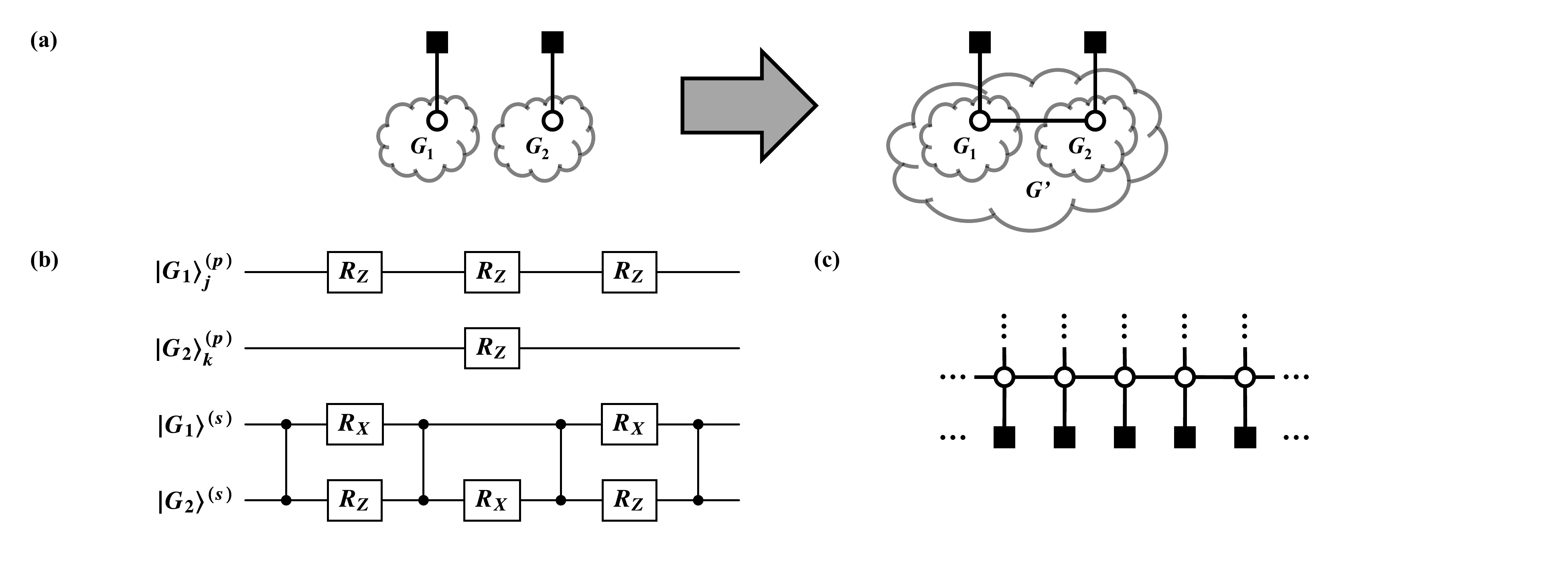}
    \caption{\label{fig:Patch} Patching-subroutine for attaching two subgraphs $G_1$ and $G_2$ by a common edge between photons. (a) The graph transformation depicting the patching. (b) The corresponding circuit diagram. Additional two-qubit spin-spin entangling gates are required for this subroutine, over the passing-subroutine. Rotations follow the same notation as in Fig. \ref{fig:HPSEpump}. (c) A 2D cluster state built on an array of auxiliary spins. Edges can be generated between photons in the layer neighboring the array of spins with this patching.}
\end{figure*}
Application of these subroutines to the construction of the graph state discussed in Sec. \ref{sec:MPC} is straightforward. The state $\ket{G}$, consisting of $n_p=12$ photons, labeled by an emission ordering depicted in Fig. \ref{fig:G}, can be built following the sequence of steps in the Build below. This procedure requires $17$ two-qubit spin-spin entangling gates in total: $13$ from passing operations, and $4$ from a single patching steps. It is known that the sequential nature of photon emission events imparts an ordering on the graph state, limiting the kinds of photonic graphs accessible by construction on a single quantum emitter \cite{Schon-2005a}. As $\mathcal{P}$ involves conditioning measurement bases on previous outcomes (steps (I.5) and (I.6)), this sets a nontrivial emission ordering on $\ket{G}$, which, following the results of \cite{Li-2022a}, requires at least two auxiliary spins to construct. 

\begin{build}{\label{prot:g1} $\mathbf{\ket{G}:}$}

    \noindent\textit{Input:} A photon emission order labeling photons $(p_{1}),\cdots,(p_{12})$, corresponding to the $n_p=12$ photons in $\ket{G}$, an emitting spin, and two auxiliary spins, labeled $(s_1)$ and $(s_2)$.
    \begin{enumerate}
        \item Pass photons $(p_{1})$ through $(p_{4})$ to $(s_1)$. Apply the join-subroutine for $(p_{4})$ and the extend-subroutine for the rest.
        \item Pass photons $(p_{5})$ through $(p_{8})$ to $(s_2)$. Apply the join-subroutine for $(p_{5})$ and the extend-subroutine for the rest.
        \item Pass the subgraph on $(s_2)$ to $(s_1)$ with the join-subroutine.
        \item Pass photons $(p_{9})$ through $(p_{12})$ to $(s_2)$. Apply the join-subroutine for $(p_{9})$ and the extend-subroutine for the rest.
         \item Patch the subgraph on $(s_2)$ with $(s_1)$. Measure out both $(s_1)$ and $(s_2)$. 
    \end{enumerate}
\end{build}

\section{\label{app:APP} Generalized emit-then-add scheme including experimentally realizable fidelity calculations}

We now consider an experimental realization of our general emit-then-add scheme involving QND measurements. While the nature of our scheme is device-independent, we assume here trapped ions as our quantum emitters and auxiliary spins, due to their high-fidelity benchmarks for two-qubit spin-spin entangling gates \cite{Wright-2019a, Clark-2021a}, exceedingly long coherence times \cite{Wang-2021a}, and reasonable collection efficiency with high-NA optics \cite{Shu-2011a, Maiwald-2012a, Chou-2017a}. We outline a fidelity model for a version of our scheme that involves heralding spin-photon Bell states, based on entanglement swapping and nonlinear optics. Note, the experimental implementation here is applicable for generating photonic graph states for arbitrary applications.

At the end of this section, we briefly demonstrate how the fidelity estimates we employ in the main text directly follow from this model. Recall, in the main text, we consider a simplified version of our emit-then-add scheme that does not require a form of QND measurement or photonic memory. This scheme is applicable for measurement-based computation, only when certain conditions are met (see Sec. \ref{sec:EtA}). Assuming ions or neutral atoms as emitters and auxiliary spins, the estimates we provide establish that. with emit-then-add, moderately size photonic graph states are possible, even with poor emitter photon collection efficiency ($\lesssim10\%$).

\subsection{\label{app:APPa} Experimental apparatus} 

For the general scheme, heralding a Bell state between an emitter and photon is implemented via entanglement swapping between a photon from an emitter and a photon from a nonlinear pair source. We use spontaneous parametric down conversion (SPDC) as our pair source for the heralding. SPDC is an optical process based on a crystal with $\chi^{(2)}$ nonlinearity whereby a photon from a pump is converted to a signal and idler photon pair of lower energy. It is often used to produce entangled photon pairs in various degrees of freedom, or as a heralded single photon source \cite{Schneeloch-2019a}. The joint measurement apparatus required for the entanglement swapping in the heralding scheme we describe here can be accomplished via a generalized Hong-Ou-Mandel interferometer \cite{Braunstein-1995a} depicted in Fig. \ref{fig:HSPE_BSM_pic}(a), where the emitter photon and signal photon from the pair source are sent to a 50:50 beamsplitter. Elements from these experiments are modeled in our estimates of the fidelity to successfully produce graph states.

\begin{figure*}
    \centering
    \includegraphics[width=1.0\textwidth]{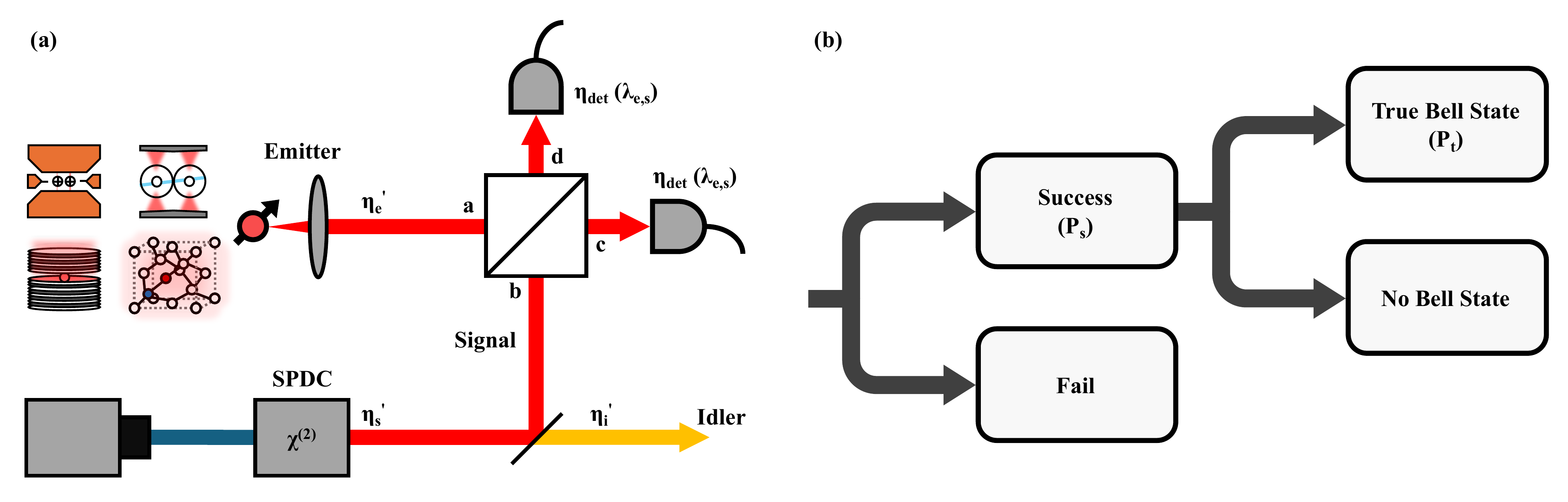}
    \caption{\label{fig:HSPE_BSM_pic} Entanglement swapping in the HoS scheme. (a) Schematic of a generalized Hong-Ou-Mandel interferometer applied as a joint measurement apparatus. A photon entangled with an emitter is sent through port $a$ of the beamsplitter, and a photon from a spontaneous parametric down conversion (SPDC) photon pair source is sent through port $b$. Detection of two single-photon orthonormal states $\ket{u}$, $\ket{v}$ in either output ports $c$ or $d$ results in entanglement between the emitter and the idler photon from the pair source. Various loss factors denoted by $\eta$ are modeled. Additional elements for qubit rotations are not shown. (b) Probability tree showing the outcomes of a Bell state measurement attempt. The ``success'' probability $P_s$ for the Bell state measurement denotes the probability of measuring a set of specific click patterns that indicate entanglement swapping. $P_t$ denotes the probability of projecting the emitter and idler onto a Bell state per measurement attempt.}
\end{figure*}

We briefly consider the set of states employed in heralding a Bell state in a device-independent form. The internal state of an emitter $(e)$ is entangled with a photon $(p_e)$ in an arbitrary basis, and an entangled photon pair is produced from the SPDC source labeled as the signal $(p_s)$ and idler $(p_i)$ photons. We can express both the idealized entangled states $\ket{\Psi_E}$ and $\ket{\Psi_P}$, respectively, as
\begin{eqnarray}
    \ket{\Psi_E}&&=\dfrac{1}{\sqrt{2}}(\ket{0}^{(e)}\ket{1;0}^{(p_e)}-\ket{1}^{(e)}\ket{0;1}^{(p_e)}), \label{eq:psi_e}\\
    \ket{\Psi_P}&&=\dfrac{1}{\sqrt{2}}(\ket{1;0}^{(p_s)}\ket{0;1}^{(p_i)}-\ket{0;1}^{(p
    _s)}\ket{1;0}^{(p_i)}), \label{eq:psi_p}
\end{eqnarray}
where $\ket{0}^{(e)}$ and $\ket{1}^{(e)}$ is the computational basis for the emitter, and the notation $\ket{m;n}$ is used to denote the photonic Fock state representation for a particular basis $\{\ket{u},\ket{v}\}$, such as orthogonal polarizations, with $m$ in $\ket{u}$ and $n$ in $\ket{v}$. The relationship between the input modes $a$ and $b$ and output modes $c$ and $d$ of the 50:50 beamsplitter can be described by the unitary transformation on the creation operators $\hat{a}^\dagger=\frac{1}{\sqrt{2}}(\hat{c}^\dagger+\hat{d}^\dagger)$ and $\hat{b}^\dagger=\frac{1}{\sqrt{2}}(\hat{c}^\dagger-\hat{d}^\dagger)$. With the emitter photon entering port $a$, and the signal photon entering port $b$, the overall state after applying the transformation can be expressed in the form

\begin{equation} \label{eq:psi_bs}
    \begin{split}
    \ket{\Psi_E}\ket{\Psi_P}=&\dfrac{(\hat{c}^\dagger_u)^2-(\hat{d}^\dagger_u)^2}{4}\ket{vac}_{c,d}\ket{0}^{(e)}\ket{0;1}^{(p_i)}
    +\dfrac{\hat{d}^\dagger_u\hat{d}^\dagger_v-\hat{c}^\dagger_u\hat{c}^\dagger_v}{2\sqrt{2}}\ket{vac}_{c,d}\ket{\Psi^+}^{(e,p_i)}\\
    +&\dfrac{\hat{c}^\dagger_u\hat{d}^\dagger_v-\hat{c}^\dagger_v\hat{d}^\dagger_u}{2\sqrt{2}}\ket{vac}_{c,d}\ket{\Psi^-}^{(e,p_i)}
    +\dfrac{(\hat{c}^\dagger_v)^2-(\hat{d}^\dagger_v)^2}{4}\ket{vac}_{c,d}\ket{1}^{(e)}\ket{1;0}^{(p_i)},
    \end{split}
\end{equation}
where
\begin{equation} \label{eq:psi_ent}
    \ket{\Psi^\pm}^{(e,p_i)}=\dfrac{1}{\sqrt{2}}(\ket{0}^{(e)}\ket{1;0}^{(p_i)}\pm\ket{1}^{(e)}\ket{0;1}^{(p_i)}).
\end{equation}

We are interested in projecting onto either $\ket{\Psi^+}^{(e,p_i)}$ or $\ket{\Psi^-}^{(e,p_i)}$ which can be realized by detecting specific click patterns on photon counting detectors at the output ports capable of resolving the orthonormal components. From Eq. \eqref{eq:psi_bs}, we can see that detecting orthonormal one-photon states in either ports $c$ or $d$ results in a projection onto $\ket{\Psi^+}^{(e,p_i)}$ while detection in opposite ports results in $\ket{\Psi^-}^{(e,p_i)}$. The overall probability of projection onto either of the Bell states is $\frac{1}{2}$ for each heralded measurement attempt. Upon the detection of $\ket{\Psi^-}^{(e,p_i)}$, we can perform a qubit rotation to produce the state $\ket{\Psi^+}^{(e,p_i)}$ for phase consistency. We can then perform a series of one- and two-qubit gates to add a photon to a graph. Upon failure to detect the correct detection pattern, we reinitialize the emitter and photon pair and attempt the measurement again. We assume number-resolving detectors at the output ports of the beamsplitter capable of detecting photons in the basis $\{\ket{u},\ket{v}\}$. For example, two polarizing beamsplitters and four detectors would be used after the output ports for photons entangled in a polarization basis.

The initial state of the emitter can be described by the state defined in Eq. \eqref{eq:psi_e}. The output state of SPDC is a multi-mode squeezed state where the photon pair production probability is dependent on the pump amplitude. Assuming a classical pump, we take the Hamiltonian for the SPDC process to approximately be of the form \cite{Lamas-Linares-2001a, Kok-2000a}
\begin{equation}\label{eq:spdc_hamiltonian}
    \hat{H}=e^{i\phi}\kappa\hat{K}^\dagger + e^{-i\phi}\kappa\hat{K},
\end{equation}
where $\hat{K}^\dagger=\hat{a}_{u;s}^\dagger\hat{a}_{v;i}^\dagger-\hat{a}_{v;s}^\dagger\hat{a}_{u;i}^\dagger$ represents the creation of entangled signal and idler pairs denoted $s$ and $i$ in the basis $\{\ket{u},\ket{v}\}$. The time evolution operator $\hat{U}(t)=\exp(i\hat{H}t/\hbar)$ yields the resulting state
\begin{equation}\label{eq:psi_spdc}
    \ket{\Psi_{SPDC}}=N\sum_{n=0}^{\infty}\tanh^n{r}\sum_{m=0}^{n}(-1)^m\ket{n-m;m}^{(p_s)}\ket{m;n-m}^{(p_i)},
\end{equation}
where $r=\kappa t/\hbar$ is the interaction parameter that is dependent on the pump field amplitude, $t$ is the interaction time of the pump through the crystal, and $N$ is the normalization constant with $N=1-\tanh^2{r}$. By tuning the interaction parameter $r$, we can change the probability of producing a single pair ($n=1$ in Eq. \eqref{eq:psi_spdc} leading to Eq. \eqref{eq:psi_p}) to optimize the fidelity in this scheme. Note that increasing the pump power also increases the relative contribution of the higher order terms ($n>1$), which increases the probability of detecting multiple photon pairs.

The above is applicable for polarization-entangled photon pairs directly produced in type-II SPDC described in \cite{Lamas-Linares-2001a, Kok-2000a}; the output state and choice of degree-of-freedom for entanglement will of course be dependent on the specific properties of the photon pair source and pump. However, we note that we are assuming the pump can be treated as a classical, single-mode source, and we use a simplified Hamiltonian and output state for the photon pairs entangled in $\{\ket{u},\ket{v}\}$. Moreover, factors such as the multi-modal nature of the pump, phase differences or instabilities between the orthonormal components in the optical path, or distinguishability between the photons from the emitter and the SPDC source can degrade the fidelity of the scheme. 

\subsection{\label{app:APPb} Sources of infidelity}

Several experimentally relevant sources of infidelity are considered in our model. Photonic loss and detector dark counts can induce false heralding events, leading to a failure to resolve a true Bell state and subsequent addition of a photon to the graph. Decoherence across the emitter and any auxiliary spins utilized in building the graph will produce errors on any computation done on the graph state. The fidelity of each two-qubit spin-spin entangling gate performed in the construction is considered as well. We can also consider the fidelity of the initial entanglement between the emitter and its photon, as well as the entanglement of the photon pair, used in the entanglement swapping procedure. 

\subsubsection{\label{app:APPbi} Photonic loss}

We use the standard approach to modeling photonic losses, which is to introduce a fictitious beamsplitter with transmittance $\eta$ and reflectance $1 - \eta$ on the lossy channel. This is equivalent to applying the transformation on the creation operator $\hat{a}^\dagger=\sqrt{\eta}\hat{a}_T^\dagger+\sqrt{1-\eta}\hat{a}_R^\dagger$ \cite{Leonhardt-1995a} where $T$ and $R$ denote the transmitted and reflected paths, respectively. The Fock state $\ket{n}$ serving as the input transforms as $\ket{n}\rightarrow\sum_{k=0}^{n}\sqrt{\binom{n}{k}\eta^{n-k}(1-\eta)^k}\ket{n-k}_T\ket{k}_R$ and the reflected path is traced out. We assume, for simplicity, that $\eta$ is independent of the photonic degrees of freedom for each loss channel modeled. 
Applying this to an emitter with an initial state given by Eq. \eqref{eq:psi_e} and performing the partial trace over the reflected path, the resulting density matrix $\hat{\rho}_{E}$ for the emitter-photon pair incorporating losses is
\begin{eqnarray}
    \hat{\rho}_{E}=&&\dfrac{1}{2}(1-\eta_e)\op{0}{0}^{(e)}\otimes\op{vac}{vac}^{(p_e)}+\dfrac{1}{2}(1-\eta_e)\op{1}{1}^{(e)}\otimes\op{vac}{vac}^{(p_e)} \nonumber\\
    &&\quad\quad +\dfrac{1}{2}\eta_e\op{0}{0}^{(e)}\otimes\op{1;0}{1;0}^{(p_e)}+\dfrac{1}{2}\eta_e\op{1}{1}^{(e)}\otimes\op{0;1}{0;1}^{(p_e)} \nonumber\\
    &&\quad\quad -\dfrac{1}{2}\eta_e\op{0}{1}^{(e)}\otimes\op{1;0}{0;1}^{(p_e)}-\dfrac{1}{2}\eta_e\op{1}{0}^{(e)}\otimes\op{0;1}{1;0}^{(p_e)}. \label{eq:rho_e}
\end{eqnarray}

Similarly, the density matrix for the SPDC source $\hat{\rho}_{SPDC}$ becomes

\begin{equation}\label{eq:rho_spdc}
    \begin{split}
    \hat{\rho}_{SPDC}=&\sum_{n,n^*=0}^{\infty}\sum_{m=0}^{n}\sum_{m^*=0}^{n^*}\sum_{k_s,k^{'}_i=0}^{\text{min}\{n-m,n^*-m^*\}}\sum_{k^{'}_s,k_i=0}^{\text{min}\{m,m^*\}} c_{n,m,k_s,k^{'}_s,k_i,k^{'}_i}c^*_{n^*,m^*,k_s,k^{'}_s,k_i,k^{'}_i}\\
    &\ket{n-m-k_s;m-k^{'}_s}\bra{n^*-m^*-k_s;m^*-k^{'}_s}^{(p_s)}\\
    \otimes&\ket{m-k_i;n-m-k^{'}_i}\bra{m^*-k_i;n^*-m^*-k^{'}_i}^{(p_i)},
    \end{split}
\end{equation}
where
\begin{equation}\label{eq:rho_spdc_coeff}
    \begin{split}
    c_{n,m,k_s,k^{'}_s,k_i,k^{'}_i}=(1-\xi)(\sqrt{\xi})^n(-1)^m\sqrt{\binom{n-m }{ k_s}\binom{m }{ k_s^{'} }\binom{m }{ k_i} \binom{n-m }{ k^{'}_i}\eta_s^{n-k_s-k^{'}_s}(1-\eta_s)^{k_s+k^{'}_s}\eta_i^{n-k_i-k^{'}_i}(1-\eta_i)^{k_i+k^{'}_i}},
    \end{split}
\end{equation}
and we have used $\xi=\tanh^2{r}$. Here, we have made the additional substitutions $\eta_e=\eta^{'}_e\eta_{det}(\lambda_e)$ and $\eta_s=\eta^{'}_s\eta_{det}(\lambda_s)$, where $\eta_e$ and $\eta_s$ are the combined collection and photodetection efficiencies for the emitter and signal photons, respectively, in the interferometry setup. $\eta'$ accounts for the collection efficiency of the optical path as shown in Fig.~\ref{fig:HSPE_BSM_pic}(a), and $\eta_{det}(\lambda)$ is the photodetection efficiency which may be wavelength dependent (in the joint measurement, we require $\lambda_e=\lambda_s$). $\eta_i = \eta_i^{'}$ is the collection efficiency of the idler for the SPDC setup, neglecting user-specific losses such as transmission losses through an optical fiber network and imperfect detection from the user. Since we are interested in retrieving the photon from the emitter and a single pair from the SPDC source for entanglement swapping, under the loss model the resulting probabilities for $\ket{\Psi_E}$ and $\ket{\Psi_P}$ as defined in Eq. \eqref{eq:psi_e} and Eq. \eqref{eq:psi_p} are
\begin{eqnarray}
    \bra{\Psi_E}\hat{\rho}_E\ket{\Psi_E}&&=\eta_e,  \label{eq:prob_e}\\
    \bra{\Psi_P}\hat{\rho}_{SPDC}\ket{\Psi_P}&&=\dfrac{\eta_s\eta_i\xi(\bar{\eta}_s\bar{\eta}_i\xi+2)(1-\xi)^2}{(1-\bar{\eta}_s\bar{\eta}_i\xi)^4}, \label{eq:prob_spdc1}
\end{eqnarray}
where $\bar{\eta}_e=1-\eta_e$, $\bar{\eta}_s=1-\eta_s$, and $\bar{\eta}_i=1-\eta_i$.

\subsubsection{\label{app:APPbii} Photodetector metrics}

In this work, we provide practical estimates of the fidelity achievable using photon number-resolving detectors for entanglement swapping. We assume each detector in the setup has some efficiency $\eta_{det}$ and also model dark counts which arise due to inherent electronic noise in a detector. Each dark count event is assumed to be independent and generated at a constant rate $R_d$ cps, and typically dark count rates can be low ($<10^{-2}$ cps) for single-photon detectors such as Transition Edge Sensors \cite{Shah-2022a}. Furthermore, we assume the exposure time of the detectors, $t_{\text{exp}}$ is much longer than the time scale over which dark counts emerge. The probability for $n_d$ dark counts on a detector during each measurement cycle can then be approximated by a Poisson distribution \cite{Lee-2004a}
\begin{equation}\label{eq:prob_dark}
    P_d(n_d)=e^{-R_dt_{\text{exp}}}\dfrac{(R_dt_{\text{exp}})^{n_d}}{n_d!}.
\end{equation}
The joint measurement in the heralding involves the use of detectors in output ports $c$ and $d$ in Fig. \ref{fig:HSPE_BSM_pic}. Therefore, we need to consider the effect of four detection outcomes for the herald -- two detectors each with two orthonormal results, and the probability of dark counts for each. We assume identical dark count distributions for the detectors used in the herald.

\subsubsection{\label{app:APPbiii} Entanglement swapping and heralded success probability}

A successful herald occurs at a probability that we define as $P_s$, which here is the probability of detecting the set of detector click patterns requisite for the entanglement swap. An extra source of infidelity for the heralded Bell state measurement occurs from the joint measurement itself. We define $P_t$ as the overall probability of projecting the idler photon and emitter onto a Bell state for each joint measurement attempt. The probabilities are indicated in Fig. \ref{fig:HSPE_BSM_pic}(b). This will be a function of the detection efficiencies of the photons as well as the probability of zero dark count events detected. From Eq. \eqref{eq:prob_e}, Eq. \eqref{eq:prob_spdc1}, and Eq. \eqref{eq:prob_dark}, $P_t$ is given by
\begin{eqnarray} \label{eq:prob_true}
    P_t(\eta_e,\eta_s,\eta_i,\xi) &&=\dfrac{1}{2}\bra{\Psi_E}\hat{\rho}_E\ket{\Psi_E}\bra{\Psi_P}\hat{\rho}_{SPDC}\ket{\Psi_P} P^4_d(0) \nonumber\\
     &&=\dfrac{1}{2}\dfrac{\eta_e\eta_s\eta_i\xi(\bar{\eta}_s\bar{\eta}_i\xi+2)}{(1-\bar{\eta}_s\bar{\eta}_i\xi)^4}(1-\xi)^2P^4_d(0).
\end{eqnarray}

The heralded success probability $P_s$ for the joint measurement is conditioned on a detection pattern representing the orthonormality of the photon states after the beamsplitter. This is a source of infidelity as the joint detection of states such as $\ket{0;0}^{(p_e)}\ket{1;1}^{(p_s)}$, representing two signal photons that are orthonormal to each other from the photon pair source and no photons from the emitter, possible due to detection or collection losses, would not result in entanglement swapping. We can identify the set of states that would result in a ``success'':
\begin{equation}\label{eq:set_success1}
    S=\{\ket{1;0}^{(p_e)}\ket{0;1}^{(p_s)},\ket{0;1}^{(p_e)}\ket{1;0}^{(p_s)},\ket{0;0}^{(p_e)}\ket{1;1}^{(p_s)}\}
\end{equation}
For each state in the set $S$, dark counts can lead to a false detection of the state. For example, the state $\ket{1;0}^{(p_e)}\ket{0;1}^{(p_s)}$ with zero dark counts is indistinguishable from the state $\ket{1;0}^{(p_e)}\ket{0;0}^{(p_s)}$ with one dark count falsely attributed to the measurement of a signal photon in the state $\ket{0;1}^{(p_s)}$. The former has an additional associated dark count probability of $P^4_{d}(0)$ and the latter $P_{d}(1)P^3_{d}(0)$. Thus, we can compute $P_s$ by considering the probability of measuring all states in $S$ and all dark count combinations and photon number states that reconstruct each element of $S$. By taking the partial trace of the emitter and idler photon from Eq. \eqref{eq:rho_e} and Eq. \eqref{eq:rho_spdc}, and considering the probabilities mentioned in the overall set of joint measurement outcomes, we arrive at the probability $P_s$ for the joint measurement,
\begin{eqnarray}
    P_s(\eta_e,\eta_s,\xi)=&&\dfrac{3\bar{\eta}_e}{(1-\bar{\eta}_s\xi)^2}(1-\xi)^2P^2_{d}(1)P^2_{d}(0)\nonumber\\
    +&&\left[\dfrac{\eta_e}{(1-\bar{\eta}_s\xi)^2}+\dfrac{4\bar{\eta}_e\eta_s\xi}{(1-\bar{\eta}_s\xi)^3}\right](1-\xi)^2P_{d}(1)P^3_{d}(0) \nonumber\\
    +&&\left[\dfrac{\eta_e\eta_s\xi}{(1-\bar{\eta}_s\xi)^3}+\dfrac{\bar{\eta}_e\eta^2_s\xi^2}{(1-\bar{\eta}_s\xi)^4}\right](1-\xi)^2P^4_{d}(0).\label{eq:prob_success}
\end{eqnarray} 
In this paper, we consider the case where heralding is repeated until a ``success'' is flagged for each idler photon added to the graph. Thus, the fidelity of the entanglement swapping procedure is $F_{\text{swap}} = P_t/P_s$.

\subsubsection{\label{app:APPbiv} Decoherence}

The emitter used to generate photons, as well as any auxiliary spins entangled with an existing graph, will dephase across the duration until they are measured and projected back onto a known state. For an emitter or auxiliary spin represented by a density matrix $\hat{\rho}$, we model the dephasing process via the map
\begin{equation}\label{eq:decoherence_map2}
    \hat{\rho}\to\mathcal{D}(\hat{\rho};t,\tau)=\frac{1}{2}(1+e^{-t/\tau})\hat{\rho}+\frac{1}{2}(1-e^{-t/\tau})Z\hat{\rho}Z,
\end{equation}
where $\tau$ is the coherence time of the emitter or auxiliary spin and $t$ is a timescale for the dephasing process. Clearly this process leaves the state invariant with probability $\tfrac{1}{2}(1+e^{-t/\tau})$, or conjugates it by $Z$ with probability  $\tfrac{1}{2}(1-e^{-t/\tau})$. It is straightforward to verify that this map describes a Markovian process such that
\begin{equation}
    \mathcal{D}(\hat{\rho};t_1 + t_2,\tau) = \mathcal{D}(\hat{\rho};t_2,\tau) \circ \mathcal{D}(\hat{\rho};t_1,\tau).
\end{equation}

We assume that each attempted herald happens on a short, regular timescale $t_{\text{rep}}=R_{\text{rep}}^{-1}$, which is simply the inverse repetition rate of the experiment. Furthermore, we assume each qubit gate operation time is near-instantaneous compared to $t_{\text{rep}}$, so that the dephasing takes place during the attempted herald only. As the emitter is measured out and reinitialized with each cycle of the experiment, it is evident that it will only dephase for at most $t_{\text{rep}}$. This implies that for each photon added to the graph, the contribution to the graph state fidelity from the decoherence of the emitter is $\tfrac{1}{2}(1+e^{-t_{\text{rep}}/\tau_e})$, such that
\begin{equation}
    F_D^{(e)}(n_p) = \left(\frac{1}{2}\left(1+e^{-t_{\text{rep}}/\tau_e}\right)\right)^{n_p}, \label{eq:emt-dephasing}
\end{equation} 
is the total contribution towards building an $n_p$-photon graph state, and $\tau_e$ is the coherence time for the emitter.

We now model the contribution to the final state fidelity from any auxiliary spins in the system, which dephase for the entirety of their presence in the graph. Consider a mixed state density operator $\hat{\rho}_n=\sum_{\lambda}p_\lambda\op{\Phi_\lambda}{\Phi_\lambda}$, described by a convex combination of $n$-qubit pure stabilizer states $\ket{\Phi_\lambda}$, where $p_\lambda$ are the corresponding classical probabilities. This could represent the mixture of states one expects for an existing graph state built on a set of auxiliary spins, when under the action of its environment---including the prior action of the dephasing map itself. The system is initially stabilized by a set of generators $S_{\ket{\Phi_\lambda}}=\langle g_{1,\lambda}, g_{2,\lambda}, \cdots, g_{n,\lambda} \rangle$. Under the dephasing map the state evolves as
\begin{equation}
    \mathcal{D}^{(s)}(\hat{\rho}_n;t,\tau_s)=\frac{1}{2}(1+e^{-t/\tau_s})\sum_{\lambda}p_\lambda\op{\Phi_\lambda}{\Phi_\lambda}+\frac{1}{2}(1-e^{-t/\tau_s})\sum_{\lambda}p_\lambda Z^{(s)}\op{\Phi_\lambda}{\Phi_\lambda}Z^{(s)},
\end{equation}
which is once again a convex combination of stabilizer states with generators
\begin{eqnarray}
    S_{\ket{\Phi_\lambda}}&&=\langle g_{1,\lambda}, g_{2,\lambda}, \cdots, g_{n,\lambda}\rangle, \nonumber\\
    S_{Z^{(s)}\ket{\Phi_\lambda}}&&=\langle Z^{(s)}g_{1,\lambda}Z^{(s)}, Z^{(s)}g_{2,\lambda}Z^{(s)}, \cdots, Z^{(s)}g_{n,\lambda}Z^{(s)}\rangle.
\end{eqnarray}
$\tau_s$ is the coherence time of the auxiliary spins in the system, which we treat as identical. Since we are interested in graph states, $\hat{\rho}_{G=(V,E)}=\op{G}{G}$, whose stabilizers are generated by the operators $\{K_a\}_{a\in V}$ defined in Eq. \eqref{eq:correlators}, the dephasing map only acts on the generators proportional to $X^{(s)}$, for any auxiliary spin $(s)$ in the system. Hence, a graph state consisting of $n_p$ photons and $n_s$ auxiliary spins would 
evolve under the dephasing map as the mixture
\begin{equation}
     \mathcal{D}^{(s)}(\hat{\rho}_G;t,\tau_s)=\sum_{\mathbf{b}\in\mathbb{Z}_{2}^{n_s}}p(\mathbf{b})\op{G_{\mathbf{b}}}{G_{\mathbf{b}}},
\end{equation}
where $\mathbf{b}=(b_{1},b_{2},\cdots,b_{n_s})\in\mathbb{Z}_{2}^{n_s}$ and
\begin{equation}
    p(\mathbf{b})=\frac{1}{2^{n_s}}\prod_{i=1}^{n_s}(1+(-1)^{b_{i}}e^{-t/\tau_s}).
\end{equation}
This mixture is stabilized by the set of generators,
\begin{equation}
    S_{\ket{G_{\mathbf{b}}}}=\langle K_1^{(p)},\cdots,K_{n_p}^{(p)},(-1)^{b_{1}}K_{1}^{(s)},\cdots,(-1)^{b_{n_s}}K_{n_s}^{(s)}\rangle.
\end{equation}

The size of the system's stabilizer grows by one with each new photon $m$ added to the graph. For the passing-subroutines we employ in Appendix \ref{app:CONb}, we apply this rule, however in choosing either variation of the subroutine we choose whether to rotate the auxiliary spin out of the basis of the dephasing map, and consequentially project the system onto a new mixed state as described above. That is for the $m^{\text{th}}$ photon, we generate a new subgroup of the stabilizer as either

\begin{subequations}
    \begin{eqnarray}
        S_{\ket{G},\text{join}-m}&&=\langle (-1)^{b_k}\cdots Z_k^{(p)}\cdots Z_m^{(p)}X^{(s)}, X_m^{(p)}Z^{(s)}\rangle, \\
        S_{\ket{G},\text{extend}-m}&&=\langle (-1)^{b_k}\cdots Z_k^{(p)}\cdots X_m^{(p)}Z^{(s)}, Z_m^{(p)}X^{(s)}\rangle.
    \end{eqnarray}
\end{subequations}
Here, the bit conditioning phase $b_k$ is applied to the generator $K^{(s)}$ during the creation of the $k^{\text{th}}$ photon, with $k<m$. If we allow the system to continue dephasing, these subgroups evolve, distinctly, as
\begin{subequations}
    \begin{eqnarray}
        S_{\ket{G_\mathbf{b}},\text{join}-m}&&=\langle (-1)^{b_k + b_m}\cdots Z_k^{(p)}\cdots Z_m^{(p)}X^{(s)}, X_m^{(p)}Z^{(s)}\rangle, \\
        S_{\ket{G_\mathbf{b}},\text{extend}-m}&&=\langle (-1)^{b_k}\cdots Z_k^{(p)}\cdots X_m^{(p)}Z^{(s)}, (-1)^{b_m}Z_m^{(p)}X^{(s)}\rangle.
    \end{eqnarray}
\end{subequations}
We see that in the case of executing the extend-subroutine on $m^{\text{th}}$ photon, the number of states in the mixture doubles, whereas in the case of the join-subroutine, the symmetry between states where either $b_k=b_m$ or $b_k\neq b_m$ leads to an effective mixture the same size as it was prior to the $m^{\text{th}}$ photon's addition. Furthermore, we can express the probability for each state in the mixture in terms of $b_k$ and $b_m$ as
\begin{equation}
    P(b_k,b_m)=\frac{1}{4}\left(1+(-1)^{b_k}e^{-t_k/\tau_s}\right)\left(1+(-1)^{b_m}e^{-t_m/\tau_s}\right).
\end{equation}
where we have defined $t_k$ and $t_m$ as arbitrary times from the probabilistic nature of the herald. For the join-subroutine, the graph state remains unchanged when $b_k, b_m = 0$ or $b_k, b_m = 1$, whereas, for the extend-subroutine, the graph state remains unchanged only when $b_k, b_m = 0$. Hence, the fidelity under the dephasing model scales as $\frac{1}{2}(1 + e^{-(t_k + t_m)/\tau_s})$ for the join-subroutine and $\frac{1}{4}(1 + e^{-t_k/\tau_s})(1 + e^{-t_m/\tau_s})$ for the extend-subroutine for iterations $k$ and $m$. This is a subtle distinction that arises from the Markovian nature of the dephasing map when implementing the join-subroutine. It implies the final state fidelity when constructing graph states on a set of auxiliary spins will in general depend on the sequence describing how each photon is passed into the system. In contrast, we assume $t_{\text{rep}}\ll\tau_s$ in our estimates, such that this distinction is not necessary, as $\frac{1}{4}(1 + e^{-t_k/\tau_s})(1 + e^{-t_m/\tau_s}) \approx \frac{1}{2}(1 + e^{-(t_k + t_m)/\tau_s})$. It is nonetheless important to discuss, when considering auxiliary spins with shorter coherence times or constructing exceedingly large graph states, where the time to build the graph is of order the coherence time. 

Lastly, we account for the probabilistic nature of the herald. Building graph states using either scheme we propose can be viewed as a Bernoulli trial with success probability $P_s$ occurring at regular intervals $t_{\text{rep}}$. The cumulative probability that $m$ trials yields $r$ successes is given by the distribution
\begin{equation}
    h(m,r,P_s)=\binom{m-1}{r-1}P_s^r(1-P_s)^{m-r},
\end{equation}
where $m=r, r+1, r+2, \cdots$. For any auxiliary spins in the graph, any failure to herald adds an additional $t_{\text{rep}}$ to the dephasing time. Therefore, we model the mean contribution to the state fidelity for an auxiliary spin with $r$ photons passed to it under either passing-subroutine as
\begin{eqnarray}
    \langle F_D^{(s)}(r,P_s)\rangle &&= \sum_{m=r}^{\infty}h(m,r,P_s)\left(\frac{1}{2}\left(1+e^{-mt_{\text{rep}}/\tau_s}\right)\right) \nonumber\\
    &&=\dfrac{1}{2}\left(1+\left(\dfrac{P_s}{P_s+e^{t_{\text{rep}}/\tau_s}-1}\right)^{r}\right) \nonumber\\
    &&\approx\dfrac{1}{2}\left(1 + e^{-rt_{\text{rep}}/P_s\tau_s}\right), \label{eq:aux-dephasing}
\end{eqnarray}
where $\langle\cdots\rangle$ here denotes a classical expectation value, and $t_{\text{rep}}/P_s$ is the average time for a successful herald. For an $n_p$-photon graph state, we take $r$ up to the number of photons that a given auxiliary spin remains a part of the graph. Here, we consider the same coherence times for the emitter and auxiliary spins, $\tau_e = \tau_s = \tau$.

\subsubsection{\label{app:APPbvi} Gates}

Single-qubit gates applied to photons, the emitter and any auxiliary spins are assumed to be perfect in our model. Two-qubit spin-spin entangling gates are implemented in our model with a fidelity $F_{CZ}$. We assume the time to implement any single-qubit or two-qubit gate, $t_{\text{gate}}\ll t_{\text{rep}}$, and can therefore be treated as instantaneous \cite{Schafer-2018a}.

\
\subsubsection{\label{app:APPbvii} Mid-circuit measurement/reset (MCMR)}

When the additional timescales imparted by MCMR operations on the emitting spin are relevant ($t_{\text{rep}} \ll t_{\text{add}}$), the effects on the dephasing of the emitter and auxiliary spins can be handled by modifying $t_{\text{rep}}\to t_{\text{rep}}+ t_{\text{add}}$ in Eq. \ref{eq:emt-dephasing}, and $t_{\text{rep}}\to t_{\text{rep}}+ P_st_{\text{add}}$ in Eq. \ref{eq:aux-dephasing} (as MCMR operations are conditioned only on a successful herald). We assume that MCMR operations occur with a fidelity $F_{\text{MCMR}}$.

\subsubsection{\label{app:APPbviii} Initial entanglement fidelity}

In practice, there will be a non-unit fidelity on initializing the emitter onto the state described in Eq. \eqref{eq:psi_e} as well as Eq. \eqref{eq:psi_p} for the pair source. For example, the main contribution to the degradation of polarization-based entanglement fidelity for a single photon pair in SPDC comes from the spatial-temporal profile of the pump \cite{Rangarajan-2009a} and its interaction with the specific nonlinear medium, in which different spatial and spectral or temporal components impart different relative phases to the entangled pair described by Eq. \eqref{eq:psi_p}. The fidelity can be optimized by using compensating crystals or filtering techniques; however, the latter also lowers the collection efficiency. In terms of overall fidelity estimations, we can introduce additional scaling terms similar to the gate fidelity described above. We denote the initial entanglement fidelities for the emitter photon and photon-photon pair as $F_e$ and $F_p$, respectively. We focus primarily on the fidelity from the SPDC source to highlight the additional infidelity from our scheme, and assume $F_e=1$.

\subsection{\label{app:APPc} Pair generation rate optimization}

Given the infidelity from false heralds, defined in Eqs. \eqref{eq:prob_true} and \eqref{eq:prob_success}, there exists a trade-off between the rate and fidelity at which large graph states can be generated in the general scheme, as set by the dimensionless parameter $\xi$, controlling the pair generation rate of the SPDC source. Decreasing $\xi$ improves the probability of heralding a true Bell state, at the cost of longer average times between successful heralds. Conversely, the effects of decoherence on any auxiliary spins in the system, through Eq \eqref{eq:aux-dephasing}, reduces the fidelity of the graph state, as the overall construction time increases. Hence, there is an optimum rate at which this scheme can run, controlled by $\xi$ and parameterized by the relevant total collection and detection efficiencies $\{\eta_e,\eta_s,\eta_i\}$, dark count rate, $R_d$, dephasing timescale, $t_{\text{rep}}/\tau$, number of auxiliary spins, $n_s$, and total number of photons, $n_p$, required by the graph. Given a fixed set of parameters, this is a straightforward scalar maximization problem that can be accomplished numerically. If necessary, one can optimize within some $\epsilon$ of the optimum in order to achieve a speed-up in the rate.

\subsection{\label{app:APPd} Example applications}

We briefly demonstrate how the sources of infidelity, presented in Appendix \ref{app:APPb}, can be combined to produce fidelity estimates for constructing certain graph states. We note from the above that this fidelity need be optimized over $\xi$, which controls the pair generation rate of the photon pair source. Below, we assume a negligible dark count rate, and let $\eta_s=\eta_i=1$. In practice the signal photon collection efficiency can be near unity \cite{Ramelow-2013a}, and only makes a minor impact on the rate and fidelity that does not affect the overall scaling of the fidelity, relative to the other relevant parameters. We let $F_{\text{add}} = F_{p}F_{CZ}F_{\text{MCMR}}$ be the collective fidelity associated with adding a photon to an existing graph state in our scheme. Hence, the fidelity to build any graph state on a single auxiliary spin follows a straightforward form in this model,
\begin{equation}
    F_{n_s=1}(\xi;n_p,\eta_e) = \left(F_{\text{add}}F_{\text{swap}}(\xi;\eta_e)\right)^{n_p}F_{D}^{(e)}(n_p)\langle F_{D}^{(s)} \left(n_p,P_s(\eta_e,\xi)\right)\rangle. \label{eq:fid-cluster1D}
\end{equation}
Eq. \ref{eq:fid-cluster1D} applies to the construction of star graph states and linear cluster states, and can compounded appropriately to build more complicated geometries with additional spin-spin entangling gates.  

In Sec. \ref{sec:MPC}, we introduce a 12-qubit graph state $\ket{G}$, depicted in Fig. \ref{fig:G}, that is consumed in our multi-party computation protocol. As all the measurements in our protocol are Clifford, these states can built with a much simpler form of our emit-then-add scheme, which forgoes the photon pair source and entanglement swapping apparatus discussed above. In this case, the probability of a successful herald of the virtual photon is set directly by the emitter collection efficiency, such that $P_s=\eta_e$. The fidelity to produce a single copy of $\ket{G}$ in this scheme is
\begin{equation}
     F_{G} = F_{\text{add}}^{12}F_{CZ}^{5}F_D^{(e)}(12) \langle F_D^{(s_1)}(12,\eta_e)\rangle \langle F_D^{(s_2)}(4,\eta_e) \rangle^2, \label{eq:fid-G}
\end{equation}
where the superscript $(s_1)$ and $(s_2)$ distinguish between the two auxiliary spins required to build the graph. We employ this estimate in Fig. \ref{fig:RateVsET} to determine a conservative estimate of the bit error rate on each round of our protocol. There, we neglect the effect of dephasing in Eq. \ref{eq:fid-G}, in assuming $t_{\text{rep}}\ll \tau$, and let $F_{\text{add}}=F_{CZ}$ for simplicity.

\section{\label{app:SEC} Secure multi-party computation}

We prove the security of our multi-party computation protocol detailed in Sec. \ref{sec:MPC} via a simulation based argument, using the preliminaries of Appendix $\ref{app:GSP}$. The measurement sequence described in Stage I along with the classical post-processing of Stage II together accomplish the computation of any Boolean function up to quadratic order in the parties' inputs. We verify this by tracing through the transformation of the graph state $\ket{G}$ (see Fig. \ref{fig:G}) in the stabilizer formalism, demonstrating the information that is transmitted along the graph in the form of phase flips conditioned on both the parties' inputs and their measurements.  We establish security by demonstrating that the classically communicated information in the protocol reveals no new information to any of the other parties about their input, other than what they would learn from the final output of the computation. Subsequently, we detail how classical error correction can be utilized to suppress a bit error on the output of the computation. Finally, we discuss how self-testing can be employed to further secure against a possibly corrupt Source.

\subsection{\label{app:SECa} Security}

\begin{figure}
    \centering
    \includegraphics[width=0.85\linewidth]{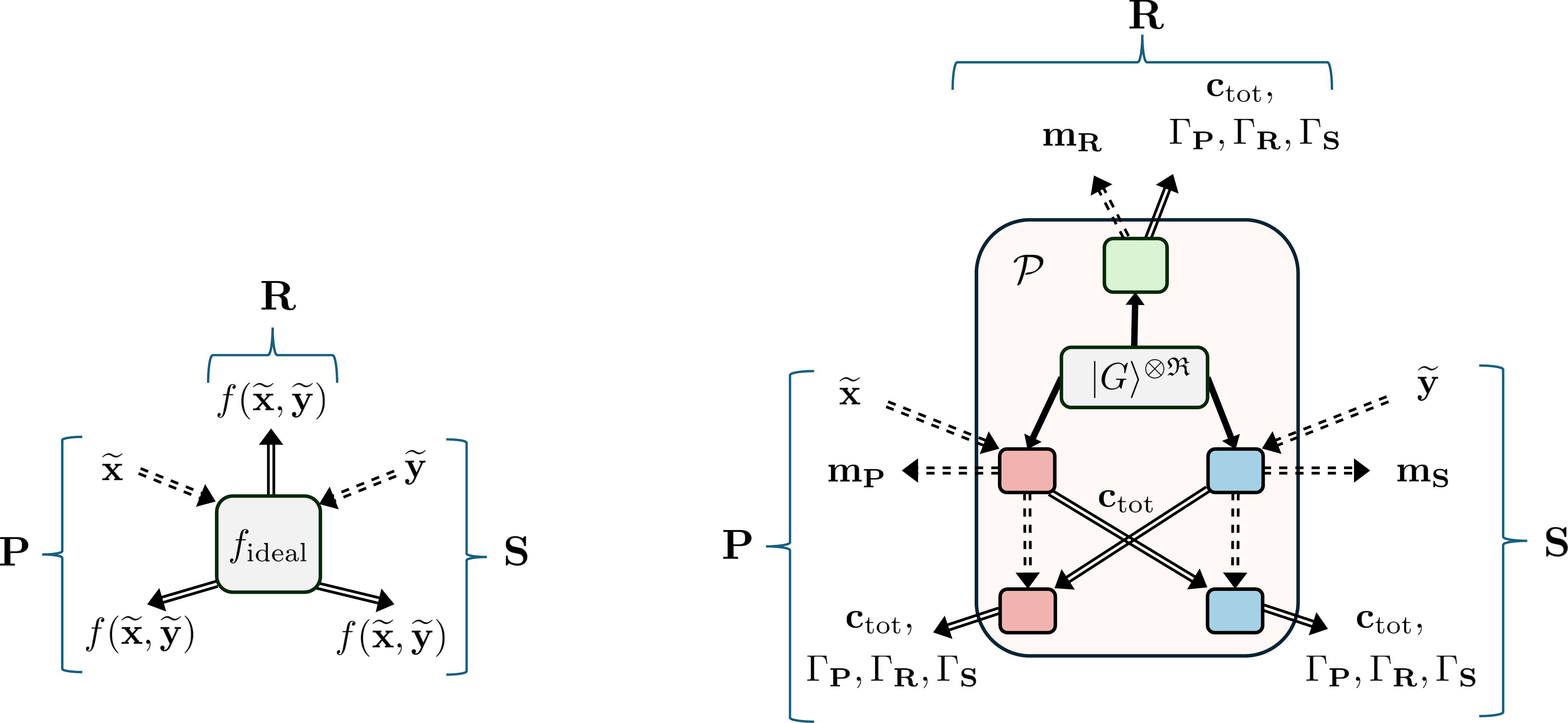}
    \caption{\label{fig:SecuritySetting} Diagram of the security setting for $\mathcal{P}$, as described in Sec. \ref{sec:MPC}. An ideal device (left) publicly broadcasts the function value $f(\Tilde{\mathbf{x}},\Tilde{\mathbf{y}})$ for private inputs $(\Tilde{\mathbf{x}},\Tilde{\mathbf{y}})$. A high-level depiction of $\mathcal{P}$ (right), with $\mathbf{m}_{\mathbf{X}}$ denoting the collective local measurement outcomes for party $\mathbf{X}\in\{\mathbf{P},\mathbf{S},\mathbf{R}\}$ and $\mathbf{c}_{\text{tot}}$ denoting all the public communication in the first part of Stage II. The boxes (red, blue, green) indicate local processing by a given party ($\mathbf{P},\mathbf{S},\mathbf{R}$). Solid (dashed) double lines indicate public (private) classical information. Solid single lines indicate distributed quantum information.} 
\end{figure}

We return to the security of our protocol $\mathcal{P}$, introduced in Sec. \ref{sec:MPC}, and present a formal proof. We adopt a simulation based notion of security, in which we present an idealized evaluation of $f$ that receives inputs privately and returns only the correct function evaluation. Fig. \ref{fig:SecuritySetting} depicts this security setting in both the ideal and real worlds. We show that all the classical and quantum information held by an adversary in any attack on the real protocol $\mathcal{P}$ can equivalently be simulated from the ideal evaluation of $f$.  Hence all the information an adversary extracts about an honest party's input through $\mathcal{P}$ is already accessible through an ideal evaluation of $f$. In this paper we address two forms of malicious attacks, depicted in Fig. \ref{fig:MaliciousAttack}, where we assume without loss of generality that $\mathbf{P}$ is the party playing honestly.

We begin by formalizing our security definition in the settings that follow. To clarify further, first consider when $\mathbf{S}$ potentially deviates from the protocol. The most general physical action can be formally described by a quantum instrument $\mathcal{I}$ that might depend on some classical side information $\lambda$ \cite{Davies-1970a}. For every $\Tilde{\mathbf{x}}$ of $\mathbf{P}$, the instrument will generate for $\mathbf{S}$ a classical-quantum (cq) state $\rho_{\text{cq}}^{\mathbf{S}}(\Tilde{\mathbf{x}},\lambda)$, which contains all of $\mathbf{S}$'s quantum registers along with classical registers recording all of the public communication $(\mathbf{c}_{\text{tot}},\Gamma_{\mathbf{P}},\Gamma_{\mathbf{S}},\Gamma_{\mathbf{R}})$. To maintain consistency with the protocol, we assume $\mathbf{S}$ still broadcasts some classical message when called for in the honest protocol, although this message can be generated in an arbitrary way. On the other hand, consider a case when $\mathbf{R}$ deviates from the instruction of $\mathcal{P}$. Whenever $\mathbf{P}$ and $\mathbf{S}$ follow the protocol honestly on any pair of inputs $(\Tilde{\mathbf{x}},\Tilde{\mathbf{y}})$, they will generate for $\mathbf{R}$ the cq state $\rho_{\text{cq}}^{\mathbf{R}}(\Tilde{\mathbf{x}},\Tilde{\mathbf{y}})$, from which any operation could be performed by $\mathbf{R}$ to extract information.

\textbf{Theorem 1}. \textit{For any two-party function $f(\Tilde{\mathbf{x}},\Tilde{\mathbf{y}})=z_{\mathbf{P}}+z_{\mathbf{S}}+\sum_{i=1}^{\mathfrak{R}}a_ib_i$}, the protocol $\mathcal{P}$ satisfies the following security conditions.
\begin{enumerate}
    \item \textit{(Correctness.) If all the parties are honest, then for every input $(\Tilde{\mathbf{x}},\Tilde{\mathbf{y}})$ the output of the ideal protocol can be locally computed from the outputs of $\mathcal{P}$.}
    \begin{equation}
        f(\Tilde{\mathbf{x}},\Tilde{\mathbf{y}})=\Gamma_{\mathbf{P}}+\Gamma_{\mathbf{S}}+\Gamma_{\mathbf{R}}
    \end{equation}
    \item \textit{(Secure if only $\mathbf{S}$ cheats.) Suppose $\mathbf{P}$ and $\mathbf{R}$ follow $\mathcal{P}$ honestly, but $\mathbf{S}$ potentially deviates. Let $\rho_{\text{cq}}^{\mathbf{S}}(\Tilde{\mathbf{x}},\lambda)$ denote the total cq state held by $\mathbf{S}$ at the end of $\mathcal{P}$, given input $\Tilde{\mathbf{x}}$  of $\mathbf{P}$ and side information $\lambda$. Then, there exists a simulator $\mathbb{S}$ for $\mathbf{S}$ interacting with the ideal functionality $f_{\text{ideal}}$ that exactly reproduces $\rho_{\text{cq}}^{\mathbf{S}}(\Tilde{\mathbf{x}},\lambda)$ for all $\Tilde{\mathbf{x}}$ and $\lambda$.}
    \item \textit{(Secure if only $\mathbf{R}$ cheats.) Suppose $\mathbf{P}$ and $\mathbf{S}$ follow $\mathcal{P}$ honestly, but $\mathbf{R}$ potentially deviates. Let $\rho_{\text{cq}}^{\mathbf{R}}(\Tilde{\mathbf{x}},\Tilde{\mathbf{y}})$ denote the total cq state held by $\mathbf{R}$ at the end of $\mathcal{P}$, given inputs $(\Tilde{\mathbf{x}},\Tilde{\mathbf{y}})$ of $\mathbf{P}$ and $\mathbf{S}$, respectively. Then, there exists a simulator $\mathbb{S}$ for $\mathbf{R}$ interacting with the ideal functionality $f_{\text{ideal}}$ that exactly reproduces $\rho_{\text{cq}}^{\mathbf{R}}(\Tilde{\mathbf{x}},\Tilde{\mathbf{y}})$ for all $(\Tilde{\mathbf{x}},\Tilde{\mathbf{y}})$.}
\end{enumerate}

\textbf{Scenario I: Honest $\mathbf{P}$ and $\mathbf{R}$}. The first scenario we consider involves $\mathbf{P}$ and $\mathbf{R}$ participating honestly in $\mathcal{P}$, while $\mathbf{S}$ is a potentially malicious adversary. $\mathcal{P}$ involves independent local measurements on each copy of $\ket{G_{i}}$ by the honest participants. Let $\Tilde{\mathbf{x}}$ be any fixed input of the function $f$ for $\mathbf{P}$. We consider two subcases where $\mathbf{S}$ takes on the role of either $\mathbf{A}_i$ or $\mathbf{B}_i$ in any arbitrary iteration $i$ in Stage I of $\mathcal{P}$.
\begin{enumerate}
    \item[(a)] \textbf{Honest $\mathbf{R}$ and $\mathbf{B}_i$}. Following $\mathcal{P}$, the post-measurement state of $\ket{G_{i}}$ can be easily expressed in the stabilizer formalism. Once qubits $(1, 6, 10)$ are measured by $\mathbf{B}_{i}$ and $(3, 7, 9)$ by $\mathbf{R}$, the post-measurment state following $\mathbf{B}_i$'s rotation $Z_{8}^{m_{i,6}}$ is described by a stabilizer with generators
    \begin{eqnarray}
        S_{\mathbf{A}_i,8,12}&&=\langle (-1)^{m_{i,1}+m_{i,3}}X_{2},\,(-1)^{m_{i,6}+m_{i,7}}Z_{5},\nonumber\\ 
        &&\quad\quad\quad\quad (-1)^{m_{i,3}}X_{4}Z_{12},\,(-1)^{m_{i,6}+m_{i,7}}X_{8}Z_{12},\,(-1)^{m_{i,9}}X_{11}Z_{12},\,Z_{4}Z_{8}Z_{11}X_{12}\rangle.
    \end{eqnarray}
    Note that subscript on $S$ implies none of $\mathbf{A}_i$ qubits have been measured yet, along with qubits $(8,12)$ of the honest parties. Next, $\mathbf{B_i}$ and $\mathbf{R}$ measure qubits $8$ and $12$, respectively. Their choice of measurement basis depends their common measurement outcome $s_{i}\equiv m_{i,9}=m_{i,10}$. The result is that $\mathbf{A}_i$ is left with a state described by the generators
    \begin{equation}
        S_{\mathbf{A}_i}=\langle (-1)^{m_{i,1}+m_{i,3}}X_{2},\, (-1)^{m_{i,6}+m_{i,7}}Z_{5},\, (-1)^{m_{i,3}+s_{i}}X_{4}X_{11},\, (-1)^{(m_{i,6}+m_{i,7})s_{i}+m_{i,8}+m_{i,12}}Z_{4}Z_{11} \rangle. \label{eq:StabA}
    \end{equation}
    Thus, the state of $\mathbf{A}_i$'s qubits are completely determined by the correlations
    \begin{subequations}
        \begin{eqnarray}
            \hat{m}_{i,2}&&\equiv m_{i,1}+m_{i,3}, \\
            \hat{m}_{i,4}&&\equiv m_{i,3}+s_{i}, \\
            \hat{m}_{i,5}&&\equiv m_{i,6}+m_{i,7} \\
            \hat{m}_{i,11}&&\equiv (m_{i,6}+m_{i,7})s_{i} + m_{i,8} + m_{i,12},
        \end{eqnarray}
    \end{subequations}
    where we have used the notation, $\hat{m}$, to indicate the measurement outcomes $\mathbf{A}_i$ could learn upon the appropriate stabilizer measurements. In the first part of Stage II, an honest $\mathbf{B}_i$ broadcasts the message $c_{i,B}=m_{i,1}+b_{i}+1$. We can therefore express $\hat{m}_{i,11}$ in terms of $\mathbf{B}_i$'s input and measurement outcomes as
    \begin{equation}
        \hat{m}_{i,11}=\hat{m}_{i,5}(\hat{m}_{i,2}+\hat{m}_{i,4}+c_{i,B}+b_{i}+1) + m_{i,8} + m_{i,12},
    \end{equation}
    where we have made the substitution $s_{i}=\hat{m}_{i,2}+\hat{m}_{i,4}+c_{i,B}+b_{i}+1$. Finally, let $\hat{c}_{i,A}\in\{0,1\}$ denote the message broadcast by a possibly malicious $\mathbf{A}_i$ in this first part of Stage II, which can be computed in some non-specified way. For example, it could depend on some side information $\lambda$ or any of the values in $\{\hat{m}_{i,2},\hat{m}_{i,4},\hat{m}_{i,5},\hat{m}_{i,11}\}$. Honest $\mathbf{B}_i$ then computes
    \begin{eqnarray}
        \beta_{i}&&=\hat{c}_{i,A}s_{i}+m_{i,8}\nonumber\\
        &&=\hat{c}_{i,A}(\hat{m}_{i,2}+\hat{m}_{i,4}+c_{i,B}+b_{i}+1)+m_{i,8}.
    \end{eqnarray}
    The critical observation here is that due to the nature of measuring only Pauli observables on stabilizer states, the measurement outcomes $(m_{i,1},m_{i,3},m_{i,6},m_{i,7},s_{i},m_{i,8},m_{i,12})$ are independent and uniformly random bits. \textit{Consequentially, the bit values} $(\hat{m}_{i,2},\hat{m}_{i,4},\hat{m}_{i,5},\hat{m}_{i,11},c_{i,B},\beta_{i})$ \textit{are also independent and uniformly random bits.} Hence, when $\mathbf{S}$ plays the role of $\mathbf{A}_i$ in $\mathcal{P}$, up to the first part of Stage II, the bit values they obtain are subsequently uncorrelated from $\mathbf{B}_i$'s input.
    
    \item[(b)] \textbf{Honest $\mathbf{R}$ and $\mathbf{A}_i$}. Once qubits $(2, 5)$ are measured by $\mathbf{A}_i$ and qubits $(3, 7, 9)$ are measured by $\mathbf{R}$, the post-measurement state following $\mathbf{A}_i$'s rotation $Z_{4}^{m_{i,2}}$ is described by a stabilizer with generators
    \begin{eqnarray}
        S_{\mathbf{B}_i,4,11,12}&&=\langle (-1)^{m_{i,2}+m_{i,3}}Z_{1},\, (-1)^{m_{i,5}+m_{i,7}}X_{6},\,(-1)^{m_{i,9}}X_{10}, \nonumber\\
        &&\quad\quad\quad\quad 
        (-1)^{m_{i,2}+m_{i,3}}X_{4}Z_{12},\,(-1)^{m_{i,7}}X_{8}Z_{12},\,(-1)^{m_{i,9}}X_{11}Z_{12},\,    Z_{4}Z_{8}Z_{11}X_{12} \rangle.
    \end{eqnarray}
    Next, $\mathbf{A}_i$ and $\mathbf{R}$ measure qubits $(4,11)$ and $12$, respectively. $\mathbf{A}_i$'s measurement basis depends on her input $a_i$ here, while the choice of $\mathbf{R}$'s depends on $s_{i}\equiv m_{i,9}$. The result is that $\mathbf{B}_i$ is left with a state described by the generators
    \begin{eqnarray}
    S_{\mathbf{B}_{i}}&&=\langle(-1)^{m_{i,2}+m_{i,3}}Z_{1},\, (-1)^{m_{i,5}+m_{i,7}}X_{6},\,(-1)^{m_{i,9}}X_{10},\nonumber\\ 
    &&\quad\quad\quad\quad (-1)^{a_{i}(m_{i,2}+m_{i,3}+s_{i}+1)+m_{i,7}s_{i}+m_{i,4}+m_{i,11}+m_{i,12}}W^{s_{i}}Z_{8}(W^{\dagger})^{s_{i}}\rangle. \label{eq:StabB}
    \end{eqnarray}
    $\mathbf{B}_i$'s qubits are completely determined by the correlations
    \begin{subequations}
        \begin{eqnarray}
            \hat{m}_{i,1}&&\equiv m_{i,2}+m_{i,3},\\
            \hat{m}_{i,6}&&\equiv m_{i,5}+m_{i,7},\\
            \hat{m}_{i,8}&&\equiv a_{i}(m_{i,2}+m_{i,3}+s_{i}+1)+m_{i,7}s_{i}+m_{i,4}+m_{i,11}+m_{i,12}, \\
            \hat{m}_{i,10}&&\equiv s_{i},
        \end{eqnarray}
    \end{subequations}
    where the measurement outcomes $(m_{i,2},m_{i,3},m_{i,4},m_{i,5},m_{i,7},s_{i},m_{i,11},m_{i,12})$ are independent and uniformly random bits. In the first part of Stage II, honest $\mathbf{A}_i$ broadcasts the message $c_{i,A}=m_{i,5}+a_{i}$. Let $\hat{c}_{i,B}\in\{0,1\}$ denote the message $\mathbf{B}_i$ broadcasts, which could depend on side information $\lambda$ or any other information accessible to $\mathbf{B}_{i}$, and let $\alpha_i=a_{i}\hat{c}_{i,B}+m_{i,4}+m_{i,11}$. As before, \textit{the bit values} $(\hat{m}_{i,1},\hat{m}_{i,6},\hat{m}_{i,8},\hat{m}_{i,10},c_{i,A},\alpha_{i})$ \textit{are independent and uniformly random bits.} Hence, when $\mathbf{S}$ plays the role of $\mathbf{B}_i$ in $\mathcal{P}$, up to the first part of Stage II, the bit values they obtain are subsequently uncorrelated from $\mathbf{A}_i$'s input.
\end{enumerate}

We now consider the second opening of Stage II, which combines all of the information $\mathbf{S}$ has obtained from playing either $\mathbf{A}_i$ or $\mathbf{B}_i$ for various $i$. Let $\mathcal{A}\subset\{1,\cdots,\mathfrak{R}\}$ denote the set of conjunctions in which $\mathbf{P}$ assumed the role $\mathbf{A}_i$, and similarly, let $\mathcal{B}\subset \{1,\cdots,\mathfrak{R}\}$ denote the set of conjunctions in which $\mathbf{P}$ assumed the role of $\mathbf{B}_i$. Honest $\mathbf{R}$ will broadcast the message $\Gamma_{\mathbf{R}}=\sum_{i=1}^{\mathfrak{R}}m_{i,12}$, while honest $\mathbf{P}$ will broadcast
\begin{eqnarray}
    \Gamma_{\mathbf{P}}&&\equiv z_{\mathbf{p}}+\sum_{i\in\mathcal{A}}\alpha_i + \sum_{i\in\mathcal{B}}\beta_i\nonumber\\
    &&=z_{\mathbf{p}}+\sum_{i\in\mathcal{A}}(a_{i}\hat{c}_{i,B}+m_{i,4}+m_{i,11})+\sum_{i\in\mathcal{B}}(\hat{c}_{i,A}(\hat{m}_{i,2}+\hat{m}_{i,4}+c_{i,B}+b_i+1)+m_{i,8}).
\end{eqnarray}
In total, all of the information (both classical and quantum) imparted to party $\mathbf{S}$ at the end of $\mathcal{P}$ is characterized by the bit values
\begin{equation}
    \{\hat{m}_{i,1},\hat{m}_{i,6},\hat{m}_{i,8},\hat{m}_{i,10},c_{i,A}|i\in\mathcal{A}\}\cup\{\hat{m}_{i,2},\hat{m}_{i,4},\hat{m}_{i,5},\hat{m}_{i,11},c_{i,B}|i\in\mathcal{B}\}\cup\{\Gamma_{\mathbf{P}},\Gamma_{\mathbf{R}}\}
\end{equation}
As highlighted above, all the bits from the former two sets along with $\Gamma_{\mathbf{P}}$ are independent and uniformly random. However, $\Gamma_{\mathbf{R}}$ is completely determined by these bits and the inputs of $\mathbf{P}$ due to the correlation
\begin{eqnarray}
   \Gamma_{\mathbf{P}}+\Gamma_{\mathbf{R}}+\sum_{i\in\mathcal{A}}\hat{m}_{i,11}+\sum_{i\in\mathcal{B}}\hat{m}_{i,8} &&= z_{\mathbf{P}} + \sum_{i\in\mathcal{A}}(a_{i}(\hat{m}_{i,1}+\hat{c}_{i,B}+1)+(\hat{m}_{i,6}+c_{i,A})s_{i})\nonumber\\
   &&\quad\quad\quad\quad+\sum_{i\in\mathcal{B}}(\hat{m}_{i,5}+\hat{c}_{i,A})(\hat{m}_{i,2}+\hat{m}_{i,4}+c_{i,B}+b_{i}+1).
\end{eqnarray}
In other words, we have
\begin{equation}
    \Gamma_{\mathbf{R}}=\sum_{i\in\mathcal{A}}\hat{m}_{i,11}+\sum_{i\in\mathcal{B}}\hat{m}_{i,8}=\left[z_{\mathbf{P}}+\sum_{i\in\mathcal{A}}a_{i}(\hat{m}_{i,1}+\hat{c}_{i,B}+1)+\sum_{i\in\mathcal{B}}(\hat{m}_{i,5}+\hat{c}_{i,A})b_{i}\right]+\zeta, \label{eq:correlation-honest-P-R}
\end{equation}
where
\begin{equation}
    \zeta=\sum_{i\in\mathcal{A}}(\hat{m}_{i,6}+c_{i,A})s_{i}+\sum_{i\in\mathcal{B}}(\hat{m}_{i,5}+\hat{c}_{i,A})(\hat{m}_{i,2}+\hat{m}_{i,4}+c_{i,b}+1). \label{eq:zeta}
\end{equation}
Observe that the term in brackets is precisely $f(\hat{\mathbf{x}},\hat{\mathbf{e}})$, where $\hat{\mathbf{e}}=\{\hat{m}_{i,1}+\hat{c}_{i,B}+1|i\in\mathcal{A}\}\cup\{\hat{m}_{i,5}+\hat{c}_{i,A}|i\in\mathcal{B}\}$.

The simulation realized in Fig. \ref{fig:MaliciousAttack}(a) is now straightforward. \st{$\mathbf{S}$} \MG{$\mathbb{S}$} generates independently and uniformly random bits
\begin{equation}
    \{\hat{m}_{i,1},\hat{m}_{i,6},\hat{m}_{i,8},\hat{m}_{i,10},c_{i,A}|i\in\mathcal{A}\}\cup\{\hat{m}_{i,2},\hat{m}_{i,4},\hat{m}_{i,5},\hat{m}_{i,11},c_{i,B}|i\in\mathcal{B}\}\cup\{\Gamma_{\mathbf{P}}\}.
\end{equation}
For each such set of values, $\mathbb{S}$ prepares quantum states with stabilizers described by Eqs. \eqref{eq:StabA} and \eqref{eq:StabB}. It then operate a quantum instrument $\mathcal{I}$, set by the specific attack enacted by $\mathbf{S}$, and generates the variables $\{\hat{c}_{i,B}|i\in\mathcal{A}\}$ and $\{\hat{c}_{i,A}|i\in\mathcal{B}\}$ just as $\mathbf{S}$ would when cheating in $\mathcal{P}$, using their prepared quantum states and whatever other side information $\lambda$ they have. What remains for $\mathbb{S}$ is to generate a $\Gamma_{\mathbf{R}}$ that has the same correlations with other variables in $\mathcal{P}$. To accomplish this, $\mathbb{S}$ first computes the value $\zeta$ in Eq \eqref{eq:zeta} using its sampled variables. It then computes the adversary's input $\Tilde{\mathbf{e}}$ from this information and uses this as its input to $f$, thereby obtaining $f(\Tilde{\mathbf{x}},\Tilde{\mathbf{e}})$, and allowing $\mathbb{S}$ to return $\rho_{\text{cq}}^{\mathbf{S}}(\Tilde{\mathbf{x}},\lambda)$ to the adversary. From Eq \eqref{eq:correlation-honest-P-R}, the correct value of $\Gamma_{\mathbf{R}}$ is given by the sum $f(\Tilde{\mathbf{x}},\Tilde{\mathbf{e}})+\zeta$.

\textbf{Scenario II: Honest $\mathbf{P}$ and $\mathbf{S}$}. We now consider the post-measurement state left to $\mathbf{R}$, when $\mathbf{P}$ and $\mathbf{S}$ follow $\mathcal{P}$ honestly. Once qubits $(2,8)$ and $(1,6,9)$ are measured on a given copy of $\ket{G_{i}}$ and the appropriate rotations are performed by $\mathbf{A}_i$ and $\mathbf{B}_i$, respectively, this post-measurement state is described by generators
\begin{eqnarray}
    S_{\mathbf{R},4,8,11}&&=\langle(-1)^{m_{i,2}}X_{3},\,(-1)^{m_{i,6}}X_{7},\,(-1)^{s_{i}}X_{9}, \nonumber\\
    &&\quad\quad\quad\quad(-1)^{m_{i,1}}X_{4}Z_{12},\,(-1)^{m_{i,5}}X_{8}Z_{12},\, (-1)^{s_{i}}X_{11}Z_{12},\, Z_{4}Z_{8}Z_{11}X_{12} \rangle, 
\end{eqnarray}
where $s_{i}\equiv m_{i,10}$. Next, $\mathbf{A}_i$ and $\mathbf{B}_i$ measure qubits $(4,11)$ and $8$, respectively, following the prescription of $\mathcal{P}$. The state of $\mathbf{R}$ is then described by generators
\begin{equation}
    S_{\mathbf{R}}=\langle(-1)^{m_{i,2}}X_{3},\,(-1)^{m_{i,6}}X_{7},\,(-1)^{s_{i}}X_{9},\,(-1)^{a_{i}(m_{i,1}+s_i+1)+m_{i,5}s_{i}+m_{i,4}+m_{i,8}+m_{i,11}}V^{s_{i}}X_{12}(V^{\dagger})^{s_{i}}\rangle.\label{eq:stabR}
\end{equation}
As before, $\mathbf{R}$'s qubits are completely determined by the correlations
\begin{subequations}
    \begin{eqnarray}
        \hat{m}_{i,3}&&\equiv m_{i,2},\\
        \hat{m}_{i,7}&&\equiv m_{i,6},\\
        \hat{m}_{i,9}&&\equiv s_{i},\\
        \hat{m}_{i,12}&&\equiv a_{i}(m_{i,1}+s_i+1)+m_{i,5}s_{i}+m_{i,4}+m_{i,8}+m_{i,11}.
    \end{eqnarray}
\end{subequations}
In the first part of Stage II, $\mathbf{R}$ learns the values $c_{i,A}$ and $c_{i,B}$, as they are defined in $\mathcal{P}$. In the second part, $\mathbf{R}$ learns
\begin{subequations}
    \begin{eqnarray}
        \Gamma_{\mathbf{P}}&&=z_{\mathbf{P}}+\sum_{i\in\mathcal{A}}(a_{i}c_{i,B}+m_{i,4}+m_{i,11})+\sum_{i\in\mathcal{B}}c_{i,A}s_{i}+m_{i,8},\\
        \Gamma_{\mathbf{S}}&&=z_{\mathbf{S}}+\sum_{i\in\mathcal{A}}c_{i,A}s_{i}+m_{i,8}+\sum_{i\in\mathcal{B}}(a_{i}c_{i,B}+m_{i,4}+m_{i,11}).
    \end{eqnarray}
\end{subequations}
The variables $\{\hat{m}_{i,3},\hat{m}_{i,7},\hat{m}_{i,9},\hat{m}_{i,12}|i=1,\cdots,\mathfrak{R}\}$ are jointly independent with $\Gamma_{\mathbf{P}}$. However, $\Gamma_{\mathbf{S}}$ is completely determine by them and the inputs of $\mathbf{P}$ and $\mathbf{S}$, as there exists the correlation
\begin{equation}
    \Gamma_{\mathbf{P}}+\Gamma_{\mathbf{S}}+\sum_{i=1}^{\mathfrak{R}}\hat{m}_{i,12}=z_{\mathbf{p}}+z_{\mathbf{S}}+\sum_{i=1}^{\mathfrak{R}}a_{i}b_{i}=f(\Tilde{\mathbf{x}},\Tilde{\mathbf{y}}). \label{eq:correlation-honest-P-S}
\end{equation}
Hence, the simulation realizing Fig. \ref{fig:MaliciousAttack}(b) is also straightforward. $\mathbb{S}$ first generates independent and uniformly random bits $\{\hat{m}_{i,3},\hat{m}_{i,7},\hat{m}_{i,9},\hat{m}_{i,12}|i=1,\cdots,\mathfrak{R}\}\cup\{\Gamma_{\mathbf{P}}\}$ and prepares the state with the generators in Eq \eqref{eq:stabR}. Then for any pair of inputs $(\Tilde{\mathbf{x}},\Tilde{\mathbf{y}})$ from $\mathbf{P}$ and $\mathbf{S}$ and public evaluation of the function $f(\Tilde{\mathbf{x}},\Tilde{\mathbf{y}})$, \st{$\mathbf{R}$} \MG{$\mathbb{S}$} computes $\Gamma_{\mathbf{S}}$ from Eq \eqref{eq:correlation-honest-P-S}, and returns $\rho_{\text{cq}}^{\textbf{R}}(\Tilde{\mathbf{x}},\Tilde{\mathbf{y}})$ to the adversary.

\subsection{\label{app:SECb} Including error correction}
In practice, each bit conjunction computed in Stage I of the will have some error. In this section we describe a basic error correction method that can be employed to suppress the overall computational error as much as desired. 

Let $\epsilon_{*}$ denote the largest probability of a bit error in each iteration of Stage I, and let $\Xi_i$ be the indicator variable for the occurrence of an error in iteration $i$.  Thus, while the noiseless protocol generates bit values $a_ib_i=m_{i,12} + \alpha_i + \beta_i$ in iteration $i$, the actual implementation generates bit values
\begin{equation}
   a_ib_i=m_{i,12}+\alpha_i+\beta_i+\Xi_i.
\end{equation}
To deal with this, the Referee can employ a simple repetition code.  That is, each iteration $i$ is repeated $K$ times, from which the Referee obtains values $m_{(i,1),12},m_{(i,2),12},\cdots, m_{(i,K),12}$, where 
\begin{equation}
    m_{(i,j),12}=a_ib_i+\alpha_{(i,j)}+\beta_{(i,j)}+\Xi_{(i,j)}.
\end{equation}
Recalling that $\alpha_{(i,j)}$ and$\beta_{(i,j)}$ are Alice and Bob's private bits, for each $j=2,\cdots,K$, Alice and Bob open the messages $\alpha_{(i,1)}+\alpha_{(i,j)}$ and $\beta_{(i,1)}+\beta_{(i,j)}$, respectively.  The Referee adds these to each corresponding $m_{(i,j),12}$ so that his $K$ bit values then have the form
\begin{equation}
    m_{(i,j),12}=a_ib_i+\alpha_{(i,1)}+\beta_{(i,1)} +\Xi_{(i,j)}.
\end{equation}
Note that each $\alpha_{(i,j)}$ and $\beta_{(i,j)}$ is an independent private random bit for Alice and Bob, so the announcements of $\alpha_{(i,1)}+\alpha_{(i,j)}$ and $\beta_{(i,1)}+\beta_{(i,j)}$ do not reveal any information about $\alpha_{(i,1)}$ and $\beta_{(i,1)}$.

The Referee then performs majority voting on the bit values $m_{(i,j),12}$, accepting the value appearing the most among the $K$ sub-iterations. By Hoeffding's inequality, this value will be $m_{i,12}=a_ib_i+\alpha_i+\beta_i$ with a bit error rate $\delta\leq \exp[-2K(\tfrac{1}{2}-\epsilon_{*}-\tfrac{1}{2K})^2]$.  Then, the combined bit error rate in computing $\Gamma_{\mathbf{R}}=\sum_{i=1}^{\mathfrak{R}}m_{i,12}$ is equal to the probability that an odd number of bit errors occur among the $m_{(i,j),12}$.  A counting argument shows this probability to be 
\begin{eqnarray}
    \epsilon&&=\frac{1}{2}[1-(1-2\delta)^{\mathfrak{R}}]\leq \delta \mathfrak{R} \leq \mathfrak{R} \cdot \exp[-2K(\tfrac{1}{2}-\epsilon_{*}-\tfrac{1}{2K})^2].
\end{eqnarray}
Note that to suppress the bit error rate, the number of repetitions $K$ needed for each iteration $i$ is exponentially smaller than the total number of iterations $\mathfrak{R}\leq \binom{N}{2}(M-1)^2$. In particular, to obtain a bit error probability $\epsilon_f$ on the computation of $f$ for fixed $\epsilon_{*}$, it suffices to take
\begin{equation}
    K= \left\lceil\frac{1}{(\frac{1}{2}-\epsilon_{*})^2}\ln\sqrt{\frac{\mathfrak{R}_1}{\epsilon_f}}\right\rceil \leq \left\lceil\frac{1}{(\frac{1}{2}-\epsilon_{*})^2}\ln\left(\frac{(M-1)N}{\sqrt{2\epsilon_f}}\right)\right\rceil.
\end{equation}

In total, the $N$-party computation of $f$ on each party's $M$-bit input can be implemented with error correction in our protocol at a lower bound unit rate of
\begin{equation}
    \frac{R}{R_{0}}\geq \left(6(M-1)^2N^2\left\lceil\frac{\ln\left((M-1)N/\sqrt{2\epsilon_f}\right)}{\left(\tfrac{1}{2}-\epsilon_{*}\right)^2}\right\rceil\right)^{-1},
\end{equation}
where $R_0$ is the scheme-dependent average rate to add a new photon to a graph state. For the simpler emit-then-add scheme we propose, employing only destructive photon measurements, $R_0=\eta_e R_{\text{rep}}$.

\subsection{\label{app:SECc} Self-testing of graph states}

We can further secure the computation against an untrustworthy Source, via a self-testing procedure, which allows the parties and the Referee to validate the state distributed by the Source. If the parties indeed have the correct state, measuring the independent components of a uniformly randomly chosen stabilizer of the state will lead to a correlated outcome, hence implementing a partial test of the state's validity. By repeating the protocol and partitioning rounds between these stabilizer tests, discarded rounds of computation, and a single accepted target round of computation, the parties can upper bound the probability of receiving the incorrect state during the target round \cite{Takeuchi-2019a, Baccari-2020a}. In implementing our MPC within the version of our experiemental scheme without photonic memory, we require the set of instructions, detailing which rounds to test, compute, and ultimately accept, be a private source of shared randomness between the parties and the Referee. This could be generated initially from a set of GHZ states distributed to the participants in the computation \cite{Yang-2013a, Xu-2014a}. Without knowledge of which rounds are stabilizer tests and which are computations, the Source is prevented from modifying the state they distribute in order to fool the participants during rounds of self-testing. Conversely, the general scheme, involving QND measurements, allows for the requisite graph states to be produced prior to the measurement-based computation protocol. Hence, the instruction set directing the participants when to self-test can be decided after the state has already been generated. In these schemes, we therefore do not require a distinct Source and Referee, at the cost of necessitating photonic memory---in requiring the full set of photonic states be generated in advance, as well as additional rounds of self-testing \cite{Unnikrishnan-2022a}.

\end{widetext}

\bibliography{ref}

\providecommand{\noopsort}[1]{}\providecommand{\singleletter}[1]{#1}%
\begin{thebibliography}{79}%
\makeatletter
\providecommand \@ifxundefined [1]{%
 \@ifx{#1\undefined}
}%
\providecommand \@ifnum [1]{%
 \ifnum #1\expandafter \@firstoftwo
 \else \expandafter \@secondoftwo
 \fi
}%
\providecommand \@ifx [1]{%
 \ifx #1\expandafter \@firstoftwo
 \else \expandafter \@secondoftwo
 \fi
}%
\providecommand \natexlab [1]{#1}%
\providecommand \enquote  [1]{``#1''}%
\providecommand \bibnamefont  [1]{#1}%
\providecommand \bibfnamefont [1]{#1}%
\providecommand \citenamefont [1]{#1}%
\providecommand \href@noop [0]{\@secondoftwo}%
\providecommand \href [0]{\begingroup \@sanitize@url \@href}%
\providecommand \@href[1]{\@@startlink{#1}\@@href}%
\providecommand \@@href[1]{\endgroup#1\@@endlink}%
\providecommand \@sanitize@url [0]{\catcode `\\12\catcode `\$12\catcode `\&12\catcode `\#12\catcode `\^12\catcode `\_12\catcode `\%12\relax}%
\providecommand \@@startlink[1]{}%
\providecommand \@@endlink[0]{}%
\providecommand \url  [0]{\begingroup\@sanitize@url \@url }%
\providecommand \@url [1]{\endgroup\@href {#1}{\urlprefix }}%
\providecommand \urlprefix  [0]{URL }%
\providecommand \Eprint [0]{\href }%
\providecommand \doibase [0]{https://doi.org/}%
\providecommand \selectlanguage [0]{\@gobble}%
\providecommand \bibinfo  [0]{\@secondoftwo}%
\providecommand \bibfield  [0]{\@secondoftwo}%
\providecommand \translation [1]{[#1]}%
\providecommand \BibitemOpen [0]{}%
\providecommand \bibitemStop [0]{}%
\providecommand \bibitemNoStop [0]{.\EOS\space}%
\providecommand \EOS [0]{\spacefactor3000\relax}%
\providecommand \BibitemShut  [1]{\csname bibitem#1\endcsname}%
\let\auto@bib@innerbib\@empty
\bibitem [{\citenamefont {Raussendorf}\ and\ \citenamefont {Briegel}(2001)}]{Raussendorf-2001a}%
  \BibitemOpen
  \bibfield  {author} {\bibinfo {author} {\bibfnamefont {R.}~\bibnamefont {Raussendorf}}\ and\ \bibinfo {author} {\bibfnamefont {H.~J.}\ \bibnamefont {Briegel}},\ }\bibfield  {title} {\bibinfo {title} {A one-way quantum computer},\ }\href {https://doi.org/10.1103/PhysRevLett.86.5188} {\bibfield  {journal} {\bibinfo  {journal} {Phys. Rev. Lett.}\ }\textbf {\bibinfo {volume} {86}},\ \bibinfo {pages} {5188} (\bibinfo {year} {2001})}\BibitemShut {NoStop}%
\bibitem [{\citenamefont {Browne}\ and\ \citenamefont {Rudolph}(2005)}]{Browne-2005a}%
  \BibitemOpen
  \bibfield  {author} {\bibinfo {author} {\bibfnamefont {D.~E.}\ \bibnamefont {Browne}}\ and\ \bibinfo {author} {\bibfnamefont {T.}~\bibnamefont {Rudolph}},\ }\bibfield  {title} {\bibinfo {title} {Resource-efficient linear optical quantum computation},\ }\href {https://doi.org/10.1103/PhysRevLett.95.010501} {\bibfield  {journal} {\bibinfo  {journal} {Phys. Rev. Lett.}\ }\textbf {\bibinfo {volume} {95}},\ \bibinfo {pages} {010501} (\bibinfo {year} {2005})}\BibitemShut {NoStop}%
\bibitem [{\citenamefont {Lindner}\ and\ \citenamefont {Rudolph}(2009)}]{Lindner-2009a}%
  \BibitemOpen
  \bibfield  {author} {\bibinfo {author} {\bibfnamefont {N.~H.}\ \bibnamefont {Lindner}}\ and\ \bibinfo {author} {\bibfnamefont {T.}~\bibnamefont {Rudolph}},\ }\bibfield  {title} {\bibinfo {title} {Proposal for pulsed on-demand sources of photonic cluster state strings},\ }\href {https://doi.org/10.1103/PhysRevLett.103.113602} {\bibfield  {journal} {\bibinfo  {journal} {Phys. Rev. Lett.}\ }\textbf {\bibinfo {volume} {103}},\ \bibinfo {pages} {113602} (\bibinfo {year} {2009})}\BibitemShut {NoStop}%
\bibitem [{\citenamefont {Economou}\ \emph {et~al.}(2010)\citenamefont {Economou}, \citenamefont {Lindner},\ and\ \citenamefont {Rudolph}}]{Economou-2010a}%
  \BibitemOpen
  \bibfield  {author} {\bibinfo {author} {\bibfnamefont {S.~E.}\ \bibnamefont {Economou}}, \bibinfo {author} {\bibfnamefont {N.}~\bibnamefont {Lindner}},\ and\ \bibinfo {author} {\bibfnamefont {T.}~\bibnamefont {Rudolph}},\ }\bibfield  {title} {\bibinfo {title} {Optically generated 2-dimensional photonic cluster state from coupled quantum dots},\ }\href {https://doi.org/10.1103/PhysRevLett.105.093601} {\bibfield  {journal} {\bibinfo  {journal} {Phys. Rev. Lett.}\ }\textbf {\bibinfo {volume} {105}},\ \bibinfo {pages} {093601} (\bibinfo {year} {2010})}\BibitemShut {NoStop}%
\bibitem [{\citenamefont {Buterakos}\ \emph {et~al.}(2017)\citenamefont {Buterakos}, \citenamefont {Barnes},\ and\ \citenamefont {Economou}}]{Buterakos-2017a}%
  \BibitemOpen
  \bibfield  {author} {\bibinfo {author} {\bibfnamefont {D.}~\bibnamefont {Buterakos}}, \bibinfo {author} {\bibfnamefont {E.}~\bibnamefont {Barnes}},\ and\ \bibinfo {author} {\bibfnamefont {S.~E.}\ \bibnamefont {Economou}},\ }\bibfield  {title} {\bibinfo {title} {Deterministic generation of all-photonic quantum repeaters from solid-state emitters},\ }\href {https://doi.org/10.1103/PhysRevX.7.041023} {\bibfield  {journal} {\bibinfo  {journal} {Phys. Rev. X}\ }\textbf {\bibinfo {volume} {7}},\ \bibinfo {pages} {041023} (\bibinfo {year} {2017})}\BibitemShut {NoStop}%
\bibitem [{\citenamefont {Russo}\ \emph {et~al.}(2019)\citenamefont {Russo}, \citenamefont {Barnes},\ and\ \citenamefont {Economou}}]{Russo-2019a}%
  \BibitemOpen
  \bibfield  {author} {\bibinfo {author} {\bibfnamefont {A.}~\bibnamefont {Russo}}, \bibinfo {author} {\bibfnamefont {E.}~\bibnamefont {Barnes}},\ and\ \bibinfo {author} {\bibfnamefont {S.~E.}\ \bibnamefont {Economou}},\ }\bibfield  {title} {\bibinfo {title} {Generation of arbitrary all-photonic graph states from quantum emitters},\ }\href {https://doi.org/10.1088/1367-2630/ab193d} {\bibfield  {journal} {\bibinfo  {journal} {New Journal of Physics}\ }\textbf {\bibinfo {volume} {21}},\ \bibinfo {pages} {055002} (\bibinfo {year} {2019})}\BibitemShut {NoStop}%
\bibitem [{\citenamefont {Hilaire}\ \emph {et~al.}(2023)\citenamefont {Hilaire}, \citenamefont {Vidro}, \citenamefont {Eisenberg},\ and\ \citenamefont {Economou}}]{Hilaire-2022a}%
  \BibitemOpen
  \bibfield  {author} {\bibinfo {author} {\bibfnamefont {P.}~\bibnamefont {Hilaire}}, \bibinfo {author} {\bibfnamefont {L.}~\bibnamefont {Vidro}}, \bibinfo {author} {\bibfnamefont {H.~S.}\ \bibnamefont {Eisenberg}},\ and\ \bibinfo {author} {\bibfnamefont {S.~E.}\ \bibnamefont {Economou}},\ }\bibfield  {title} {\bibinfo {title} {Near-deterministic hybrid generation of arbitrary photonic graph states using a single quantum emitter and linear optics},\ }\href@noop {} {\bibfield  {journal} {\bibinfo  {journal} {Quantum}\ }\textbf {\bibinfo {volume} {7}},\ \bibinfo {pages} {992} (\bibinfo {year} {2023})}\BibitemShut {NoStop}%
\bibitem [{\citenamefont {Meng}\ \emph {et~al.}(2023)\citenamefont {Meng}, \citenamefont {Faurby}, \citenamefont {Chan}, \citenamefont {Sund}, \citenamefont {Liu}, \citenamefont {Wang}, \citenamefont {Bart}, \citenamefont {Wieck}, \citenamefont {Ludwig}, \citenamefont {Midolo} \emph {et~al.}}]{Meng-2023a}%
  \BibitemOpen
  \bibfield  {author} {\bibinfo {author} {\bibfnamefont {Y.}~\bibnamefont {Meng}}, \bibinfo {author} {\bibfnamefont {C.~F.}\ \bibnamefont {Faurby}}, \bibinfo {author} {\bibfnamefont {M.~L.}\ \bibnamefont {Chan}}, \bibinfo {author} {\bibfnamefont {P.~I.}\ \bibnamefont {Sund}}, \bibinfo {author} {\bibfnamefont {Z.}~\bibnamefont {Liu}}, \bibinfo {author} {\bibfnamefont {Y.}~\bibnamefont {Wang}}, \bibinfo {author} {\bibfnamefont {N.}~\bibnamefont {Bart}}, \bibinfo {author} {\bibfnamefont {A.~D.}\ \bibnamefont {Wieck}}, \bibinfo {author} {\bibfnamefont {A.}~\bibnamefont {Ludwig}}, \bibinfo {author} {\bibfnamefont {L.}~\bibnamefont {Midolo}}, \emph {et~al.},\ }\bibfield  {title} {\bibinfo {title} {Photonic fusion of entangled resource states from a quantum emitter},\ }\href@noop {} {\bibfield  {journal} {\bibinfo  {journal} {arXiv preprint arXiv:2312.09070}\ } (\bibinfo {year} {2023})}\BibitemShut {NoStop}%
\bibitem [{\citenamefont {Thomas}\ \emph {et~al.}(2024)\citenamefont {Thomas}, \citenamefont {Ruscio}, \citenamefont {Morin},\ and\ \citenamefont {Rempe}}]{Thomas-2024a}%
  \BibitemOpen
  \bibfield  {author} {\bibinfo {author} {\bibfnamefont {P.}~\bibnamefont {Thomas}}, \bibinfo {author} {\bibfnamefont {L.}~\bibnamefont {Ruscio}}, \bibinfo {author} {\bibfnamefont {O.}~\bibnamefont {Morin}},\ and\ \bibinfo {author} {\bibfnamefont {G.}~\bibnamefont {Rempe}},\ }\bibfield  {title} {\bibinfo {title} {Fusion of deterministically generated photonic graph states},\ }\href@noop {} {\bibfield  {journal} {\bibinfo  {journal} {Nature}\ }\textbf {\bibinfo {volume} {629}},\ \bibinfo {pages} {567} (\bibinfo {year} {2024})}\BibitemShut {NoStop}%
\bibitem [{\citenamefont {Zhan}\ and\ \citenamefont {Sun}(2020)}]{Zhan-2020a}%
  \BibitemOpen
  \bibfield  {author} {\bibinfo {author} {\bibfnamefont {Y.}~\bibnamefont {Zhan}}\ and\ \bibinfo {author} {\bibfnamefont {S.}~\bibnamefont {Sun}},\ }\bibfield  {title} {\bibinfo {title} {Deterministic generation of loss-tolerant photonic cluster states with a single quantum emitter},\ }\href@noop {} {\bibfield  {journal} {\bibinfo  {journal} {Physical Review Letters}\ }\textbf {\bibinfo {volume} {125}},\ \bibinfo {pages} {223601} (\bibinfo {year} {2020})}\BibitemShut {NoStop}%
\bibitem [{\citenamefont {Schwartz}\ \emph {et~al.}(2016)\citenamefont {Schwartz}, \citenamefont {Cogan}, \citenamefont {Schmidgall}, \citenamefont {Don}, \citenamefont {Gantz}, \citenamefont {Kenneth}, \citenamefont {Lindner},\ and\ \citenamefont {Gershoni}}]{Schwartz-2016a}%
  \BibitemOpen
  \bibfield  {author} {\bibinfo {author} {\bibfnamefont {I.}~\bibnamefont {Schwartz}}, \bibinfo {author} {\bibfnamefont {D.}~\bibnamefont {Cogan}}, \bibinfo {author} {\bibfnamefont {E.~R.}\ \bibnamefont {Schmidgall}}, \bibinfo {author} {\bibfnamefont {Y.}~\bibnamefont {Don}}, \bibinfo {author} {\bibfnamefont {L.}~\bibnamefont {Gantz}}, \bibinfo {author} {\bibfnamefont {O.}~\bibnamefont {Kenneth}}, \bibinfo {author} {\bibfnamefont {N.~H.}\ \bibnamefont {Lindner}},\ and\ \bibinfo {author} {\bibfnamefont {D.}~\bibnamefont {Gershoni}},\ }\bibfield  {title} {\bibinfo {title} {Deterministic generation of a cluster state of entangled photons},\ }\href {https://doi.org/10.1126/science.aah4758} {\bibfield  {journal} {\bibinfo  {journal} {Science}\ }\textbf {\bibinfo {volume} {354}},\ \bibinfo {pages} {434–437} (\bibinfo {year} {2016})}\BibitemShut {NoStop}%
\bibitem [{\citenamefont {Schupp}\ \emph {et~al.}(2021)\citenamefont {Schupp}, \citenamefont {Krcmarsky}, \citenamefont {Krutyanskiy}, \citenamefont {Meraner}, \citenamefont {Northup},\ and\ \citenamefont {Lanyon}}]{Schupp-2021a}%
  \BibitemOpen
  \bibfield  {author} {\bibinfo {author} {\bibfnamefont {J.}~\bibnamefont {Schupp}}, \bibinfo {author} {\bibfnamefont {V.}~\bibnamefont {Krcmarsky}}, \bibinfo {author} {\bibfnamefont {V.}~\bibnamefont {Krutyanskiy}}, \bibinfo {author} {\bibfnamefont {M.}~\bibnamefont {Meraner}}, \bibinfo {author} {\bibfnamefont {T.}~\bibnamefont {Northup}},\ and\ \bibinfo {author} {\bibfnamefont {B.}~\bibnamefont {Lanyon}},\ }\bibfield  {title} {\bibinfo {title} {Interface between trapped-ion qubits and traveling photons with close-to-optimal efficiency},\ }\href {https://doi.org/10.1103/PRXQuantum.2.020331} {\bibfield  {journal} {\bibinfo  {journal} {PRX Quantum}\ }\textbf {\bibinfo {volume} {2}},\ \bibinfo {pages} {020331} (\bibinfo {year} {2021})}\BibitemShut {NoStop}%
\bibitem [{\citenamefont {Thomas}\ \emph {et~al.}(2022)\citenamefont {Thomas}, \citenamefont {Ruscio}, \citenamefont {Morin},\ and\ \citenamefont {Rempe}}]{Thomas-2022a}%
  \BibitemOpen
  \bibfield  {author} {\bibinfo {author} {\bibfnamefont {P.}~\bibnamefont {Thomas}}, \bibinfo {author} {\bibfnamefont {L.}~\bibnamefont {Ruscio}}, \bibinfo {author} {\bibfnamefont {O.}~\bibnamefont {Morin}},\ and\ \bibinfo {author} {\bibfnamefont {G.}~\bibnamefont {Rempe}},\ }\bibfield  {title} {\bibinfo {title} {Efficient generation of entangled multiphoton graph states from a single atom},\ }\href@noop {} {\bibfield  {journal} {\bibinfo  {journal} {Nature}\ }\textbf {\bibinfo {volume} {608}},\ \bibinfo {pages} {677} (\bibinfo {year} {2022})}\BibitemShut {NoStop}%
\bibitem [{\citenamefont {Cogan}\ \emph {et~al.}(2023)\citenamefont {Cogan}, \citenamefont {Su}, \citenamefont {Kenneth},\ and\ \citenamefont {Gershoni}}]{Cogan-2023a}%
  \BibitemOpen
  \bibfield  {author} {\bibinfo {author} {\bibfnamefont {D.}~\bibnamefont {Cogan}}, \bibinfo {author} {\bibfnamefont {Z.-E.}\ \bibnamefont {Su}}, \bibinfo {author} {\bibfnamefont {O.}~\bibnamefont {Kenneth}},\ and\ \bibinfo {author} {\bibfnamefont {D.}~\bibnamefont {Gershoni}},\ }\bibfield  {title} {\bibinfo {title} {Deterministic generation of indistinguishable photons in a cluster state},\ }\href {https://doi.org/10.1038/s41566-022-01152-2} {\bibfield  {journal} {\bibinfo  {journal} {Nature Photonics}\ }\textbf {\bibinfo {volume} {17}},\ \bibinfo {pages} {324–329} (\bibinfo {year} {2023})}\BibitemShut {NoStop}%
\bibitem [{\citenamefont {Munro}\ \emph {et~al.}(2005)\citenamefont {Munro}, \citenamefont {Nemoto}, \citenamefont {Beausoleil},\ and\ \citenamefont {Spiller}}]{Munro-2005a}%
  \BibitemOpen
  \bibfield  {author} {\bibinfo {author} {\bibfnamefont {W.~J.}\ \bibnamefont {Munro}}, \bibinfo {author} {\bibfnamefont {K.}~\bibnamefont {Nemoto}}, \bibinfo {author} {\bibfnamefont {R.~G.}\ \bibnamefont {Beausoleil}},\ and\ \bibinfo {author} {\bibfnamefont {T.~P.}\ \bibnamefont {Spiller}},\ }\bibfield  {title} {\bibinfo {title} {High-efficiency quantum-nondemolition single-photon-number-resolving detector},\ }\href {https://doi.org/10.1103/PhysRevA.71.033819} {\bibfield  {journal} {\bibinfo  {journal} {Phys. Rev. A}\ }\textbf {\bibinfo {volume} {71}},\ \bibinfo {pages} {033819} (\bibinfo {year} {2005})}\BibitemShut {NoStop}%
\bibitem [{\citenamefont {Xiao}\ \emph {et~al.}(2008)\citenamefont {Xiao}, \citenamefont {\c{S}ahin Kaya~\"{O}zdemir}, \citenamefont {Gaddam}, \citenamefont {Dong}, \citenamefont {Imoto},\ and\ \citenamefont {Yang}}]{Xiao-2008a}%
  \BibitemOpen
  \bibfield  {author} {\bibinfo {author} {\bibfnamefont {Y.-F.}\ \bibnamefont {Xiao}}, \bibinfo {author} {\bibnamefont {\c{S}ahin Kaya~\"{O}zdemir}}, \bibinfo {author} {\bibfnamefont {V.}~\bibnamefont {Gaddam}}, \bibinfo {author} {\bibfnamefont {C.-H.}\ \bibnamefont {Dong}}, \bibinfo {author} {\bibfnamefont {N.}~\bibnamefont {Imoto}},\ and\ \bibinfo {author} {\bibfnamefont {L.}~\bibnamefont {Yang}},\ }\bibfield  {title} {\bibinfo {title} {Quantum nondemolition measurement of photon number via optical kerr effect in an ultra-high-q microtoroid cavity},\ }\href {https://doi.org/10.1364/OE.16.021462} {\bibfield  {journal} {\bibinfo  {journal} {Opt. Express}\ }\textbf {\bibinfo {volume} {16}},\ \bibinfo {pages} {21462} (\bibinfo {year} {2008})}\BibitemShut {NoStop}%
\bibitem [{\citenamefont {Yanagimoto}\ \emph {et~al.}(2023)\citenamefont {Yanagimoto}, \citenamefont {Nehra}, \citenamefont {Hamerly}, \citenamefont {Ng}, \citenamefont {Marandi},\ and\ \citenamefont {Mabuchi}}]{Yanagimoto-2023a}%
  \BibitemOpen
  \bibfield  {author} {\bibinfo {author} {\bibfnamefont {R.}~\bibnamefont {Yanagimoto}}, \bibinfo {author} {\bibfnamefont {R.}~\bibnamefont {Nehra}}, \bibinfo {author} {\bibfnamefont {R.}~\bibnamefont {Hamerly}}, \bibinfo {author} {\bibfnamefont {E.}~\bibnamefont {Ng}}, \bibinfo {author} {\bibfnamefont {A.}~\bibnamefont {Marandi}},\ and\ \bibinfo {author} {\bibfnamefont {H.}~\bibnamefont {Mabuchi}},\ }\bibfield  {title} {\bibinfo {title} {Quantum nondemolition measurements with optical parametric amplifiers for ultrafast universal quantum information processing},\ }\href@noop {} {\bibfield  {journal} {\bibinfo  {journal} {PRX Quantum}\ }\textbf {\bibinfo {volume} {4}},\ \bibinfo {pages} {010333} (\bibinfo {year} {2023})}\BibitemShut {NoStop}%
\bibitem [{\citenamefont {Bock}\ \emph {et~al.}(2018)\citenamefont {Bock}, \citenamefont {Eich}, \citenamefont {Kucera}, \citenamefont {Kreis}, \citenamefont {Lenhard}, \citenamefont {Becher},\ and\ \citenamefont {Eschner}}]{Bock-2018a}%
  \BibitemOpen
  \bibfield  {author} {\bibinfo {author} {\bibfnamefont {M.}~\bibnamefont {Bock}}, \bibinfo {author} {\bibfnamefont {P.}~\bibnamefont {Eich}}, \bibinfo {author} {\bibfnamefont {S.}~\bibnamefont {Kucera}}, \bibinfo {author} {\bibfnamefont {M.}~\bibnamefont {Kreis}}, \bibinfo {author} {\bibfnamefont {A.}~\bibnamefont {Lenhard}}, \bibinfo {author} {\bibfnamefont {C.}~\bibnamefont {Becher}},\ and\ \bibinfo {author} {\bibfnamefont {J.}~\bibnamefont {Eschner}},\ }\bibfield  {title} {\bibinfo {title} {High-fidelity entanglement between a trapped ion and a telecom photon via quantum frequency conversion},\ }\href@noop {} {\bibfield  {journal} {\bibinfo  {journal} {Nature communications}\ }\textbf {\bibinfo {volume} {9}},\ \bibinfo {pages} {1} (\bibinfo {year} {2018})}\BibitemShut {NoStop}%
\bibitem [{\citenamefont {Economou}\ \emph {et~al.}(2006)\citenamefont {Economou}, \citenamefont {Sham}, \citenamefont {Wu},\ and\ \citenamefont {Steel}}]{Economou-2006a}%
  \BibitemOpen
  \bibfield  {author} {\bibinfo {author} {\bibfnamefont {S.~E.}\ \bibnamefont {Economou}}, \bibinfo {author} {\bibfnamefont {L.~J.}\ \bibnamefont {Sham}}, \bibinfo {author} {\bibfnamefont {Y.}~\bibnamefont {Wu}},\ and\ \bibinfo {author} {\bibfnamefont {D.~G.}\ \bibnamefont {Steel}},\ }\bibfield  {title} {\bibinfo {title} {Proposal for optical u(1) rotations of electron spin trapped in a quantum dot},\ }\href {https://doi.org/10.1103/PhysRevB.74.205415} {\bibfield  {journal} {\bibinfo  {journal} {Phys. Rev. B}\ }\textbf {\bibinfo {volume} {74}},\ \bibinfo {pages} {205415} (\bibinfo {year} {2006})}\BibitemShut {NoStop}%
\bibitem [{\citenamefont {Gimeno-Segovia}\ \emph {et~al.}(2019)\citenamefont {Gimeno-Segovia}, \citenamefont {Rudolph},\ and\ \citenamefont {Economou}}]{Gimeno-Segovia-2019a}%
  \BibitemOpen
  \bibfield  {author} {\bibinfo {author} {\bibfnamefont {M.}~\bibnamefont {Gimeno-Segovia}}, \bibinfo {author} {\bibfnamefont {T.}~\bibnamefont {Rudolph}},\ and\ \bibinfo {author} {\bibfnamefont {S.~E.}\ \bibnamefont {Economou}},\ }\bibfield  {title} {\bibinfo {title} {Deterministic generation of large-scale entangled photonic cluster state from interacting solid state emitters},\ }\href {https://doi.org/10.1103/PhysRevLett.123.070501} {\bibfield  {journal} {\bibinfo  {journal} {Phys. Rev. Lett.}\ }\textbf {\bibinfo {volume} {123}},\ \bibinfo {pages} {070501} (\bibinfo {year} {2019})}\BibitemShut {NoStop}%
\bibitem [{\citenamefont {Riel{\"a}nder}\ \emph {et~al.}(2016)\citenamefont {Riel{\"a}nder}, \citenamefont {Lenhard}, \citenamefont {Mazzera},\ and\ \citenamefont {De~Riedmatten}}]{Rielander-2016a}%
  \BibitemOpen
  \bibfield  {author} {\bibinfo {author} {\bibfnamefont {D.}~\bibnamefont {Riel{\"a}nder}}, \bibinfo {author} {\bibfnamefont {A.}~\bibnamefont {Lenhard}}, \bibinfo {author} {\bibfnamefont {M.}~\bibnamefont {Mazzera}},\ and\ \bibinfo {author} {\bibfnamefont {H.}~\bibnamefont {De~Riedmatten}},\ }\bibfield  {title} {\bibinfo {title} {Cavity enhanced telecom heralded single photons for spin-wave solid state quantum memories},\ }\href@noop {} {\bibfield  {journal} {\bibinfo  {journal} {New Journal of Physics}\ }\textbf {\bibinfo {volume} {18}},\ \bibinfo {pages} {123013} (\bibinfo {year} {2016})}\BibitemShut {NoStop}%
\bibitem [{\citenamefont {Covey}\ \emph {et~al.}(2023)\citenamefont {Covey}, \citenamefont {Weinfurter},\ and\ \citenamefont {Bernien}}]{Covey-2023a}%
  \BibitemOpen
  \bibfield  {author} {\bibinfo {author} {\bibfnamefont {J.~P.}\ \bibnamefont {Covey}}, \bibinfo {author} {\bibfnamefont {H.}~\bibnamefont {Weinfurter}},\ and\ \bibinfo {author} {\bibfnamefont {H.}~\bibnamefont {Bernien}},\ }\bibfield  {title} {\bibinfo {title} {Quantum networks with neutral atom processing nodes},\ }\bibfield  {journal} {\bibinfo  {journal} {npj Quantum Information}\ }\textbf {\bibinfo {volume} {9}},\ \href {https://doi.org/10.1038/s41534-023-00759-9} {10.1038/s41534-023-00759-9} (\bibinfo {year} {2023})\BibitemShut {NoStop}%
\bibitem [{\citenamefont {Ward}\ and\ \citenamefont {Keller}(2022)}]{Ward-2022a}%
  \BibitemOpen
  \bibfield  {author} {\bibinfo {author} {\bibfnamefont {T.}~\bibnamefont {Ward}}\ and\ \bibinfo {author} {\bibfnamefont {M.}~\bibnamefont {Keller}},\ }\bibfield  {title} {\bibinfo {title} {Generation of time-bin-encoded photons in an ion-cavity system},\ }\href {https://doi.org/10.1088/1367-2630/aca9ee} {\bibfield  {journal} {\bibinfo  {journal} {New Journal of Physics}\ }\textbf {\bibinfo {volume} {24}},\ \bibinfo {pages} {123028} (\bibinfo {year} {2022})}\BibitemShut {NoStop}%
\bibitem [{\citenamefont {Krutyanskiy}\ \emph {et~al.}(2023)\citenamefont {Krutyanskiy}, \citenamefont {Galli}, \citenamefont {Krcmarsky}, \citenamefont {Baier}, \citenamefont {Fioretto}, \citenamefont {Pu}, \citenamefont {Mazloom}, \citenamefont {Sekatski}, \citenamefont {Canteri}, \citenamefont {Teller}, \citenamefont {Schupp}, \citenamefont {Bate}, \citenamefont {Meraner}, \citenamefont {Sangouard}, \citenamefont {Lanyon},\ and\ \citenamefont {Northup}}]{Krutyanskiy-2023a}%
  \BibitemOpen
  \bibfield  {author} {\bibinfo {author} {\bibfnamefont {V.}~\bibnamefont {Krutyanskiy}}, \bibinfo {author} {\bibfnamefont {M.}~\bibnamefont {Galli}}, \bibinfo {author} {\bibfnamefont {V.}~\bibnamefont {Krcmarsky}}, \bibinfo {author} {\bibfnamefont {S.}~\bibnamefont {Baier}}, \bibinfo {author} {\bibfnamefont {D.~A.}\ \bibnamefont {Fioretto}}, \bibinfo {author} {\bibfnamefont {Y.}~\bibnamefont {Pu}}, \bibinfo {author} {\bibfnamefont {A.}~\bibnamefont {Mazloom}}, \bibinfo {author} {\bibfnamefont {P.}~\bibnamefont {Sekatski}}, \bibinfo {author} {\bibfnamefont {M.}~\bibnamefont {Canteri}}, \bibinfo {author} {\bibfnamefont {M.}~\bibnamefont {Teller}}, \bibinfo {author} {\bibfnamefont {J.}~\bibnamefont {Schupp}}, \bibinfo {author} {\bibfnamefont {J.}~\bibnamefont {Bate}}, \bibinfo {author} {\bibfnamefont {M.}~\bibnamefont {Meraner}}, \bibinfo {author} {\bibfnamefont {N.}~\bibnamefont {Sangouard}}, \bibinfo {author} {\bibfnamefont {B.~P.}\ \bibnamefont {Lanyon}},\ and\ \bibinfo {author} {\bibfnamefont {T.~E.}\
  \bibnamefont {Northup}},\ }\bibfield  {title} {\bibinfo {title} {Entanglement of trapped-ion qubits separated by 230 meters},\ }\href {https://doi.org/10.1103/PhysRevLett.130.050803} {\bibfield  {journal} {\bibinfo  {journal} {Phys. Rev. Lett.}\ }\textbf {\bibinfo {volume} {130}},\ \bibinfo {pages} {050803} (\bibinfo {year} {2023})}\BibitemShut {NoStop}%
\bibitem [{\citenamefont {Jayakumar}\ \emph {et~al.}(2014)\citenamefont {Jayakumar}, \citenamefont {Predojević}, \citenamefont {Kauten}, \citenamefont {Huber}, \citenamefont {Solomon},\ and\ \citenamefont {Weihs}}]{Jayakumar-2014a}%
  \BibitemOpen
  \bibfield  {author} {\bibinfo {author} {\bibfnamefont {H.}~\bibnamefont {Jayakumar}}, \bibinfo {author} {\bibfnamefont {A.}~\bibnamefont {Predojević}}, \bibinfo {author} {\bibfnamefont {T.}~\bibnamefont {Kauten}}, \bibinfo {author} {\bibfnamefont {T.}~\bibnamefont {Huber}}, \bibinfo {author} {\bibfnamefont {G.~S.}\ \bibnamefont {Solomon}},\ and\ \bibinfo {author} {\bibfnamefont {G.}~\bibnamefont {Weihs}},\ }\bibfield  {title} {\bibinfo {title} {Time-bin entangled photons from a quantum dot},\ }\bibfield  {journal} {\bibinfo  {journal} {Nature Communications}\ }\textbf {\bibinfo {volume} {5}},\ \href {https://doi.org/10.1038/ncomms5251} {10.1038/ncomms5251} (\bibinfo {year} {2014})\BibitemShut {NoStop}%
\bibitem [{\citenamefont {Senellart}\ \emph {et~al.}(2017)\citenamefont {Senellart}, \citenamefont {Solomon},\ and\ \citenamefont {White}}]{Senellart-2017a}%
  \BibitemOpen
  \bibfield  {author} {\bibinfo {author} {\bibfnamefont {P.}~\bibnamefont {Senellart}}, \bibinfo {author} {\bibfnamefont {G.}~\bibnamefont {Solomon}},\ and\ \bibinfo {author} {\bibfnamefont {A.}~\bibnamefont {White}},\ }\bibfield  {title} {\bibinfo {title} {High-performance semiconductor quantum-dot single-photon sources},\ }\href {https://doi.org/10.1038/nnano.2017.218} {\bibfield  {journal} {\bibinfo  {journal} {Nature Nanotechnology}\ }\textbf {\bibinfo {volume} {12}},\ \bibinfo {pages} {1026–1039} (\bibinfo {year} {2017})}\BibitemShut {NoStop}%
\bibitem [{\citenamefont {Anand}\ \emph {et~al.}(2024)\citenamefont {Anand}, \citenamefont {Bradley}, \citenamefont {White}, \citenamefont {Ramesh}, \citenamefont {Singh},\ and\ \citenamefont {Bernien}}]{Anand-2024a}%
  \BibitemOpen
  \bibfield  {author} {\bibinfo {author} {\bibfnamefont {S.}~\bibnamefont {Anand}}, \bibinfo {author} {\bibfnamefont {C.~E.}\ \bibnamefont {Bradley}}, \bibinfo {author} {\bibfnamefont {R.}~\bibnamefont {White}}, \bibinfo {author} {\bibfnamefont {V.}~\bibnamefont {Ramesh}}, \bibinfo {author} {\bibfnamefont {K.}~\bibnamefont {Singh}},\ and\ \bibinfo {author} {\bibfnamefont {H.}~\bibnamefont {Bernien}},\ }\bibfield  {title} {\bibinfo {title} {A dual-species rydberg array},\ }\href@noop {} {\bibfield  {journal} {\bibinfo  {journal} {Nature Physics}\ }\textbf {\bibinfo {volume} {20}},\ \bibinfo {pages} {1} (\bibinfo {year} {2024})}\BibitemShut {NoStop}%
\bibitem [{\citenamefont {Bruzewicz}\ \emph {et~al.}(2019)\citenamefont {Bruzewicz}, \citenamefont {McConnell}, \citenamefont {Stuart}, \citenamefont {Sage},\ and\ \citenamefont {Chiaverini}}]{Bruzewicz-2019a}%
  \BibitemOpen
  \bibfield  {author} {\bibinfo {author} {\bibfnamefont {C.}~\bibnamefont {Bruzewicz}}, \bibinfo {author} {\bibfnamefont {R.}~\bibnamefont {McConnell}}, \bibinfo {author} {\bibfnamefont {J.}~\bibnamefont {Stuart}}, \bibinfo {author} {\bibfnamefont {J.}~\bibnamefont {Sage}},\ and\ \bibinfo {author} {\bibfnamefont {J.}~\bibnamefont {Chiaverini}},\ }\bibfield  {title} {\bibinfo {title} {Dual-species, multi-qubit logic primitives for ca+/sr+ trapped-ion crystals},\ }\href@noop {} {\bibfield  {journal} {\bibinfo  {journal} {npj Quantum Information}\ }\textbf {\bibinfo {volume} {5}},\ \bibinfo {pages} {102} (\bibinfo {year} {2019})}\BibitemShut {NoStop}%
\bibitem [{\citenamefont {Bradley}\ \emph {et~al.}(2019)\citenamefont {Bradley}, \citenamefont {Randall}, \citenamefont {Abobeih}, \citenamefont {Berrevoets}, \citenamefont {Degen}, \citenamefont {Bakker}, \citenamefont {Markham}, \citenamefont {Twitchen},\ and\ \citenamefont {Taminiau}}]{Bradley-2019a}%
  \BibitemOpen
  \bibfield  {author} {\bibinfo {author} {\bibfnamefont {C.~E.}\ \bibnamefont {Bradley}}, \bibinfo {author} {\bibfnamefont {J.}~\bibnamefont {Randall}}, \bibinfo {author} {\bibfnamefont {M.~H.}\ \bibnamefont {Abobeih}}, \bibinfo {author} {\bibfnamefont {R.~C.}\ \bibnamefont {Berrevoets}}, \bibinfo {author} {\bibfnamefont {M.~J.}\ \bibnamefont {Degen}}, \bibinfo {author} {\bibfnamefont {M.~A.}\ \bibnamefont {Bakker}}, \bibinfo {author} {\bibfnamefont {M.}~\bibnamefont {Markham}}, \bibinfo {author} {\bibfnamefont {D.~J.}\ \bibnamefont {Twitchen}},\ and\ \bibinfo {author} {\bibfnamefont {T.~H.}\ \bibnamefont {Taminiau}},\ }\bibfield  {title} {\bibinfo {title} {A ten-qubit solid-state spin register with quantum memory up to one minute},\ }\href@noop {} {\bibfield  {journal} {\bibinfo  {journal} {Physical Review X}\ }\textbf {\bibinfo {volume} {9}},\ \bibinfo {pages} {031045} (\bibinfo {year} {2019})}\BibitemShut {NoStop}%
\bibitem [{\citenamefont {Varnava}\ \emph {et~al.}(2006)\citenamefont {Varnava}, \citenamefont {Browne},\ and\ \citenamefont {Rudolph}}]{Varnava-2006a}%
  \BibitemOpen
  \bibfield  {author} {\bibinfo {author} {\bibfnamefont {M.}~\bibnamefont {Varnava}}, \bibinfo {author} {\bibfnamefont {D.~E.}\ \bibnamefont {Browne}},\ and\ \bibinfo {author} {\bibfnamefont {T.}~\bibnamefont {Rudolph}},\ }\bibfield  {title} {\bibinfo {title} {Loss tolerance in one-way quantum computation via counterfactual error correction},\ }\bibfield  {journal} {\bibinfo  {journal} {Physical Review Letters}\ }\textbf {\bibinfo {volume} {97}},\ \href {https://doi.org/10.1103/physrevlett.97.120501} {10.1103/physrevlett.97.120501} (\bibinfo {year} {2006})\BibitemShut {NoStop}%
\bibitem [{\citenamefont {Morimae}\ and\ \citenamefont {Fujii}(2012)}]{Morimae-2012a}%
  \BibitemOpen
  \bibfield  {author} {\bibinfo {author} {\bibfnamefont {T.}~\bibnamefont {Morimae}}\ and\ \bibinfo {author} {\bibfnamefont {K.}~\bibnamefont {Fujii}},\ }\bibfield  {title} {\bibinfo {title} {Blind topological measurement-based quantum computation},\ }\bibfield  {journal} {\bibinfo  {journal} {Nature Communications}\ }\textbf {\bibinfo {volume} {3}},\ \href {https://doi.org/10.1038/ncomms2043} {10.1038/ncomms2043} (\bibinfo {year} {2012})\BibitemShut {NoStop}%
\bibitem [{\citenamefont {Morley-Short}\ \emph {et~al.}(2017)\citenamefont {Morley-Short}, \citenamefont {Bartolucci}, \citenamefont {Gimeno-Segovia}, \citenamefont {Shadbolt}, \citenamefont {Cable},\ and\ \citenamefont {Rudolph}}]{Morley-Short-2017a}%
  \BibitemOpen
  \bibfield  {author} {\bibinfo {author} {\bibfnamefont {S.}~\bibnamefont {Morley-Short}}, \bibinfo {author} {\bibfnamefont {S.}~\bibnamefont {Bartolucci}}, \bibinfo {author} {\bibfnamefont {M.}~\bibnamefont {Gimeno-Segovia}}, \bibinfo {author} {\bibfnamefont {P.}~\bibnamefont {Shadbolt}}, \bibinfo {author} {\bibfnamefont {H.}~\bibnamefont {Cable}},\ and\ \bibinfo {author} {\bibfnamefont {T.}~\bibnamefont {Rudolph}},\ }\bibfield  {title} {\bibinfo {title} {Physical-depth architectural requirements for generating universal photonic cluster states},\ }\href {https://doi.org/10.1088/2058-9565/aa913b} {\bibfield  {journal} {\bibinfo  {journal} {Quantum Science and Technology}\ }\textbf {\bibinfo {volume} {3}},\ \bibinfo {pages} {015005} (\bibinfo {year} {2017})}\BibitemShut {NoStop}%
\bibitem [{\citenamefont {Pant}\ \emph {et~al.}(2019)\citenamefont {Pant}, \citenamefont {Towsley}, \citenamefont {Englund},\ and\ \citenamefont {Guha}}]{Pant-2019a}%
  \BibitemOpen
  \bibfield  {author} {\bibinfo {author} {\bibfnamefont {M.}~\bibnamefont {Pant}}, \bibinfo {author} {\bibfnamefont {D.}~\bibnamefont {Towsley}}, \bibinfo {author} {\bibfnamefont {D.}~\bibnamefont {Englund}},\ and\ \bibinfo {author} {\bibfnamefont {S.}~\bibnamefont {Guha}},\ }\bibfield  {title} {\bibinfo {title} {Percolation thresholds for photonic quantum computing},\ }\href@noop {} {\bibfield  {journal} {\bibinfo  {journal} {Nature communications}\ }\textbf {\bibinfo {volume} {10}},\ \bibinfo {pages} {1070} (\bibinfo {year} {2019})}\BibitemShut {NoStop}%
\bibitem [{\citenamefont {Bartolucci}\ \emph {et~al.}(2023)\citenamefont {Bartolucci}, \citenamefont {Birchall}, \citenamefont {Bombín}, \citenamefont {Cable}, \citenamefont {Dawson}, \citenamefont {Gimeno-Segovia}, \citenamefont {Johnston}, \citenamefont {Kieling}, \citenamefont {Nickerson}, \citenamefont {Pant}, \citenamefont {Pastawski}, \citenamefont {Rudolph},\ and\ \citenamefont {Sparrow}}]{Bartolucci-2023a}%
  \BibitemOpen
  \bibfield  {author} {\bibinfo {author} {\bibfnamefont {S.}~\bibnamefont {Bartolucci}}, \bibinfo {author} {\bibfnamefont {P.}~\bibnamefont {Birchall}}, \bibinfo {author} {\bibfnamefont {H.}~\bibnamefont {Bombín}}, \bibinfo {author} {\bibfnamefont {H.}~\bibnamefont {Cable}}, \bibinfo {author} {\bibfnamefont {C.}~\bibnamefont {Dawson}}, \bibinfo {author} {\bibfnamefont {M.}~\bibnamefont {Gimeno-Segovia}}, \bibinfo {author} {\bibfnamefont {E.}~\bibnamefont {Johnston}}, \bibinfo {author} {\bibfnamefont {K.}~\bibnamefont {Kieling}}, \bibinfo {author} {\bibfnamefont {N.}~\bibnamefont {Nickerson}}, \bibinfo {author} {\bibfnamefont {M.}~\bibnamefont {Pant}}, \bibinfo {author} {\bibfnamefont {F.}~\bibnamefont {Pastawski}}, \bibinfo {author} {\bibfnamefont {T.}~\bibnamefont {Rudolph}},\ and\ \bibinfo {author} {\bibfnamefont {C.}~\bibnamefont {Sparrow}},\ }\bibfield  {title} {\bibinfo {title} {Fusion-based quantum computation},\ }\bibfield  {journal} {\bibinfo  {journal} {Nature Communications}\ }\textbf {\bibinfo
  {volume} {14}},\ \href {https://doi.org/10.1038/s41467-023-36493-1} {10.1038/s41467-023-36493-1} (\bibinfo {year} {2023})\BibitemShut {NoStop}%
\bibitem [{\citenamefont {L\"{o}bl}\ \emph {et~al.}(2024)\citenamefont {L\"{o}bl}, \citenamefont {Paesani},\ and\ \citenamefont {Sørensen}}]{Lobl-2024a}%
  \BibitemOpen
  \bibfield  {author} {\bibinfo {author} {\bibfnamefont {M.~C.}\ \bibnamefont {L\"{o}bl}}, \bibinfo {author} {\bibfnamefont {S.}~\bibnamefont {Paesani}},\ and\ \bibinfo {author} {\bibfnamefont {A.~S.}\ \bibnamefont {Sørensen}},\ }\bibfield  {title} {\bibinfo {title} {Loss-tolerant architecture for quantum computing with quantum emitters},\ }\href {https://doi.org/10.22331/q-2024-03-28-1302} {\bibfield  {journal} {\bibinfo  {journal} {Quantum}\ }\textbf {\bibinfo {volume} {8}},\ \bibinfo {pages} {1302} (\bibinfo {year} {2024})}\BibitemShut {NoStop}%
\bibitem [{\citenamefont {Schneeloch}\ \emph {et~al.}(2019)\citenamefont {Schneeloch}, \citenamefont {Knarr}, \citenamefont {Bogorin}, \citenamefont {Levangie}, \citenamefont {Tison}, \citenamefont {Frank}, \citenamefont {Howland}, \citenamefont {Fanto},\ and\ \citenamefont {Alsing}}]{Schneeloch-2019a}%
  \BibitemOpen
  \bibfield  {author} {\bibinfo {author} {\bibfnamefont {J.}~\bibnamefont {Schneeloch}}, \bibinfo {author} {\bibfnamefont {S.~H.}\ \bibnamefont {Knarr}}, \bibinfo {author} {\bibfnamefont {D.~F.}\ \bibnamefont {Bogorin}}, \bibinfo {author} {\bibfnamefont {M.~L.}\ \bibnamefont {Levangie}}, \bibinfo {author} {\bibfnamefont {C.~C.}\ \bibnamefont {Tison}}, \bibinfo {author} {\bibfnamefont {R.}~\bibnamefont {Frank}}, \bibinfo {author} {\bibfnamefont {G.~A.}\ \bibnamefont {Howland}}, \bibinfo {author} {\bibfnamefont {M.~L.}\ \bibnamefont {Fanto}},\ and\ \bibinfo {author} {\bibfnamefont {P.~M.}\ \bibnamefont {Alsing}},\ }\bibfield  {title} {\bibinfo {title} {Introduction to the absolute brightness and number statistics in spontaneous parametric down-conversion},\ }\href {https://doi.org/10.1088/2040-8986/ab05a8} {\bibfield  {journal} {\bibinfo  {journal} {Journal of Optics}\ }\textbf {\bibinfo {volume} {21}},\ \bibinfo {pages} {043501} (\bibinfo {year} {2019})}\BibitemShut {NoStop}%
\bibitem [{\citenamefont {Zhang}\ \emph {et~al.}(2021)\citenamefont {Zhang}, \citenamefont {Huang}, \citenamefont {Liu}, \citenamefont {Li},\ and\ \citenamefont {Guo}}]{Zhang-2021a}%
  \BibitemOpen
  \bibfield  {author} {\bibinfo {author} {\bibfnamefont {C.}~\bibnamefont {Zhang}}, \bibinfo {author} {\bibfnamefont {Y.-F.}\ \bibnamefont {Huang}}, \bibinfo {author} {\bibfnamefont {B.-H.}\ \bibnamefont {Liu}}, \bibinfo {author} {\bibfnamefont {C.-F.}\ \bibnamefont {Li}},\ and\ \bibinfo {author} {\bibfnamefont {G.-C.}\ \bibnamefont {Guo}},\ }\bibfield  {title} {\bibinfo {title} {Spontaneous parametric down-conversion sources for multiphoton experiments},\ }\href {https://doi.org/https://doi.org/10.1002/qute.202000132} {\bibfield  {journal} {\bibinfo  {journal} {Advanced Quantum Technologies}\ }\textbf {\bibinfo {volume} {4}},\ \bibinfo {pages} {2000132} (\bibinfo {year} {2021})}\BibitemShut {NoStop}%
\bibitem [{\citenamefont {Pan}\ \emph {et~al.}(1998)\citenamefont {Pan}, \citenamefont {Bouwmeester}, \citenamefont {Weinfurter},\ and\ \citenamefont {Zeilinger}}]{Pan-1998a}%
  \BibitemOpen
  \bibfield  {author} {\bibinfo {author} {\bibfnamefont {J.-W.}\ \bibnamefont {Pan}}, \bibinfo {author} {\bibfnamefont {D.}~\bibnamefont {Bouwmeester}}, \bibinfo {author} {\bibfnamefont {H.}~\bibnamefont {Weinfurter}},\ and\ \bibinfo {author} {\bibfnamefont {A.}~\bibnamefont {Zeilinger}},\ }\bibfield  {title} {\bibinfo {title} {Experimental entanglement swapping: entangling photons that never interacted},\ }\href@noop {} {\bibfield  {journal} {\bibinfo  {journal} {Physical review letters}\ }\textbf {\bibinfo {volume} {80}},\ \bibinfo {pages} {3891} (\bibinfo {year} {1998})}\BibitemShut {NoStop}%
\bibitem [{\citenamefont {Houshmand}\ \emph {et~al.}(2018)\citenamefont {Houshmand}, \citenamefont {Houshmand},\ and\ \citenamefont {Fitzsimons}}]{Houshmand-2018a}%
  \BibitemOpen
  \bibfield  {author} {\bibinfo {author} {\bibfnamefont {M.}~\bibnamefont {Houshmand}}, \bibinfo {author} {\bibfnamefont {M.}~\bibnamefont {Houshmand}},\ and\ \bibinfo {author} {\bibfnamefont {J.~F.}\ \bibnamefont {Fitzsimons}},\ }\bibfield  {title} {\bibinfo {title} {Minimal qubit resources for the realization of measurement-based quantum computation},\ }\href {https://doi.org/10.1103/PhysRevA.98.012318} {\bibfield  {journal} {\bibinfo  {journal} {Phys. Rev. A}\ }\textbf {\bibinfo {volume} {98}},\ \bibinfo {pages} {012318} (\bibinfo {year} {2018})}\BibitemShut {NoStop}%
\bibitem [{\citenamefont {Li}\ \emph {et~al.}(2022)\citenamefont {Li}, \citenamefont {Economou},\ and\ \citenamefont {Barnes}}]{Li-2022a}%
  \BibitemOpen
  \bibfield  {author} {\bibinfo {author} {\bibfnamefont {B.}~\bibnamefont {Li}}, \bibinfo {author} {\bibfnamefont {S.~E.}\ \bibnamefont {Economou}},\ and\ \bibinfo {author} {\bibfnamefont {E.}~\bibnamefont {Barnes}},\ }\bibfield  {title} {\bibinfo {title} {Photonic resource state generation from a minimal number of quantum emitters},\ }\bibfield  {journal} {\bibinfo  {journal} {npj Quantum Information}\ }\textbf {\bibinfo {volume} {8}},\ \href {https://doi.org/10.1038/s41534-022-00522-6} {10.1038/s41534-022-00522-6} (\bibinfo {year} {2022})\BibitemShut {NoStop}%
\bibitem [{\citenamefont {Raussendorf}\ \emph {et~al.}(2003)\citenamefont {Raussendorf}, \citenamefont {Browne},\ and\ \citenamefont {Briegel}}]{Raussendorf-2003a}%
  \BibitemOpen
  \bibfield  {author} {\bibinfo {author} {\bibfnamefont {R.}~\bibnamefont {Raussendorf}}, \bibinfo {author} {\bibfnamefont {D.~E.}\ \bibnamefont {Browne}},\ and\ \bibinfo {author} {\bibfnamefont {H.~J.}\ \bibnamefont {Briegel}},\ }\bibfield  {title} {\bibinfo {title} {Measurement-based quantum computation on cluster states},\ }\href {https://doi.org/10.1103/PhysRevA.68.022312} {\bibfield  {journal} {\bibinfo  {journal} {Phys. Rev. A}\ }\textbf {\bibinfo {volume} {68}},\ \bibinfo {pages} {022312} (\bibinfo {year} {2003})}\BibitemShut {NoStop}%
\bibitem [{\citenamefont {Browne}\ \emph {et~al.}(2007)\citenamefont {Browne}, \citenamefont {Kashefi}, \citenamefont {Mhalla},\ and\ \citenamefont {Perdrix}}]{Browne-2007a}%
  \BibitemOpen
  \bibfield  {author} {\bibinfo {author} {\bibfnamefont {D.~E.}\ \bibnamefont {Browne}}, \bibinfo {author} {\bibfnamefont {E.}~\bibnamefont {Kashefi}}, \bibinfo {author} {\bibfnamefont {M.}~\bibnamefont {Mhalla}},\ and\ \bibinfo {author} {\bibfnamefont {S.}~\bibnamefont {Perdrix}},\ }\bibfield  {title} {\bibinfo {title} {Generalized flow and determinism in measurement-based quantum computation},\ }\href {https://doi.org/10.1088/1367-2630/9/8/250} {\bibfield  {journal} {\bibinfo  {journal} {New Journal of Physics}\ }\textbf {\bibinfo {volume} {9}},\ \bibinfo {pages} {250–250} (\bibinfo {year} {2007})}\BibitemShut {NoStop}%
\bibitem [{\citenamefont {Bluvstein}\ \emph {et~al.}(2022)\citenamefont {Bluvstein}, \citenamefont {Levine}, \citenamefont {Semeghini}, \citenamefont {Wang}, \citenamefont {Ebadi}, \citenamefont {Kalinowski}, \citenamefont {Keesling}, \citenamefont {Maskara}, \citenamefont {Pichler}, \citenamefont {Greiner}, \citenamefont {Vuletić},\ and\ \citenamefont {Lukin}}]{Bluvstein-2022a}%
  \BibitemOpen
  \bibfield  {author} {\bibinfo {author} {\bibfnamefont {D.}~\bibnamefont {Bluvstein}}, \bibinfo {author} {\bibfnamefont {H.}~\bibnamefont {Levine}}, \bibinfo {author} {\bibfnamefont {G.}~\bibnamefont {Semeghini}}, \bibinfo {author} {\bibfnamefont {T.~T.}\ \bibnamefont {Wang}}, \bibinfo {author} {\bibfnamefont {S.}~\bibnamefont {Ebadi}}, \bibinfo {author} {\bibfnamefont {M.}~\bibnamefont {Kalinowski}}, \bibinfo {author} {\bibfnamefont {A.}~\bibnamefont {Keesling}}, \bibinfo {author} {\bibfnamefont {N.}~\bibnamefont {Maskara}}, \bibinfo {author} {\bibfnamefont {H.}~\bibnamefont {Pichler}}, \bibinfo {author} {\bibfnamefont {M.}~\bibnamefont {Greiner}}, \bibinfo {author} {\bibfnamefont {V.}~\bibnamefont {Vuletić}},\ and\ \bibinfo {author} {\bibfnamefont {M.~D.}\ \bibnamefont {Lukin}},\ }\bibfield  {title} {\bibinfo {title} {A quantum processor based on coherent transport of entangled atom arrays},\ }\href {https://doi.org/10.1038/s41586-022-04592-6} {\bibfield  {journal} {\bibinfo  {journal} {Nature}\ }\textbf
  {\bibinfo {volume} {604}},\ \bibinfo {pages} {451–456} (\bibinfo {year} {2022})}\BibitemShut {NoStop}%
\bibitem [{\citenamefont {Wang}\ \emph {et~al.}(2021)\citenamefont {Wang}, \citenamefont {Luan}, \citenamefont {Qiao}, \citenamefont {Um}, \citenamefont {Zhang}, \citenamefont {Wang}, \citenamefont {Yuan}, \citenamefont {Gu}, \citenamefont {Zhang},\ and\ \citenamefont {Kim}}]{Wang-2021a}%
  \BibitemOpen
  \bibfield  {author} {\bibinfo {author} {\bibfnamefont {P.}~\bibnamefont {Wang}}, \bibinfo {author} {\bibfnamefont {C.-Y.}\ \bibnamefont {Luan}}, \bibinfo {author} {\bibfnamefont {M.}~\bibnamefont {Qiao}}, \bibinfo {author} {\bibfnamefont {M.}~\bibnamefont {Um}}, \bibinfo {author} {\bibfnamefont {J.}~\bibnamefont {Zhang}}, \bibinfo {author} {\bibfnamefont {Y.}~\bibnamefont {Wang}}, \bibinfo {author} {\bibfnamefont {X.}~\bibnamefont {Yuan}}, \bibinfo {author} {\bibfnamefont {M.}~\bibnamefont {Gu}}, \bibinfo {author} {\bibfnamefont {J.}~\bibnamefont {Zhang}},\ and\ \bibinfo {author} {\bibfnamefont {K.}~\bibnamefont {Kim}},\ }\bibfield  {title} {\bibinfo {title} {Single ion qubit with estimated coherence time exceeding one hour},\ }\href@noop {} {\bibfield  {journal} {\bibinfo  {journal} {Nature communications}\ }\textbf {\bibinfo {volume} {12}},\ \bibinfo {pages} {1} (\bibinfo {year} {2021})}\BibitemShut {NoStop}%
\bibitem [{\citenamefont {Evered}\ \emph {et~al.}(2023)\citenamefont {Evered}, \citenamefont {Bluvstein}, \citenamefont {Kalinowski}, \citenamefont {Ebadi}, \citenamefont {Manovitz}, \citenamefont {Zhou}, \citenamefont {Li}, \citenamefont {Geim}, \citenamefont {Wang}, \citenamefont {Maskara}, \citenamefont {Levine}, \citenamefont {Semeghini}, \citenamefont {Greiner}, \citenamefont {Vuletić},\ and\ \citenamefont {Lukin}}]{Evered-2023a}%
  \BibitemOpen
  \bibfield  {author} {\bibinfo {author} {\bibfnamefont {S.~J.}\ \bibnamefont {Evered}}, \bibinfo {author} {\bibfnamefont {D.}~\bibnamefont {Bluvstein}}, \bibinfo {author} {\bibfnamefont {M.}~\bibnamefont {Kalinowski}}, \bibinfo {author} {\bibfnamefont {S.}~\bibnamefont {Ebadi}}, \bibinfo {author} {\bibfnamefont {T.}~\bibnamefont {Manovitz}}, \bibinfo {author} {\bibfnamefont {H.}~\bibnamefont {Zhou}}, \bibinfo {author} {\bibfnamefont {S.~H.}\ \bibnamefont {Li}}, \bibinfo {author} {\bibfnamefont {A.~A.}\ \bibnamefont {Geim}}, \bibinfo {author} {\bibfnamefont {T.~T.}\ \bibnamefont {Wang}}, \bibinfo {author} {\bibfnamefont {N.}~\bibnamefont {Maskara}}, \bibinfo {author} {\bibfnamefont {H.}~\bibnamefont {Levine}}, \bibinfo {author} {\bibfnamefont {G.}~\bibnamefont {Semeghini}}, \bibinfo {author} {\bibfnamefont {M.}~\bibnamefont {Greiner}}, \bibinfo {author} {\bibfnamefont {V.}~\bibnamefont {Vuletić}},\ and\ \bibinfo {author} {\bibfnamefont {M.~D.}\ \bibnamefont {Lukin}},\ }\bibfield  {title} {\bibinfo {title}
  {High-fidelity parallel entangling gates on a neutral-atom quantum computer},\ }\href {https://doi.org/10.1038/s41586-023-06481-y} {\bibfield  {journal} {\bibinfo  {journal} {Nature}\ }\textbf {\bibinfo {volume} {622}},\ \bibinfo {pages} {268–272} (\bibinfo {year} {2023})}\BibitemShut {NoStop}%
\bibitem [{\citenamefont {Srinivas}\ \emph {et~al.}(2021)\citenamefont {Srinivas}, \citenamefont {Burd}, \citenamefont {Knaack}, \citenamefont {Sutherland}, \citenamefont {Kwiatkowski}, \citenamefont {Glancy}, \citenamefont {Knill}, \citenamefont {Wineland}, \citenamefont {Leibfried}, \citenamefont {Wilson}, \citenamefont {Allcock},\ and\ \citenamefont {Slichter}}]{Srinivas-2021a}%
  \BibitemOpen
  \bibfield  {author} {\bibinfo {author} {\bibfnamefont {R.}~\bibnamefont {Srinivas}}, \bibinfo {author} {\bibfnamefont {S.~C.}\ \bibnamefont {Burd}}, \bibinfo {author} {\bibfnamefont {H.~M.}\ \bibnamefont {Knaack}}, \bibinfo {author} {\bibfnamefont {R.~T.}\ \bibnamefont {Sutherland}}, \bibinfo {author} {\bibfnamefont {A.}~\bibnamefont {Kwiatkowski}}, \bibinfo {author} {\bibfnamefont {S.}~\bibnamefont {Glancy}}, \bibinfo {author} {\bibfnamefont {E.}~\bibnamefont {Knill}}, \bibinfo {author} {\bibfnamefont {D.~J.}\ \bibnamefont {Wineland}}, \bibinfo {author} {\bibfnamefont {D.}~\bibnamefont {Leibfried}}, \bibinfo {author} {\bibfnamefont {A.~C.}\ \bibnamefont {Wilson}}, \bibinfo {author} {\bibfnamefont {D.~T.~C.}\ \bibnamefont {Allcock}},\ and\ \bibinfo {author} {\bibfnamefont {D.~H.}\ \bibnamefont {Slichter}},\ }\bibfield  {title} {\bibinfo {title} {High-fidelity laser-free universal control of trapped ion qubits},\ }\href {https://doi.org/10.1038/s41586-021-03809-4} {\bibfield  {journal} {\bibinfo  {journal}
  {Nature}\ }\textbf {\bibinfo {volume} {597}},\ \bibinfo {pages} {209–213} (\bibinfo {year} {2021})}\BibitemShut {NoStop}%
\bibitem [{\citenamefont {Clark}\ \emph {et~al.}(2021)\citenamefont {Clark}, \citenamefont {Tinkey}, \citenamefont {Sawyer}, \citenamefont {Meier}, \citenamefont {Burkhardt}, \citenamefont {Seck}, \citenamefont {Shappert}, \citenamefont {Guise}, \citenamefont {Volin}, \citenamefont {Fallek}, \citenamefont {Hayden}, \citenamefont {Rellergert},\ and\ \citenamefont {Brown}}]{Clark-2021a}%
  \BibitemOpen
  \bibfield  {author} {\bibinfo {author} {\bibfnamefont {C.~R.}\ \bibnamefont {Clark}}, \bibinfo {author} {\bibfnamefont {H.~N.}\ \bibnamefont {Tinkey}}, \bibinfo {author} {\bibfnamefont {B.~C.}\ \bibnamefont {Sawyer}}, \bibinfo {author} {\bibfnamefont {A.~M.}\ \bibnamefont {Meier}}, \bibinfo {author} {\bibfnamefont {K.~A.}\ \bibnamefont {Burkhardt}}, \bibinfo {author} {\bibfnamefont {C.~M.}\ \bibnamefont {Seck}}, \bibinfo {author} {\bibfnamefont {C.~M.}\ \bibnamefont {Shappert}}, \bibinfo {author} {\bibfnamefont {N.~D.}\ \bibnamefont {Guise}}, \bibinfo {author} {\bibfnamefont {C.~E.}\ \bibnamefont {Volin}}, \bibinfo {author} {\bibfnamefont {S.~D.}\ \bibnamefont {Fallek}}, \bibinfo {author} {\bibfnamefont {H.~T.}\ \bibnamefont {Hayden}}, \bibinfo {author} {\bibfnamefont {W.~G.}\ \bibnamefont {Rellergert}},\ and\ \bibinfo {author} {\bibfnamefont {K.~R.}\ \bibnamefont {Brown}},\ }\bibfield  {title} {\bibinfo {title} {High-fidelity bell-state preparation with $^{40}\text{CA}^{+}$ optical qubits},\ }\bibfield
  {journal} {\bibinfo  {journal} {Physical Review Letters}\ }\textbf {\bibinfo {volume} {127}},\ \href {https://doi.org/10.1103/physrevlett.127.130505} {10.1103/physrevlett.127.130505} (\bibinfo {year} {2021})\BibitemShut {NoStop}%
\bibitem [{\citenamefont {Gaebler}\ \emph {et~al.}(2021)\citenamefont {Gaebler}, \citenamefont {Baldwin}, \citenamefont {Moses}, \citenamefont {Dreiling}, \citenamefont {Figgatt}, \citenamefont {Foss-Feig}, \citenamefont {Hayes},\ and\ \citenamefont {Pino}}]{Gaebler-2021a}%
  \BibitemOpen
  \bibfield  {author} {\bibinfo {author} {\bibfnamefont {J.~P.}\ \bibnamefont {Gaebler}}, \bibinfo {author} {\bibfnamefont {C.~H.}\ \bibnamefont {Baldwin}}, \bibinfo {author} {\bibfnamefont {S.~A.}\ \bibnamefont {Moses}}, \bibinfo {author} {\bibfnamefont {J.~M.}\ \bibnamefont {Dreiling}}, \bibinfo {author} {\bibfnamefont {C.}~\bibnamefont {Figgatt}}, \bibinfo {author} {\bibfnamefont {M.}~\bibnamefont {Foss-Feig}}, \bibinfo {author} {\bibfnamefont {D.}~\bibnamefont {Hayes}},\ and\ \bibinfo {author} {\bibfnamefont {J.~M.}\ \bibnamefont {Pino}},\ }\bibfield  {title} {\bibinfo {title} {Suppression of midcircuit measurement crosstalk errors with micromotion},\ }\bibfield  {journal} {\bibinfo  {journal} {Physical Review A}\ }\textbf {\bibinfo {volume} {104}},\ \href {https://doi.org/10.1103/physreva.104.062440} {10.1103/physreva.104.062440} (\bibinfo {year} {2021})\BibitemShut {NoStop}%
\bibitem [{\citenamefont {Norcia}\ \emph {et~al.}(2023)\citenamefont {Norcia}, \citenamefont {Cairncross}, \citenamefont {Barnes}, \citenamefont {Battaglino}, \citenamefont {Brown}, \citenamefont {Brown}, \citenamefont {Cassella}, \citenamefont {Chen}, \citenamefont {Coxe}, \citenamefont {Crow}, \citenamefont {Epstein}, \citenamefont {Griger}, \citenamefont {Jones}, \citenamefont {Kim}, \citenamefont {Kindem}, \citenamefont {King}, \citenamefont {Kondov}, \citenamefont {Kotru}, \citenamefont {Lauigan}, \citenamefont {Li}, \citenamefont {Lu}, \citenamefont {Megidish}, \citenamefont {Marjanovic}, \citenamefont {McDonald}, \citenamefont {Mittiga}, \citenamefont {Muniz}, \citenamefont {Narayanaswami}, \citenamefont {Nishiguchi}, \citenamefont {Notermans}, \citenamefont {Paule}, \citenamefont {Pawlak}, \citenamefont {Peng}, \citenamefont {Ryou}, \citenamefont {Smull}, \citenamefont {Stack}, \citenamefont {Stone}, \citenamefont {Sucich}, \citenamefont {Urbanek}, \citenamefont {van~de Veerdonk}, \citenamefont
  {Vendeiro}, \citenamefont {Wilkason}, \citenamefont {Wu}, \citenamefont {Xie}, \citenamefont {Zhang},\ and\ \citenamefont {Bloom}}]{Norcia-2023a}%
  \BibitemOpen
  \bibfield  {author} {\bibinfo {author} {\bibfnamefont {M.~A.}\ \bibnamefont {Norcia}}, \bibinfo {author} {\bibfnamefont {W.~B.}\ \bibnamefont {Cairncross}}, \bibinfo {author} {\bibfnamefont {K.}~\bibnamefont {Barnes}}, \bibinfo {author} {\bibfnamefont {P.}~\bibnamefont {Battaglino}}, \bibinfo {author} {\bibfnamefont {A.}~\bibnamefont {Brown}}, \bibinfo {author} {\bibfnamefont {M.~O.}\ \bibnamefont {Brown}}, \bibinfo {author} {\bibfnamefont {K.}~\bibnamefont {Cassella}}, \bibinfo {author} {\bibfnamefont {C.-A.}\ \bibnamefont {Chen}}, \bibinfo {author} {\bibfnamefont {R.}~\bibnamefont {Coxe}}, \bibinfo {author} {\bibfnamefont {D.}~\bibnamefont {Crow}}, \bibinfo {author} {\bibfnamefont {J.}~\bibnamefont {Epstein}}, \bibinfo {author} {\bibfnamefont {C.}~\bibnamefont {Griger}}, \bibinfo {author} {\bibfnamefont {A.~M.~W.}\ \bibnamefont {Jones}}, \bibinfo {author} {\bibfnamefont {H.}~\bibnamefont {Kim}}, \bibinfo {author} {\bibfnamefont {J.~M.}\ \bibnamefont {Kindem}}, \bibinfo {author} {\bibfnamefont
  {J.}~\bibnamefont {King}}, \bibinfo {author} {\bibfnamefont {S.~S.}\ \bibnamefont {Kondov}}, \bibinfo {author} {\bibfnamefont {K.}~\bibnamefont {Kotru}}, \bibinfo {author} {\bibfnamefont {J.}~\bibnamefont {Lauigan}}, \bibinfo {author} {\bibfnamefont {M.}~\bibnamefont {Li}}, \bibinfo {author} {\bibfnamefont {M.}~\bibnamefont {Lu}}, \bibinfo {author} {\bibfnamefont {E.}~\bibnamefont {Megidish}}, \bibinfo {author} {\bibfnamefont {J.}~\bibnamefont {Marjanovic}}, \bibinfo {author} {\bibfnamefont {M.}~\bibnamefont {McDonald}}, \bibinfo {author} {\bibfnamefont {T.}~\bibnamefont {Mittiga}}, \bibinfo {author} {\bibfnamefont {J.~A.}\ \bibnamefont {Muniz}}, \bibinfo {author} {\bibfnamefont {S.}~\bibnamefont {Narayanaswami}}, \bibinfo {author} {\bibfnamefont {C.}~\bibnamefont {Nishiguchi}}, \bibinfo {author} {\bibfnamefont {R.}~\bibnamefont {Notermans}}, \bibinfo {author} {\bibfnamefont {T.}~\bibnamefont {Paule}}, \bibinfo {author} {\bibfnamefont {K.~A.}\ \bibnamefont {Pawlak}}, \bibinfo {author} {\bibfnamefont
  {L.~S.}\ \bibnamefont {Peng}}, \bibinfo {author} {\bibfnamefont {A.}~\bibnamefont {Ryou}}, \bibinfo {author} {\bibfnamefont {A.}~\bibnamefont {Smull}}, \bibinfo {author} {\bibfnamefont {D.}~\bibnamefont {Stack}}, \bibinfo {author} {\bibfnamefont {M.}~\bibnamefont {Stone}}, \bibinfo {author} {\bibfnamefont {A.}~\bibnamefont {Sucich}}, \bibinfo {author} {\bibfnamefont {M.}~\bibnamefont {Urbanek}}, \bibinfo {author} {\bibfnamefont {R.~J.~M.}\ \bibnamefont {van~de Veerdonk}}, \bibinfo {author} {\bibfnamefont {Z.}~\bibnamefont {Vendeiro}}, \bibinfo {author} {\bibfnamefont {T.}~\bibnamefont {Wilkason}}, \bibinfo {author} {\bibfnamefont {T.-Y.}\ \bibnamefont {Wu}}, \bibinfo {author} {\bibfnamefont {X.}~\bibnamefont {Xie}}, \bibinfo {author} {\bibfnamefont {X.}~\bibnamefont {Zhang}},\ and\ \bibinfo {author} {\bibfnamefont {B.~J.}\ \bibnamefont {Bloom}},\ }\bibfield  {title} {\bibinfo {title} {Midcircuit qubit measurement and rearrangement in a $^{171}\mathrm{Yb}$ atomic array},\ }\href
  {https://doi.org/10.1103/PhysRevX.13.041034} {\bibfield  {journal} {\bibinfo  {journal} {Phys. Rev. X}\ }\textbf {\bibinfo {volume} {13}},\ \bibinfo {pages} {041034} (\bibinfo {year} {2023})}\BibitemShut {NoStop}%
\bibitem [{\citenamefont {Yao}(1982)}]{Yao-1982a}%
  \BibitemOpen
  \bibfield  {author} {\bibinfo {author} {\bibfnamefont {A.~C.}\ \bibnamefont {Yao}},\ }\bibfield  {title} {\bibinfo {title} {Protocols for secure computations},\ }in\ \href {https://doi.org/10.1109/SFCS.1982.38} {\emph {\bibinfo {booktitle} {23rd Annual Symposium on Foundations of Computer Science (sfcs 1982)}}}\ (\bibinfo {year} {1982})\ pp.\ \bibinfo {pages} {160--164}\BibitemShut {NoStop}%
\bibitem [{\citenamefont {Goldreich}\ \emph {et~al.}(1987)\citenamefont {Goldreich}, \citenamefont {Micali},\ and\ \citenamefont {Wigderson}}]{Goldreich-1987a}%
  \BibitemOpen
  \bibfield  {author} {\bibinfo {author} {\bibfnamefont {O.}~\bibnamefont {Goldreich}}, \bibinfo {author} {\bibfnamefont {S.}~\bibnamefont {Micali}},\ and\ \bibinfo {author} {\bibfnamefont {A.}~\bibnamefont {Wigderson}},\ }\bibfield  {title} {\bibinfo {title} {How to play any mental game},\ }in\ \href {https://doi.org/10.1145/28395.28420} {\emph {\bibinfo {booktitle} {Proceedings of the nineteenth annual ACM conference on Theory of computing - STOC ’87}}},\ \bibinfo {series and number} {STOC ’87}\ (\bibinfo  {publisher} {ACM Press},\ \bibinfo {year} {1987})\BibitemShut {NoStop}%
\bibitem [{\citenamefont {Zhao}\ \emph {et~al.}(2019)\citenamefont {Zhao}, \citenamefont {Zhao}, \citenamefont {Zhao}, \citenamefont {Chen}, \citenamefont {Gao}, \citenamefont {Li},\ and\ \citenamefont {Tan}}]{Zhao-2019a}%
  \BibitemOpen
  \bibfield  {author} {\bibinfo {author} {\bibfnamefont {C.}~\bibnamefont {Zhao}}, \bibinfo {author} {\bibfnamefont {S.}~\bibnamefont {Zhao}}, \bibinfo {author} {\bibfnamefont {M.}~\bibnamefont {Zhao}}, \bibinfo {author} {\bibfnamefont {Z.}~\bibnamefont {Chen}}, \bibinfo {author} {\bibfnamefont {C.-Z.}\ \bibnamefont {Gao}}, \bibinfo {author} {\bibfnamefont {H.}~\bibnamefont {Li}},\ and\ \bibinfo {author} {\bibfnamefont {Y.-a.}\ \bibnamefont {Tan}},\ }\bibfield  {title} {\bibinfo {title} {Secure multi-party computation: Theory, practice and applications},\ }\href {https://doi.org/10.1016/j.ins.2018.10.024} {\bibfield  {journal} {\bibinfo  {journal} {Information Sciences}\ }\textbf {\bibinfo {volume} {476}},\ \bibinfo {pages} {357–372} (\bibinfo {year} {2019})}\BibitemShut {NoStop}%
\bibitem [{\citenamefont {Beaver}(1992)}]{Beaver-1992a}%
  \BibitemOpen
  \bibfield  {author} {\bibinfo {author} {\bibfnamefont {D.}~\bibnamefont {Beaver}},\ }\bibinfo {title} {Efficient multiparty protocols using circuit randomization},\ in\ \href {https://doi.org/10.1007/3-540-46766-1_34} {\emph {\bibinfo {booktitle} {Lecture Notes in Computer Science}}}\ (\bibinfo  {publisher} {Springer Berlin Heidelberg},\ \bibinfo {year} {1992})\ p.\ \bibinfo {pages} {420–432}\BibitemShut {NoStop}%
\bibitem [{\citenamefont {Nielsen}\ \emph {et~al.}(2012)\citenamefont {Nielsen}, \citenamefont {Nordholt}, \citenamefont {Orlandi},\ and\ \citenamefont {Burra}}]{Nielsen-2012a}%
  \BibitemOpen
  \bibfield  {author} {\bibinfo {author} {\bibfnamefont {J.~B.}\ \bibnamefont {Nielsen}}, \bibinfo {author} {\bibfnamefont {P.~S.}\ \bibnamefont {Nordholt}}, \bibinfo {author} {\bibfnamefont {C.}~\bibnamefont {Orlandi}},\ and\ \bibinfo {author} {\bibfnamefont {S.~S.}\ \bibnamefont {Burra}},\ }\bibinfo {title} {A new approach to practical active-secure two-party computation},\ in\ \href {https://doi.org/10.1007/978-3-642-32009-5_40} {\emph {\bibinfo {booktitle} {Advances in Cryptology – CRYPTO 2012}}}\ (\bibinfo  {publisher} {Springer Berlin Heidelberg},\ \bibinfo {year} {2012})\ p.\ \bibinfo {pages} {681–700}\BibitemShut {NoStop}%
\bibitem [{\citenamefont {Damgård}\ \emph {et~al.}(2012)\citenamefont {Damgård}, \citenamefont {Pastro}, \citenamefont {Smart},\ and\ \citenamefont {Zakarias}}]{Damgard-2012a}%
  \BibitemOpen
  \bibfield  {author} {\bibinfo {author} {\bibfnamefont {I.}~\bibnamefont {Damgård}}, \bibinfo {author} {\bibfnamefont {V.}~\bibnamefont {Pastro}}, \bibinfo {author} {\bibfnamefont {N.}~\bibnamefont {Smart}},\ and\ \bibinfo {author} {\bibfnamefont {S.}~\bibnamefont {Zakarias}},\ }\bibinfo {title} {Multiparty computation from somewhat homomorphic encryption},\ in\ \href {https://doi.org/10.1007/978-3-642-32009-5_38} {\emph {\bibinfo {booktitle} {Advances in Cryptology – CRYPTO 2012}}}\ (\bibinfo  {publisher} {Springer Berlin Heidelberg},\ \bibinfo {year} {2012})\ p.\ \bibinfo {pages} {643–662}\BibitemShut {NoStop}%
\bibitem [{\citenamefont {Choudhury}\ and\ \citenamefont {Patra}(2017)}]{Choudhury-2017a}%
  \BibitemOpen
  \bibfield  {author} {\bibinfo {author} {\bibfnamefont {A.}~\bibnamefont {Choudhury}}\ and\ \bibinfo {author} {\bibfnamefont {A.}~\bibnamefont {Patra}},\ }\bibfield  {title} {\bibinfo {title} {An efficient framework for unconditionally secure multiparty computation},\ }\href {https://doi.org/10.1109/tit.2016.2614685} {\bibfield  {journal} {\bibinfo  {journal} {IEEE Transactions on Information Theory}\ }\textbf {\bibinfo {volume} {63}},\ \bibinfo {pages} {428–468} (\bibinfo {year} {2017})}\BibitemShut {NoStop}%
\bibitem [{\citenamefont {Takeuchi}\ \emph {et~al.}(2019)\citenamefont {Takeuchi}, \citenamefont {Mantri}, \citenamefont {Morimae}, \citenamefont {Mizutani},\ and\ \citenamefont {Fitzsimons}}]{Takeuchi-2019a}%
  \BibitemOpen
  \bibfield  {author} {\bibinfo {author} {\bibfnamefont {Y.}~\bibnamefont {Takeuchi}}, \bibinfo {author} {\bibfnamefont {A.}~\bibnamefont {Mantri}}, \bibinfo {author} {\bibfnamefont {T.}~\bibnamefont {Morimae}}, \bibinfo {author} {\bibfnamefont {A.}~\bibnamefont {Mizutani}},\ and\ \bibinfo {author} {\bibfnamefont {J.}~\bibnamefont {Fitzsimons}},\ }\bibfield  {title} {\bibinfo {title} {Resource-efficient verification of quantum computing using serfling's bound},\ }\bibfield  {journal} {\bibinfo  {journal} {npj Quantum Information}\ }\textbf {\bibinfo {volume} {5}},\ \href {https://doi.org/10.1038/s41534-019-0142-2} {10.1038/s41534-019-0142-2} (\bibinfo {year} {2019})\BibitemShut {NoStop}%
\bibitem [{\citenamefont {Unnikrishnan}\ and\ \citenamefont {Markham}(2022)}]{Unnikrishnan-2022a}%
  \BibitemOpen
  \bibfield  {author} {\bibinfo {author} {\bibfnamefont {A.}~\bibnamefont {Unnikrishnan}}\ and\ \bibinfo {author} {\bibfnamefont {D.}~\bibnamefont {Markham}},\ }\bibfield  {title} {\bibinfo {title} {Verification of graph states in an untrusted network},\ }\href {https://doi.org/10.1103/PhysRevA.105.052420} {\bibfield  {journal} {\bibinfo  {journal} {Phys. Rev. A}\ }\textbf {\bibinfo {volume} {105}},\ \bibinfo {pages} {052420} (\bibinfo {year} {2022})}\BibitemShut {NoStop}%
\bibitem [{\citenamefont {Davies}\ and\ \citenamefont {Lewis}(1970)}]{Davies-1970a}%
  \BibitemOpen
  \bibfield  {author} {\bibinfo {author} {\bibfnamefont {E.~B.}\ \bibnamefont {Davies}}\ and\ \bibinfo {author} {\bibfnamefont {J.~T.}\ \bibnamefont {Lewis}},\ }\bibfield  {title} {\bibinfo {title} {An operational approach to quantum probability},\ }\href {https://doi.org/10.1007/bf01647093} {\bibfield  {journal} {\bibinfo  {journal} {Communications in Mathematical Physics}\ }\textbf {\bibinfo {volume} {17}},\ \bibinfo {pages} {239–260} (\bibinfo {year} {1970})}\BibitemShut {NoStop}%
\bibitem [{\citenamefont {Gordon}\ \emph {et~al.}(2015)\citenamefont {Gordon}, \citenamefont {Liu},\ and\ \citenamefont {Shi}}]{Gordon-2015a}%
  \BibitemOpen
  \bibfield  {author} {\bibinfo {author} {\bibfnamefont {S.~D.}\ \bibnamefont {Gordon}}, \bibinfo {author} {\bibfnamefont {F.-H.}\ \bibnamefont {Liu}},\ and\ \bibinfo {author} {\bibfnamefont {E.}~\bibnamefont {Shi}},\ }\href {https://eprint.iacr.org/2015/371} {\bibinfo {title} {Constant-round {MPC} with fairness and guarantee of output delivery}},\ \bibinfo {howpublished} {Cryptology {ePrint} Archive, Paper 2015/371} (\bibinfo {year} {2015})\BibitemShut {NoStop}%
\bibitem [{\citenamefont {Damgård}\ \emph {et~al.}(2020)\citenamefont {Damgård}, \citenamefont {Magri}, \citenamefont {Ravi}, \citenamefont {Siniscalchi},\ and\ \citenamefont {Yakoubov}}]{Damgard-2020a}%
  \BibitemOpen
  \bibfield  {author} {\bibinfo {author} {\bibfnamefont {I.}~\bibnamefont {Damgård}}, \bibinfo {author} {\bibfnamefont {B.}~\bibnamefont {Magri}}, \bibinfo {author} {\bibfnamefont {D.}~\bibnamefont {Ravi}}, \bibinfo {author} {\bibfnamefont {L.}~\bibnamefont {Siniscalchi}},\ and\ \bibinfo {author} {\bibfnamefont {S.}~\bibnamefont {Yakoubov}},\ }\href {https://eprint.iacr.org/2020/1254} {\bibinfo {title} {Broadcast-optimal two round {MPC} with an honest majority}},\ \bibinfo {howpublished} {Cryptology {ePrint} Archive, Paper 2020/1254} (\bibinfo {year} {2020})\BibitemShut {NoStop}%
\bibitem [{\citenamefont {Hein}\ \emph {et~al.}(2004)\citenamefont {Hein}, \citenamefont {Eisert},\ and\ \citenamefont {Briegel}}]{Hein-2004a}%
  \BibitemOpen
  \bibfield  {author} {\bibinfo {author} {\bibfnamefont {M.}~\bibnamefont {Hein}}, \bibinfo {author} {\bibfnamefont {J.}~\bibnamefont {Eisert}},\ and\ \bibinfo {author} {\bibfnamefont {H.~J.}\ \bibnamefont {Briegel}},\ }\bibfield  {title} {\bibinfo {title} {Multiparty entanglement in graph states},\ }\href {https://doi.org/10.1103/PhysRevA.69.062311} {\bibfield  {journal} {\bibinfo  {journal} {Phys. Rev. A}\ }\textbf {\bibinfo {volume} {69}},\ \bibinfo {pages} {062311} (\bibinfo {year} {2004})}\BibitemShut {NoStop}%
\bibitem [{\citenamefont {Sch\"{o}n}\ \emph {et~al.}(2005)\citenamefont {Sch\"{o}n}, \citenamefont {Solano}, \citenamefont {Verstraete}, \citenamefont {Cirac},\ and\ \citenamefont {Wolf}}]{Schon-2005a}%
  \BibitemOpen
  \bibfield  {author} {\bibinfo {author} {\bibfnamefont {C.}~\bibnamefont {Sch\"{o}n}}, \bibinfo {author} {\bibfnamefont {E.}~\bibnamefont {Solano}}, \bibinfo {author} {\bibfnamefont {F.}~\bibnamefont {Verstraete}}, \bibinfo {author} {\bibfnamefont {J.~I.}\ \bibnamefont {Cirac}},\ and\ \bibinfo {author} {\bibfnamefont {M.~M.}\ \bibnamefont {Wolf}},\ }\bibfield  {title} {\bibinfo {title} {Sequential generation of entangled multiqubit states},\ }\bibfield  {journal} {\bibinfo  {journal} {Physical Review Letters}\ }\textbf {\bibinfo {volume} {95}},\ \href {https://doi.org/10.1103/physrevlett.95.110503} {10.1103/physrevlett.95.110503} (\bibinfo {year} {2005})\BibitemShut {NoStop}%
\bibitem [{\citenamefont {Wright}\ \emph {et~al.}(2019)\citenamefont {Wright}, \citenamefont {Beck}, \citenamefont {Debnath}, \citenamefont {Amini}, \citenamefont {Nam}, \citenamefont {Grzesiak}, \citenamefont {Chen}, \citenamefont {Pisenti}, \citenamefont {Chmielewski}, \citenamefont {Collins} \emph {et~al.}}]{Wright-2019a}%
  \BibitemOpen
  \bibfield  {author} {\bibinfo {author} {\bibfnamefont {K.}~\bibnamefont {Wright}}, \bibinfo {author} {\bibfnamefont {K.~M.}\ \bibnamefont {Beck}}, \bibinfo {author} {\bibfnamefont {S.}~\bibnamefont {Debnath}}, \bibinfo {author} {\bibfnamefont {J.}~\bibnamefont {Amini}}, \bibinfo {author} {\bibfnamefont {Y.}~\bibnamefont {Nam}}, \bibinfo {author} {\bibfnamefont {N.}~\bibnamefont {Grzesiak}}, \bibinfo {author} {\bibfnamefont {J.-S.}\ \bibnamefont {Chen}}, \bibinfo {author} {\bibfnamefont {N.}~\bibnamefont {Pisenti}}, \bibinfo {author} {\bibfnamefont {M.}~\bibnamefont {Chmielewski}}, \bibinfo {author} {\bibfnamefont {C.}~\bibnamefont {Collins}}, \emph {et~al.},\ }\bibfield  {title} {\bibinfo {title} {Benchmarking an 11-qubit quantum computer},\ }\href@noop {} {\bibfield  {journal} {\bibinfo  {journal} {Nature communications}\ }\textbf {\bibinfo {volume} {10}},\ \bibinfo {pages} {1} (\bibinfo {year} {2019})}\BibitemShut {NoStop}%
\bibitem [{\citenamefont {Shu}\ \emph {et~al.}(2011)\citenamefont {Shu}, \citenamefont {Chou}, \citenamefont {Kurz}, \citenamefont {Dietrich},\ and\ \citenamefont {Blinov}}]{Shu-2011a}%
  \BibitemOpen
  \bibfield  {author} {\bibinfo {author} {\bibfnamefont {G.}~\bibnamefont {Shu}}, \bibinfo {author} {\bibfnamefont {C.-K.}\ \bibnamefont {Chou}}, \bibinfo {author} {\bibfnamefont {N.}~\bibnamefont {Kurz}}, \bibinfo {author} {\bibfnamefont {M.~R.}\ \bibnamefont {Dietrich}},\ and\ \bibinfo {author} {\bibfnamefont {B.~B.}\ \bibnamefont {Blinov}},\ }\bibfield  {title} {\bibinfo {title} {Efficient fluorescence collection and ion imaging with the ``tack'' ion trap},\ }\href {https://doi.org/10.1364/JOSAB.28.002865} {\bibfield  {journal} {\bibinfo  {journal} {J. Opt. Soc. Am. B}\ }\textbf {\bibinfo {volume} {28}},\ \bibinfo {pages} {2865} (\bibinfo {year} {2011})}\BibitemShut {NoStop}%
\bibitem [{\citenamefont {Maiwald}\ \emph {et~al.}(2012)\citenamefont {Maiwald}, \citenamefont {Golla}, \citenamefont {Fischer}, \citenamefont {Bader}, \citenamefont {Heugel}, \citenamefont {Chalopin}, \citenamefont {Sondermann},\ and\ \citenamefont {Leuchs}}]{Maiwald-2012a}%
  \BibitemOpen
  \bibfield  {author} {\bibinfo {author} {\bibfnamefont {R.}~\bibnamefont {Maiwald}}, \bibinfo {author} {\bibfnamefont {A.}~\bibnamefont {Golla}}, \bibinfo {author} {\bibfnamefont {M.}~\bibnamefont {Fischer}}, \bibinfo {author} {\bibfnamefont {M.}~\bibnamefont {Bader}}, \bibinfo {author} {\bibfnamefont {S.}~\bibnamefont {Heugel}}, \bibinfo {author} {\bibfnamefont {B.}~\bibnamefont {Chalopin}}, \bibinfo {author} {\bibfnamefont {M.}~\bibnamefont {Sondermann}},\ and\ \bibinfo {author} {\bibfnamefont {G.}~\bibnamefont {Leuchs}},\ }\bibfield  {title} {\bibinfo {title} {Collecting more than half the fluorescence photons from a single ion},\ }\href@noop {} {\bibfield  {journal} {\bibinfo  {journal} {Physical Review A}\ }\textbf {\bibinfo {volume} {86}},\ \bibinfo {pages} {043431} (\bibinfo {year} {2012})}\BibitemShut {NoStop}%
\bibitem [{\citenamefont {Chou}\ \emph {et~al.}(2017)\citenamefont {Chou}, \citenamefont {Auchter}, \citenamefont {Lilieholm}, \citenamefont {Smith},\ and\ \citenamefont {Blinov}}]{Chou-2017a}%
  \BibitemOpen
  \bibfield  {author} {\bibinfo {author} {\bibfnamefont {C.-K.}\ \bibnamefont {Chou}}, \bibinfo {author} {\bibfnamefont {C.}~\bibnamefont {Auchter}}, \bibinfo {author} {\bibfnamefont {J.}~\bibnamefont {Lilieholm}}, \bibinfo {author} {\bibfnamefont {K.}~\bibnamefont {Smith}},\ and\ \bibinfo {author} {\bibfnamefont {B.}~\bibnamefont {Blinov}},\ }\bibfield  {title} {\bibinfo {title} {Note: Single ion imaging and fluorescence collection with a parabolic mirror trap},\ }\href@noop {} {\bibfield  {journal} {\bibinfo  {journal} {Review of Scientific Instruments}\ }\textbf {\bibinfo {volume} {88}} (\bibinfo {year} {2017})}\BibitemShut {NoStop}%
\bibitem [{\citenamefont {Braunstein}\ and\ \citenamefont {Mann}(1995)}]{Braunstein-1995a}%
  \BibitemOpen
  \bibfield  {author} {\bibinfo {author} {\bibfnamefont {S.~L.}\ \bibnamefont {Braunstein}}\ and\ \bibinfo {author} {\bibfnamefont {A.}~\bibnamefont {Mann}},\ }\bibfield  {title} {\bibinfo {title} {Measurement of the bell operator and quantum teleportation},\ }\href {https://doi.org/10.1103/PhysRevA.51.R1727} {\bibfield  {journal} {\bibinfo  {journal} {Phys. Rev. A}\ }\textbf {\bibinfo {volume} {51}},\ \bibinfo {pages} {R1727} (\bibinfo {year} {1995})}\BibitemShut {NoStop}%
\bibitem [{\citenamefont {Lamas-Linares}\ \emph {et~al.}(2001)\citenamefont {Lamas-Linares}, \citenamefont {Howell},\ and\ \citenamefont {Bouwmeester}}]{Lamas-Linares-2001a}%
  \BibitemOpen
  \bibfield  {author} {\bibinfo {author} {\bibfnamefont {A.}~\bibnamefont {Lamas-Linares}}, \bibinfo {author} {\bibfnamefont {J.~C.}\ \bibnamefont {Howell}},\ and\ \bibinfo {author} {\bibfnamefont {D.}~\bibnamefont {Bouwmeester}},\ }\bibfield  {title} {\bibinfo {title} {Stimulated emission of polarization-entangled photons},\ }\href {https://doi.org/10.1038/35091014} {\bibfield  {journal} {\bibinfo  {journal} {Nature}\ }\textbf {\bibinfo {volume} {412}},\ \bibinfo {pages} {887} (\bibinfo {year} {2001})}\BibitemShut {NoStop}%
\bibitem [{\citenamefont {Kok}\ and\ \citenamefont {Braunstein}(2000)}]{Kok-2000a}%
  \BibitemOpen
  \bibfield  {author} {\bibinfo {author} {\bibfnamefont {P.}~\bibnamefont {Kok}}\ and\ \bibinfo {author} {\bibfnamefont {S.~L.}\ \bibnamefont {Braunstein}},\ }\bibfield  {title} {\bibinfo {title} {Postselected versus nonpostselected quantum teleportation using parametric down-conversion},\ }\href {https://doi.org/10.1103/PhysRevA.61.042304} {\bibfield  {journal} {\bibinfo  {journal} {Phys. Rev. A}\ }\textbf {\bibinfo {volume} {61}},\ \bibinfo {pages} {042304} (\bibinfo {year} {2000})}\BibitemShut {NoStop}%
\bibitem [{\citenamefont {Leonhardt}\ and\ \citenamefont {Paul}(1995)}]{Leonhardt-1995a}%
  \BibitemOpen
  \bibfield  {author} {\bibinfo {author} {\bibfnamefont {U.}~\bibnamefont {Leonhardt}}\ and\ \bibinfo {author} {\bibfnamefont {H.}~\bibnamefont {Paul}},\ }\bibfield  {title} {\bibinfo {title} {Measuring the quantum state of light},\ }\href {https://doi.org/https://doi.org/10.1016/0079-6727(94)00007-L} {\bibfield  {journal} {\bibinfo  {journal} {Progress in Quantum Electronics}\ }\textbf {\bibinfo {volume} {19}},\ \bibinfo {pages} {89} (\bibinfo {year} {1995})}\BibitemShut {NoStop}%
\bibitem [{\citenamefont {Shah}\ \emph {et~al.}(2022)\citenamefont {Shah}, \citenamefont {Isleif}, \citenamefont {Januschek}, \citenamefont {Lindner},\ and\ \citenamefont {Schott}}]{Shah-2022a}%
  \BibitemOpen
  \bibfield  {author} {\bibinfo {author} {\bibfnamefont {R.}~\bibnamefont {Shah}}, \bibinfo {author} {\bibfnamefont {K.-S.}\ \bibnamefont {Isleif}}, \bibinfo {author} {\bibfnamefont {F.}~\bibnamefont {Januschek}}, \bibinfo {author} {\bibfnamefont {A.}~\bibnamefont {Lindner}},\ and\ \bibinfo {author} {\bibfnamefont {M.}~\bibnamefont {Schott}},\ }\bibfield  {title} {\bibinfo {title} {Characterising a single-photon detector for alps ii},\ }\href {https://doi.org/10.1007/s10909-022-02720-0} {\bibfield  {journal} {\bibinfo  {journal} {Journal of Low Temperature Physics}\ }\textbf {\bibinfo {volume} {209}},\ \bibinfo {pages} {355–362} (\bibinfo {year} {2022})}\BibitemShut {NoStop}%
\bibitem [{\citenamefont {Lee}\ \emph {et~al.}(2004)\citenamefont {Lee}, \citenamefont {Yurtsever}, \citenamefont {Kok}, \citenamefont {Hockney}, \citenamefont {Adami}, \citenamefont {Braunstein},\ and\ \citenamefont {Dowling}}]{Lee-2004a}%
  \BibitemOpen
  \bibfield  {author} {\bibinfo {author} {\bibfnamefont {H.}~\bibnamefont {Lee}}, \bibinfo {author} {\bibfnamefont {U.}~\bibnamefont {Yurtsever}}, \bibinfo {author} {\bibfnamefont {P.}~\bibnamefont {Kok}}, \bibinfo {author} {\bibfnamefont {G.~M.}\ \bibnamefont {Hockney}}, \bibinfo {author} {\bibfnamefont {C.}~\bibnamefont {Adami}}, \bibinfo {author} {\bibfnamefont {S.~L.}\ \bibnamefont {Braunstein}},\ and\ \bibinfo {author} {\bibfnamefont {J.~P.}\ \bibnamefont {Dowling}},\ }\bibfield  {title} {\bibinfo {title} {Towards photostatistics from photon-number discriminating detectors},\ }\href {https://doi.org/10.1080/09500340408235289} {\bibfield  {journal} {\bibinfo  {journal} {Journal of Modern Optics}\ }\textbf {\bibinfo {volume} {51}},\ \bibinfo {pages} {1517} (\bibinfo {year} {2004})}\BibitemShut {NoStop}%
\bibitem [{\citenamefont {Sch{\"a}fer}\ \emph {et~al.}(2018)\citenamefont {Sch{\"a}fer}, \citenamefont {Ballance}, \citenamefont {Thirumalai}, \citenamefont {Stephenson}, \citenamefont {Ballance}, \citenamefont {Steane},\ and\ \citenamefont {Lucas}}]{Schafer-2018a}%
  \BibitemOpen
  \bibfield  {author} {\bibinfo {author} {\bibfnamefont {V.}~\bibnamefont {Sch{\"a}fer}}, \bibinfo {author} {\bibfnamefont {C.}~\bibnamefont {Ballance}}, \bibinfo {author} {\bibfnamefont {K.}~\bibnamefont {Thirumalai}}, \bibinfo {author} {\bibfnamefont {L.}~\bibnamefont {Stephenson}}, \bibinfo {author} {\bibfnamefont {T.}~\bibnamefont {Ballance}}, \bibinfo {author} {\bibfnamefont {A.}~\bibnamefont {Steane}},\ and\ \bibinfo {author} {\bibfnamefont {D.}~\bibnamefont {Lucas}},\ }\bibfield  {title} {\bibinfo {title} {Fast quantum logic gates with trapped-ion qubits},\ }\href@noop {} {\bibfield  {journal} {\bibinfo  {journal} {Nature}\ }\textbf {\bibinfo {volume} {555}},\ \bibinfo {pages} {75} (\bibinfo {year} {2018})}\BibitemShut {NoStop}%
\bibitem [{\citenamefont {Rangarajan}\ \emph {et~al.}(2009)\citenamefont {Rangarajan}, \citenamefont {Goggin},\ and\ \citenamefont {Kwiat}}]{Rangarajan-2009a}%
  \BibitemOpen
  \bibfield  {author} {\bibinfo {author} {\bibfnamefont {R.}~\bibnamefont {Rangarajan}}, \bibinfo {author} {\bibfnamefont {M.}~\bibnamefont {Goggin}},\ and\ \bibinfo {author} {\bibfnamefont {P.}~\bibnamefont {Kwiat}},\ }\bibfield  {title} {\bibinfo {title} {Optimizing type-i polarization-entangled photons},\ }\href {https://doi.org/10.1364/OE.17.018920} {\bibfield  {journal} {\bibinfo  {journal} {Opt. Express}\ }\textbf {\bibinfo {volume} {17}},\ \bibinfo {pages} {18920} (\bibinfo {year} {2009})}\BibitemShut {NoStop}%
\bibitem [{\citenamefont {Ramelow}\ \emph {et~al.}(2013)\citenamefont {Ramelow}, \citenamefont {Mech}, \citenamefont {Giustina}, \citenamefont {Gr{\"o}blacher}, \citenamefont {Wieczorek}, \citenamefont {Beyer}, \citenamefont {Lita}, \citenamefont {Calkins}, \citenamefont {Gerrits}, \citenamefont {Nam} \emph {et~al.}}]{Ramelow-2013a}%
  \BibitemOpen
  \bibfield  {author} {\bibinfo {author} {\bibfnamefont {S.}~\bibnamefont {Ramelow}}, \bibinfo {author} {\bibfnamefont {A.}~\bibnamefont {Mech}}, \bibinfo {author} {\bibfnamefont {M.}~\bibnamefont {Giustina}}, \bibinfo {author} {\bibfnamefont {S.}~\bibnamefont {Gr{\"o}blacher}}, \bibinfo {author} {\bibfnamefont {W.}~\bibnamefont {Wieczorek}}, \bibinfo {author} {\bibfnamefont {J.}~\bibnamefont {Beyer}}, \bibinfo {author} {\bibfnamefont {A.}~\bibnamefont {Lita}}, \bibinfo {author} {\bibfnamefont {B.}~\bibnamefont {Calkins}}, \bibinfo {author} {\bibfnamefont {T.}~\bibnamefont {Gerrits}}, \bibinfo {author} {\bibfnamefont {S.~W.}\ \bibnamefont {Nam}}, \emph {et~al.},\ }\bibfield  {title} {\bibinfo {title} {Highly efficient heralding of entangled single photons},\ }\href@noop {} {\bibfield  {journal} {\bibinfo  {journal} {Optics express}\ }\textbf {\bibinfo {volume} {21}},\ \bibinfo {pages} {6707} (\bibinfo {year} {2013})}\BibitemShut {NoStop}%
\bibitem [{\citenamefont {Baccari}\ \emph {et~al.}(2020)\citenamefont {Baccari}, \citenamefont {Augusiak}, \citenamefont {Šupić}, \citenamefont {Tura},\ and\ \citenamefont {Acín}}]{Baccari-2020a}%
  \BibitemOpen
  \bibfield  {author} {\bibinfo {author} {\bibfnamefont {F.}~\bibnamefont {Baccari}}, \bibinfo {author} {\bibfnamefont {R.}~\bibnamefont {Augusiak}}, \bibinfo {author} {\bibfnamefont {I.}~\bibnamefont {Šupić}}, \bibinfo {author} {\bibfnamefont {J.}~\bibnamefont {Tura}},\ and\ \bibinfo {author} {\bibfnamefont {A.}~\bibnamefont {Acín}},\ }\bibfield  {title} {\bibinfo {title} {Scalable bell inequalities for qubit graph states and robust self-testing},\ }\bibfield  {journal} {\bibinfo  {journal} {Physical Review Letters}\ }\textbf {\bibinfo {volume} {124}},\ \href {https://doi.org/10.1103/physrevlett.124.020402} {10.1103/physrevlett.124.020402} (\bibinfo {year} {2020})\BibitemShut {NoStop}%
\bibitem [{\citenamefont {Yang}\ and\ \citenamefont {Hwang}(2013)}]{Yang-2013a}%
  \BibitemOpen
  \bibfield  {author} {\bibinfo {author} {\bibfnamefont {C.-W.}\ \bibnamefont {Yang}}\ and\ \bibinfo {author} {\bibfnamefont {T.}~\bibnamefont {Hwang}},\ }\bibfield  {title} {\bibinfo {title} {Fault tolerant quantum key distributions using entanglement swapping of ghz states over collective-noise channels},\ }\href {https://doi.org/10.1007/s11128-013-0593-x} {\bibfield  {journal} {\bibinfo  {journal} {Quantum Information Processing}\ }\textbf {\bibinfo {volume} {12}},\ \bibinfo {pages} {3207–3222} (\bibinfo {year} {2013})}\BibitemShut {NoStop}%
\bibitem [{\citenamefont {Xu}\ \emph {et~al.}(2014)\citenamefont {Xu}, \citenamefont {Wen}, \citenamefont {Gao},\ and\ \citenamefont {Qin}}]{Xu-2014a}%
  \BibitemOpen
  \bibfield  {author} {\bibinfo {author} {\bibfnamefont {G.-B.}\ \bibnamefont {Xu}}, \bibinfo {author} {\bibfnamefont {Q.-Y.}\ \bibnamefont {Wen}}, \bibinfo {author} {\bibfnamefont {F.}~\bibnamefont {Gao}},\ and\ \bibinfo {author} {\bibfnamefont {S.-J.}\ \bibnamefont {Qin}},\ }\bibfield  {title} {\bibinfo {title} {Novel multiparty quantum key agreement protocol with ghz states},\ }\href {https://doi.org/10.1007/s11128-014-0816-9} {\bibfield  {journal} {\bibinfo  {journal} {Quantum Information Processing}\ }\textbf {\bibinfo {volume} {13}},\ \bibinfo {pages} {2587–2594} (\bibinfo {year} {2014})}\BibitemShut {NoStop}%
\end{thebibliography}%

\end{document}